\documentclass[useAMS,usenatbib]{mn2e}

\usepackage{lscape}
\usepackage{epsfig}
\usepackage{eufrak}
\usepackage{rotating}

\title[A Comparison of Magellanic Cloud Star Cluster colours with SSP
  Model Predictions]{A Comparison of Optical and Near-Infrared Colours
    of Magellanic Cloud Star Clusters with Predictions of Simple
    Stellar Population Models}

\author[P. Pessev et al.]{P. M. Pessev$^{1}$\thanks{E-mails:
    pessev@stsci.edu (PMP); goudfroo@stsci.edu (PG); puziat@nrc.ca (THP); rupali.chandar@utoledo.edu (RC)}, P. Goudfrooij$^{1\star}$, T. H. Puzia$^{2\star}$ and R. Chandar$^{3,4\star}$\\ 
$^{1}$Space Telescope Science Institute, 3700 San Martin Drive,
 Baltimore, MD 21218, U.S.A.\\ 
$^{2}$Herzberg Institute of Astrophysics, 5071 West Saanich Road,
 Victoria, BC V9E 2E7, Canada\\ 
$^{3}$The Observatories of the Carnegie Institution of Washington, 813
 Santa Barbara Street, Pasadena, CA 91101-1292, U.S.A.\\ 
$^{4}$Department of Physics and Astronomy, The University of Toledo,
 2801 West Bancroft Street, Toledo, OH 43606, U.S.A.\\
}
\begin{document}

\date{Accepted .... Received ...; in original form ...}

\pagerange{\pageref{firstpage}--\pageref{lastpage}} \pubyear{2007}

\maketitle

\label{firstpage}

\begin{abstract}
We present integrated $JHK_s$ 2MASS photometry and a compilation of
integrated-light optical photoelectric measurements for 84 star
clusters in the Magellanic Clouds.  These clusters range in age from
$\approx200$~Myr to $>10$~Gyr, and have [Fe/H] values from $-2.2$ to
$-0.1$ dex.  We find a spread in the intrinsic colours of clusters
with similar ages and metallicities, at least some of which is due to
stochastic fluctuations in the number of bright stars residing in
low-mass clusters.  We use 54 clusters with the most reliable age and
metallicity estimates as test particles to evaluate the performance of
four widely used SSP models in the optical/NIR colour-colour space.
All models reproduce the reddening-corrected colours of the old ($\ge$
10 Gyr) globular clusters quite well, but model performance varies at
younger ages.  In order to account for the effects of stochastic
fluctuations in individual clusters, we provide composite $B-V$,
$B-J$, $V-J$, $V-K_s$ and $J-K_s$ colours for Magellanic Cloud
clusters in several different age intervals. 
The accumulated mass for most composite clusters are higher than
that needed to keep luminosity variations due to stochastic
fluctuations below the 10\% level.
The colours of the composite clusters are clearly distinct in
optical-NIR colour-colour space for the following intervals of age:
$>10$~Gyr, $2-9$~Gyr, $1-2$~Gyr, and $200$~Myr$-1$~Gyr.  This suggests
that a combination of optical plus NIR colours can be used to
differentiate clusters of different age and metallicity.
\end{abstract}


%%%%%%%%%%%%%%%%%%%%%%%%%%%%%%%%%%%%%%%%%%%%%%%%%%
\section{Introduction}

The most efficient method to determine the age and metallicity for
unresolved stellar systems (especially at high redshift) is by comparing
their observed colours with the predictions of evolution synthesis models
\citep[e.g.][]{bc93, bc03, worthey94, vazdekis99, maraston98, maraston05,
af03}. Thus, it is important to test the integrated colours predicted by
recent models, based on objects which have accurate ages and metallicities
determined independently. In the present paper we focus our attention on
the combination of visual and near-infrared (NIR) photometry, which has
proven to be important for breaking the age-metallicity degeneracy,
particularly in stellar populations older than 
$\approx {\rmn{a\ few\ times}} \times 100$~Myr
\citep[e.g.][]{goudfrooij01, puzia02, hempel04}.

With the advent of the {\it Spitzer Space Telescope (Spitzer)\/} and
mid-infrared (MIR) instrumentation for some large ground-based telescopes,
the NIR spectral region is now accessible at intermediate-to-high redshifts. 
In a recent paper based on {\it Spitzer\/} Infrared Array Camera
(IRAC) imaging, \cite{vanderwel2006} reported significant discrepancies
between some model predictions and the observed rest-frame $K$-band
properties of early-type galaxies at z $\approx 1$. Their results show
that the interpretation of NIR photometry is hampered by model
uncertainties. 
As a consequence the determination of masses of distant
stellar systems based on such data can have uncertainties up to a factor
2.5 \citep[see][]{bruzual07}.

Unfortunately, providing accurate model predictions in the near-infrared
is challenging, since there are limitations imposed by the current lack of
understanding of certain stages of stellar evolution (e.g., thermally
pulsing asymptotic giant branch, or TP-AGB stars). These objects 
significantly affect the spectral energy distribution (SED) in the NIR and
MIR for stellar populations with ages between $\approx200$ Myr to 3 Gyr. 
Another possible complication is that the stellar
libraries used by population synthesis models contain mostly stars from
the solar neighborhood.  These stars have a star formation history which
is not necessarily typical for extragalactic populations (e.g. relatively
little variation of [$\alpha$/Fe] ratios), and there are only a very
limited number of AGB spectra available.

The Large and Small Magellanic Clouds (LMC and SMC respectively) provide
a unique opportunity to test the accuracy of most current SSP
models, since they contain a significant population of intermediate-age 
massive star clusters which are not easily accessible in our Galaxy. 
The ages and
metallicities of these star clusters can be determined from deep
colour-magnitude diagrams (CMDs) reaching below the main sequence turn-off
(MSTO)\footnote{Obtaining photometry with sufficient quality to secure reliable age and metallicity estimates for clusters in galaxies beyond the Magellanic Clouds requires a significant investment of observing time. To date only one such cluster,  SKHB~312 in M31, has a CMD deep enough to probe the MSTO region \citep{brown2004}. The photometry for this object was obtained as a result of a program utilizing 126 {\it Hubble Space Telescope (HST)\/} orbits.
}. Medium and high-resolution spectroscopy of individual giants in these clusters also provides independent metallicity estimates.
Therefore their integrated-light properties (easily observed with small
and moderate-aperture telescopes) can be combined with the accurate
age/metallicity measurements and used to test (and calibrate) the SSP
models. 

In \cite{pessev2006} (hereafter Paper~I) we used the Two Micron All Sky
Survey (2MASS; \cite{skrutskie2006}) to derive NIR $(JHK_s)$
integrated-light magnitudes and colours for a large sample of Magellanic
Cloud star clusters, based on a homogeneous, accurately calibrated
dataset. In the present study we use the sample from Paper~I and new
photometry for 9 additional objects (forming the largest dataset of integrated
NIR magnitudes and colours of LMC/SMC star clusters to date) to test the
performance of several SSP models. We combine the 2MASS data with optical
photometry originating from the work of \cite{bica_et_al_96} and the
compilation of \cite{vdb81}. The technique adopted in Paper~I - measuring
$JHK_s$ curves of growth to large radii allows us to utilize rather
heterogeneous databases of optical photometry, usually performed with a
set of fixed apertures. We use 54 clusters from our sample as ``test
particles''. These clusters were chosen to have reliable age and
metallicity measurements, covering a wide parameter space.

\setcounter{figure}{0}
\begin{figure} 
\centering
\includegraphics[bb=14 14 256 256,width=8.15cm]{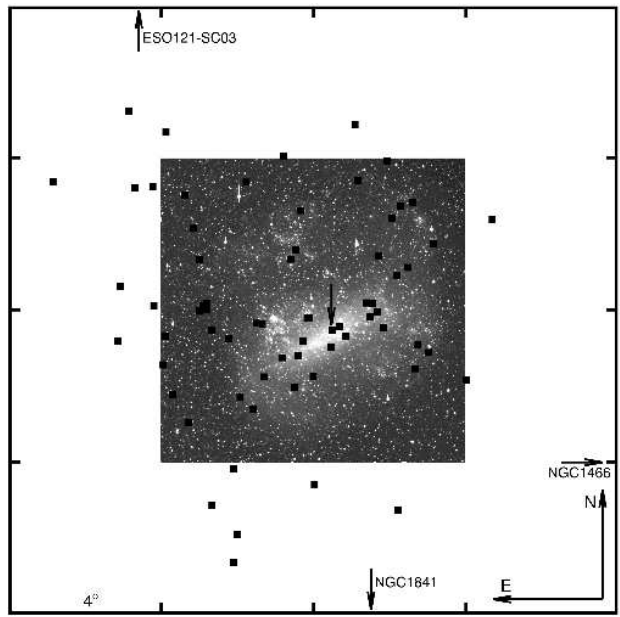}
\caption{
A finding chart of the LMC showing the clusters in our sample. The $R$-band
image is centred on $\alpha_{2000} = 05^{h} 26^{m} 37.7^{s}$ and
$\delta_{2000} = -68^{o} 56' 57.5"$. The arrow near
the centre outlines the position of NGC~1928 ($\alpha_{2000} = 05^{h}
20^{m} 57.7^{s}$ and $\delta_{2000} = -69^{\rmn{o}} 28' 40.2"$).  This cluster
is located close to the geometrical centre of the LMC bar, and was adopted
by Bica et al.\ (1996) as a reference point for the relative coordinates of
the LMC cluster system. The labeled arrows show the direction towards the
clusters lying outside the boundaries of this chart. The $R$-band image
(G. Bothun 1997, private communication) covers $8^{\rmn{o}}$x$8^{\rmn{o}}$, 
while the dimensions of the chart are $16^{\rmn{o}}$x$16^{\rmn{o}}$.}
\label{fig:lmc_map}
\end{figure}

\setcounter{figure}{1}
\begin{figure} 
\centering
\includegraphics[bb=14 14 256 256,width=8.15cm]{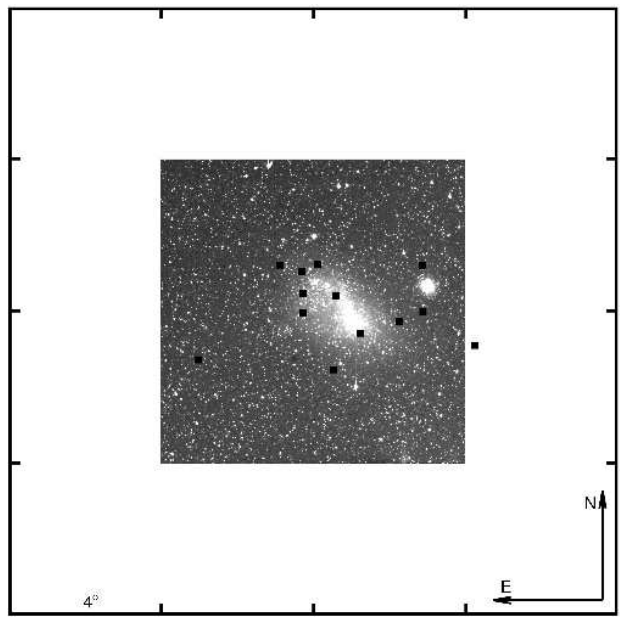}
\caption{A finding chart of the SMC showing the clusters in our sample. The $R$-band
image (G. Bothun 1997, private communication) is centred 
on $\alpha_{2000} = 01^{h} 04^{m} 42.8^{s}$ and
$\delta_{2000} = -72^{\rmn{o}} 52' 32.4''$. SMC clusters cover a smaller area than LMC objects. There is only one cluster outside the $R$ frame, but for illustrational purposes
the dimensions of this finder chart are identical to those of the LMC chart in
Figure~\ref{fig:lmc_map}.}
\label{fig:smc_map}
\end{figure}

This paper is organized as follows: in \S 2 we define our extended sample
and present the new photometry along with the compilation of visual
magnitudes and colours. Four sets of SSP models are tested in \S 3,
followed by concluding remarks in \S 4. Information about the
properties of the cluster sample is presented in Appendices A and B. Transformations between the model grids in the 
\cite{bb88} system and the photometric system of 2 MASS are provided in Appendix C.
\label{intro}

%%%%%%%%%%%%%%%%%%%%%%%%%%%%%%%%%%%%%%%%%%%%%%%%%%%
\section{Photometry of Magellanic Cloud Clusters}
\subsection[]{Extending the Sample - New Near-Infrared Integrated Photometry of Magellanic Cloud Clusters}

\setcounter{table}{0}
\begin{table*}
\centering
\begin{minipage}{140mm}
\caption{Extended Magellanic Cloud Cluster Sample}
\label{tab:ext}
\begin{tabular}{@{}lccrlllllllll@{}}
\hline
ID & $\alpha_{2000}$ & $\delta_{2000}$ & \multicolumn{4}{c}{Age, Errors \& Ref}&\multicolumn{3}{c}{[Fe/H], Err \& Ref} & $A_v$ & $\Delta A_v$ \\
(1) & (2) & (3) & (4) & (5) & (6) & (7) & (8) & (9) & (10) & (11) & (12)\\
\hline

NGC265        &   00:47:12  & --73:28:38  &  8.5\hspace{4pt} & +0.3   &--0.3   &  2  & --0.62  & $^{+0.23}_{-0.61}$  &  2  & 0.34 & 0.02\\[3pt]
NGC1644      &   04:37:39  &--66:12:00  &  9.53 & +0.05   &--0.05   &  1  & --1.4   &$\pm$0.2    &         1       & 0.39 & 0.02\\[3pt]
NGC1928      &   05:20:58  &--69:28:40  & 10.11& +0.06   &--0.08   &  5  & --1.27 &$\pm$0.14    &       5   & 0.34 & 0.06\\[3pt]
NGC1994      &   05:28:22  &--69:08:30  &  7.06 &     &   &  4&   &   &          & 0.41 & 0.02\\[3pt]
NGC2058      &   05:36:55  &--70:09:44  &  7.85 & +0.10   &--0.15   &  4  &    &  &       & 0.39 & 0.02\\[3pt]
NGC2107      &   05:43:13  &--70:38:23  &  8.55 & +0.25   &--0.24   &  3  &     &  &        & 0.36 & 0.04\\[3pt]
NGC2108      &   05:43:56  &--69:10:48  &  8.90 & +0.26   &--0.26   &  3 &     &   &       & 0.50 & 0.05\\[3pt]
NGC2134      &   05:51:56  &--71:05:54  &  8.27 &   &   &  7  & --0.4   & &         7       & 0.62 & 0.03\\[3pt]
NGC2154      &   05:57:38  &--67:15:42  &  9.16 & +0.28   &--0.28   &  3  & --0.56   &$\pm$0.2     &         6       & 0.39 & 0.03\\[5pt]
\hline
\end{tabular}
{\it Notes to Table~\ref{tab:ext}:} Column~(1): Object ID. Columns~(2) and (3): Cluster coordinates - Right Ascension (given as hours, minutes, seconds) and Declination (degrees, minutes, seconds) in J2000 retrieved from Simbad Astronomical Database. Column~(4): age of the object given as {\it log(age)} with corresponding errors in columns~(5) and (6) and literature references (see bellow) in (7). Column~(8): metallicity values retrieved from the literature with their errors (9) and references (10). Column~(11): $V$ band reddening values for the objects with corresponding errors (12) retrieved from the MCPS reddening estimation tool.

{\it References:} (1) \cite{bica86} (2) \cite{chiosi07} (3) \cite{elson_fall_85} (4) \cite{elson_fall_88} (5) \cite{mg04} (6) \cite{olszewski91} (7) \cite{vallenari94}

\end{minipage}
\end{table*}

\setcounter{table}{1}
\begin{table*}
\centering
\begin{minipage}{140mm}
\caption{2MASS Atlas Images of the Clusters in the Extended Sample.}
\label{tab:sampleAtl}
\begin{tabular}{@{}lcccc@{}}
\hline
ID & N & $J$ & $H$ & $K_s$\\
(1) & (2) & (3) & (4) & (5)\\
\hline
NGC265	&	1	& aJ\_asky\_980809s0810198.fits	&	aH\_asky\_980809s0810198.fits	&	aK\_asky\_980809s0810198.fits\\
NGC1644	&	1	& aJ\_asky\_991026s1140257.fits	&	aH\_asky\_991026s1140257.fits	&	aK\_asky\_991026s1140257.fits\\
NGC1928	&	1	& aJ\_asky\_981220s0850162.fits	&	aH\_asky\_981220s0850162.fits	&	aK\_asky\_981220s0850162.fits\\
NGC1994	&	1	& aJ\_asky\_000212s0190150.fits	&	aH\_asky\_000212s0190150.fits	&	aK\_asky\_000212s0190150.fits\\
NGC2058	&	1	& aJ\_asky\_000206s0240186.fits	&	aH\_asky\_000206s0240186.fits	&	aK\_asky\_000206s0240186.fits\\
NGC2107	&	1	& aJ\_asky\_980321s0080209.fits	&	aH\_asky\_980321s0080209.fits	&	aK\_asky\_980321s0080209.fits\\
NGC2108	&	1	& aJ\_asky\_980321s0080150.fits	&	aH\_asky\_980321s0080150.fits	&	aK\_asky\_980321s0080150.fits\\
NGC2134	&	1	& aJ\_asky\_981025s1000044.fits	&	aH\_asky\_981025s1000044.fits	&	aK\_asky\_981025s1000044.fits\\
NGC2154	&	1	& aJ\_asky\_981025s1110068.fits	&	aH\_asky\_981025s1110068.fits	&	aK\_asky\_981025s1110068.fits\\
\hline
\end{tabular}
{\it Notes to Table~\ref{tab:sampleAtl}:} Column~(1): Object ID. Column~(2): Number of image sets retrieved. Columns~(3), (4) and (5): designations of the individual $J,H$ and $K_s$ frames.
\end{minipage}
\end{table*}

We selected nine objects (see Table~\ref{tab:ext} for details) to add to our
original sample which was presented in Paper~I. 
The total sample now includes all of the old clusters in the Magellanic Clouds (with the
exception of the Reticulum cluster, which is excluded due to insufficient
depth of the available 2MASS images) and all of the "NIR-enhanced"
clusters from \cite{persson83}. Some objects were included in this
extended sample because they have new integrated-light optical magnitudes
and colours available. Overall the sample provides optimal coverage of the
age-metallicity parameter space of MC star clusters.  We intend to add
2MASS $JHK_s$ integrated photometry for additional clusters as new
information based on deep CMDs becomes available in the future.

The 2MASS \citep{skrutskie2006} atlas images were recovered through the
interactive image service available on the survey
web-page\footnote{http://irsa.ipac.caltech.edu/applications/2MASS/IM/inter- active.html} 
(see Table~\ref{tab:sampleAtl} for information about the
utilized images).  We analysed these data following the reduction scheme
presented in Paper~I. Extinction corrections were derived using the online
tools provided by the Magellanic Clouds Photometric Survey
\citep[MCPS,][]{zaritsky97}; details about extinction estimates for SMC and LMC are
provided in \cite{zaritsky02} and \cite{zaritsky04} respectively. ($A_v$ values and their uncertainties are listed in the last two columns of Table~\ref{tab:ext}.)
  
All of the MC clusters with $JHK_s$ NIR 2MASS photometry from Paper~I and
the present work are shown in the finder charts (see
Figures~\ref{fig:lmc_map} and \ref{fig:smc_map} for the LMC and SMC,
respectively). The cluster positions are marked by the centres of our
aperture sets, derived as described in Paper~I. Table~\ref{tab:phot}
contains $JHK_s$ photometry for the 9 new clusters\footnote{The listed
magnitudes are not corrected for reddening.}.
\label{samplext}

\setcounter{table}{2}
\begin{table*}
\centering
\begin{minipage}{150mm}
\caption{New NIR photometry of Magellanic Cloud clusters}
\label{tab:phot}
\begin{tabular}{@{}lccccrlrrrrrr@{}}
\hline
ID & $\alpha_{2000}$ & $\delta_{2000} $ & d & Flag & D & & $J$ & $Jerr$ & $H$ & $Herr$ & $K_s$ & ${K_s}err$\\
(1) & (2) & (3) & (4) & (5) & (6) &  & (7) & (8) & (9) & (10) & (11) & (12)\\
\hline
NGC265	&	00:47:09.9	&	-73:28:39.3	&	7.5	&	AAB &	  20  & & 13.00  &  0.03 &  12.43  &  0.03 &  12.53  &  0.05\\
        &	 		 	&				&		&		&	  40  & & 11.80  &  0.03 &  10.48  &  0.02 &  10.41  &  0.02\\
        &	 		 	&				&		&		&	  60  & & 11.54  &  0.06 &  10.31  &  0.03 &  10.27  &  0.05\\
        &	 		 	&				&		&		&	  80  & & 11.35  &  0.08 &  10.15  &  0.05 &  10.04  &  0.06\\
      	&				&				&		&		&    100  & & 11.16  &  0.11 &  10.05  &  0.07 &   9.92  &  0.09\\
      	&				&				&		&		&    120  & & 10.96  &  0.13 &   9.90  &  0.09 &   9.79  &  0.12\\
      	&				&				&		&		&    140   & & 10.78  &  0.15 &   9.84  &  0.12 &   9.71  &  0.15\\
      	&				&				&		&		&    160 &: & 10.42  &  0.14 &   9.61  &  0.12 &   9.64  &  0.18\\
      	&				&				&		&		&    180 &: & 10.32  &  0.17 &   9.52  &  0.15 &   9.50  &  0.21\\
      	&				&				&		&		&    200 &:  & 10.18  &  0.19 &   9.44  &  0.17 &   9.42  &  0.24\\[5pt]
NGC1644	&	04:37:39.8	&	-66:11:55.5	&	6.5	&	BBB &     20  & & 12.08  &  0.03 &  11.62  &  0.02 &  11.43  &  0.02\\
        &				&				&		&		&     40  & & 11.51  &  0.06 &  11.05  &  0.05 &  10.84  &  0.05\\
        &				&				&		&		&     60  & & 11.31  &  0.11 &  10.88  &  0.10 &  10.76  &  0.10\\
        &				&				&	   	&	   	&     80 &:  & 11.12  &  0.17 &  10.74  &  0.16 &  10.61  &  0.16\\
        &				&				&		&		&     90 &:  & 11.05  &  0.20 &  10.66  &  0.18 &  10.51  &  0.18\\[5pt]
 NGC1928	&	05:20:57.8	&	-69:28:41.2	&	1	&	AAB &     20  & & 11.65  &  0.03 &  11.05  &  0.03 &  11.24  &  0.04\\
        &				&				&		&		&     40  & & 11.04  &  0.05 &  10.51  &  0.06 &  10.56  &  0.09\\
        &				&				&		&		&     60  & & 10.73  &  0.08 &  10.28  &  0.11 &  10.25  &  0.15\\
       	&				&				&		&	    &     80  & & 10.54  &  0.12 &  10.10  &  0.17 &   9.34  &  0.11\\
       	&				&				&		&	    &    100  & & 10.40  &  0.17 &   9.98  &  0.24 &   9.24  &  0.16\\
       	&				&				&		&	    &    120  & & 10.17  &  0.20 &   9.73  &  0.28 &   9.06  &  0.20\\
       	&				&				&		&	    &    140  & & 10.05  &  0.25 &   9.60  &  0.35 &   8.75  &  0.21\\
       	&				&				&		&	    &    160  & &  9.73  &  0.24 &   9.16  &  0.30 &   8.69  &  0.26\\
       	&				&				&		&	    &    180  & &  9.49  &  0.25 &   8.78  &  0.27 &   8.54  &  0.29\\
       	&				&				&		&	    &    200  & &  9.35  &  0.28 &   8.70  &  0.32 &   8.47  &  0.34\\[5pt]
NGC1994	&	05:28:22.4	&	-69:08:31.3	&	2	&	BBA &     20  & &  8.33  &  0.01 &   7.63  &  0.01 &   7.45  &  0.01\\
        &				&				&		&		&     40  & &  8.28  &  0.01 &   7.59  &  0.01 &   7.40  &  0.01\\
       	&				&				&		&	    &     60  & &  8.25  &  0.01 &   7.57  &  0.01 &   7.38  &  0.01\\
        &				&				&		&		&     80  & &  8.22  &  0.02 &   7.55  &  0.01 &   7.36  &  0.01\\
      	&				&				&		&	    &    100  & &  8.14  &  0.02 &   7.47  &  0.01 &   7.35  &  0.02\\
      	&				&				&		&	    &    120  & &  7.95  &  0.03 &   7.26  &  0.02 &   7.14  &  0.02\\
       	&				&				&		&	    &    140  & &  7.80  &  0.03 &   7.10  &  0.02 &   6.95  &  0.03\\
      	&				&				&		&	    &    160  & &  7.78  &  0.04 &   7.09  &  0.02 &   6.93  &  0.04\\
      	&				&				&		&	    &    180  & &  7.76  &  0.05 &   7.07  &  0.03 &   6.92  &  0.04\\
        &				&				&		&		&    200  & &  7.74  &  0.05 &   7.06  &  0.04 &   6.91  &  0.05\\[5pt]
NGC2058	&	05:36:54.0	&	-70:09:42.0	&	10	&	BBB &     20  & & 10.72  &  0.02 &  10.15  &  0.01 &  10.08 &   0.02\\
       	&				&				&		&	                  &     40  & &  9.82  &  0.03  &  9.22 &   0.02  &  9.17 &   0.03\\
       	&				&				&		&	                  &     60  & &  9.57  &  0.04  &  9.03  &  0.04  &  8.95 &   0.05\\
        &				&				&		&		        &     80  & &  9.26   & 0.06  &  8.76  &  0.06  &  8.60  &  0.07\\[5pt]
NGC2107	&	05:43:13.3	&	-70:38:29.8	&	2.5	&	BBB &     20  & & 10.81  &  0.03 &   9.98  &  0.02 &   9.70  &  0.02\\
       	&				&				&		&	    &     40  & & 10.21  &  0.07 &   9.54  &  0.04 &   9.31  &  0.05\\
        &				&				&		&		&     60  & & 10.05  &  0.13 &   9.43  &  0.08 &   9.20  &  0.11\\
       	&				&				&		&	    &     80  & &  9.90  &  0.20 &   9.35  &  0.14 &   9.12  &  0.18\\
       	&				&				&		&	    &    100  & &  9.70  &  0.26 &   9.14  &  0.18 &   8.92  &  0.24\\
        &				&				&		&		&    120  & &  9.47  &  0.31 &   8.87  &  0.20 &   8.63  &  0.27\\
       	&				&				&		&	    &    140  & &  9.34  &  0.39 &   8.72  &  0.24 &   8.32  &  0.27\\[5pt]
NGC2108	&	05:43:56.7	&	-69:10:49.9	&	4	&	BBB &     20  & & 12.31  &  0.02 &  11.23  &  0.01 &  10.97  &  0.01\\
        &				&				&		&		&     40  & & 10.78  &  0.01 &   9.88  &  0.01 &   9.36  &  0.01\\
        &				&				&		&		&     60  & & 10.48  &  0.02 &   9.71  &  0.02 &   9.25  &  0.01\\
        &				&				&		&		&     80  & & 10.35  &  0.03 &   9.63  &  0.03 &   9.20  &  0.02\\
        &				&				&		&		&    100  & & 10.26  &  0.03 &   9.56  &  0.04 &   9.15  &  0.03\\
        &				&				&		&		&    120  & & 10.17  &  0.04 &   9.47  &  0.05 &   9.10  &  0.05\\
       	&				&				&		&	    &    140  & & 10.00  &  0.05 &   9.31  &  0.06 &   9.00  &  0.06\\
       	&				&				&		&	    &    160  & &  9.55  &  0.05 &   8.83  &  0.05 &   8.45  &  0.04\\
       	&				&				&		&	    &    180  & &  9.50  &  0.05 &   8.81  &  0.06 &   8.41  &  0.06\\
       	&				&				&		&	    &    200  & &  9.45  &  0.06 &   8.77  &  0.08 &   8.39  &  0.07\\[5pt]

\hline
\end{tabular}
{\it Notes to Table~\ref{tab:phot}}. See next page
\end{minipage}
\end{table*}

\setcounter{table}{2}
\begin{table*}
\centering
\begin{minipage}{150mm}
\caption{Continued}
\begin{tabular}{@{}lccccrlrrrrrr@{}}
\hline
ID & $\alpha_{2000}$ & $\delta_{2000} $ & d & Flag & D & & $J$ & $Jerr$ & $H$ & $Herr$ & $K_s$ & ${K_s}err$\\
(1) & (2) & (3) & (4) & (5) & (6) & & (7) & (8) & (9) & (10) & (11) & (12)\\
\hline
NGC2134	&	05:51:56.5	&	-71:05:50.4	&	4.5	&	BBB &     20  & & 10.80  &  0.01 &  10.16  &  0.01 &  10.10  &  0.01\\
        &				&				&		&		&     40  & & 10.18  &  0.02 &   9.68  &  0.01 &   9.65  &  0.01\\
        &				&				&		&		&     60  & &  9.94  &  0.03 &   9.50  &  0.02 &   9.43  &  0.02\\
        &				&				&		&		&     80  & &  9.78  &  0.04 &   9.39  &  0.03 &   9.25  &  0.03\\
       	&				&				&		&	    &    100  & &  9.46  &  0.04 &   8.96  &  0.03 &   8.77  &  0.03\\
       	&				&				&		&	    &    120  & &  9.42  &  0.06 &   8.93  &  0.05 &   8.74  &  0.05\\
       	&				&				&		&	    &    140  & &  8.95  &  0.05 &   8.31  &  0.04 &   8.01  &  0.03\\
       	&				&				&		&	    &    160  & &  8.86  &  0.06 &   8.22  &  0.04 &   7.89  &  0.04\\
       	&				&				&		&	    &    180  & &  8.82  &  0.08 &   8.16  &  0.05 &   7.85  &  0.05\\
       	&				&				&		&	    &    200  & &  8.82  &  0.10 &   8.18  &  0.07 &   7.86  &  0.06\\[5pt]
NGC2154	&	05:57:37.9	&	-67:15:43.7	&	2.0	&	BAB &     20  & & 11.24  &  0.01 &  10.38  &  0.01 &   9.86  &  0.01\\
        &				&				&		&		&     40  & & 10.44  &  0.01 &   9.64  &  0.01 &   9.18  &  0.01\\
       	&				&				&		&	    &     60  & & 10.16  &  0.02 &   9.38  &  0.01 &   8.92  &  0.01\\
        &				&				&		&		&     80  & &  9.94  &  0.02 &   9.22  &  0.02 &   8.79  &  0.01\\
        &				&				&		&		&    100  & &  9.80  &  0.03 &   9.11  &  0.02 &   8.68  &  0.02\\
        &				&				&		&		&    110  & &  9.78  &  0.04 &   9.11  &  0.03 &   8.68  &  0.02\\
\hline
\end{tabular}
{\it Notes to Table~\ref{tab:phot}:} Column~(1) is the cluster
designation, (2) and (3) are the right ascension and declination of
the position used to centre the apertures for the integral photometry
((hh:mm:ss.s) and (dd:mm:ss.s) respectively). Column~(4) is the offset
(in arcseconds) measured on 2MASS images between that position and the
cluster coordinates retrieved from SIMBAD. Column~(5) contains a flag,
providing information about the age (first letter), metalicity
estimates (second letter) and the photometry (third letter) for each
cluster. A is corresponding to a reliable age, metallicity and
photometry, B denotes the cases when the age and metallicity values
are uncertain and when used in the third position B stands for the
cases described in Section~\ref{notes} and/or the photometry was
provided in aperture sizes smaller than 200\arcsec. The aperture
diameters (arcseconds) used for each measurement are listed in
column~(6), they are denoted with colon in case complications with
photometry were suspected. The photometry information (uncorrected for
reddening) is given in columns~(7) -- (12), in the order: $J$
magnitude, photometric uncertainty in $J$, and the same information for
the other two survey bands $H$ and $K_s$. The photometric uncertainty in
each band is calculated as the quadrature sum of the zero point uncertainty,
internal uncertainty of the photometry, and the uncertainty due to stochastic
fluctuations of the background stellar population.
\end{minipage}
\end{table*}

\setcounter{table}{3}
\begin{table*}
\centering
\begin{minipage}{160mm}
\caption{Compilation of Optical Photometry}
\label{tab:comp}
\begin{tabular}{@{}lcrrrrrcrrr@{}}
\hline
ID & Galaxy & \multicolumn{1}{c}{D}& \multicolumn{1}{c}{$V$}  & \multicolumn{1}{c}{$(B-V)$} & \multicolumn{1}{c}{$(U-B)$}  & Ref. & $A_{V}$ & Ref. & SWB & Note\\
(1) & (2) & (3) & \multicolumn{1}{c}{(4)} & \multicolumn{1}{c}{(5)} & \multicolumn{1}{c}{(6)} & (7) & (8) & (9) &  \multicolumn{1}{c}{(10)} & (11) \\
\hline
NGC121				& SMC &	 62	& $11.24\pm0.01$ & $0.76\pm0.00$ & $ 0.10\pm0.01$ &  1 & $0.16\pm0.09$ & 11 & VII &  \\
NGC152				& SMC &	 62	& $12.94\pm0.04$ & $0.70\pm0.02$ & $ 0.18\pm0.02$ &  1 & $0.16\pm0.03$ &  7 & V   &  \\
NGC265				& SMC &	 62	& $12.13\pm0.03$ & $0.30\pm0.01$ & $-0.11\pm0.05$ &  1 & $0.19^{\ +\ 0.15}_{\ -\ 0.13}$ &  6 & III &  \\
NGC330				& SMC &	 62	& $ 9.60\pm0.01$ & $0.15\pm0.01$ & $-0.46\pm0.00$ &  1 & $0.37\pm0.02$ &  - & I   &  \\
NGC339				& SMC &	 62	& $12.84\pm0.04$ & $0.73\pm0.01$ & $ 0.03\pm0.00$ &  1 & $0.09\pm0.12$ & 11 & VII &  \\
NGC361				& SMC &	 62	& $12.24\pm0.01$ & $0.78\pm0.00$ & $ 0.14\pm0.02$ &  1 & $0.22\pm0.09$ & 11 & VII &  \\
NGC411				& SMC &	 62	& $12.21\pm0.04$ & $0.62\pm0.01$ & $ 0.24\pm0.02$ &  1 & $0.37\pm0.03$ &  7 & V   &  \\
NGC416				& SMC &	 62	& $11.42\pm0.00$ & $0.73\pm0.00$ & $ 0.12\pm0.00$ &  1 & $0.25\pm0.09$ & 10 & VII &  \\
NGC419				& SMC &	 62	& $10.61\pm0.01$ & $0.65\pm0.01$ & $ 0.24\pm0.01$ &  1 & $0.20\pm0.02$ &  - & V   &  \\
NGC458				& SMC &	 62	& $11.73\pm0.03$ & $0.16\pm0.00$ & $-0.17\pm0.01$ &  1 & $0.32\pm0.02$ &  - & V   &  \\
NGC1466	   	   	   	& LMC &	 60	& $11.59\pm0.03$ & $0.68\pm0.02$ & $ 0.13\pm0.03$ & 15 & $0.28\pm0.06$ & 16 & VII &  \\
NGC1644	 	 	 	& LMC &	 60	& $12.89\pm0.02$ & $0.60\pm0.03$ & $ 0.19\pm0.02$ & 15 & $0.39\pm0.02$ &  - & V   &  \\
NGC1651				& LMC &	100	& $12.28\pm0.04$ & $0.71\pm0.04$ & $ 0.28\pm0.04$ &  4 & $0.34\pm0.03$ &  8 & V   &  \\
NGC$1711_{\rm (B)}$	& LMC &	 50	& $12.50\pm0.04$ & $0.27\pm0.04$ & $-0.04\pm0.04$ &  4 & $0.56\pm0.01$ &  - & III & 1\\
NGC$1711_{\rm (v)}$	& LMC &	 60	& $10.11\pm0.03$ & $0.12\pm0.01$ & $-0.37\pm0.02$ & 15 &               &    & III & 1\\
NGC1718		        & LMC &	 62	& $12.25\pm0.01$ & $0.76\pm0.01$ & $ 0.26\pm0.02$ &  2 & $0.31\pm0.09$ &  8 & VI  &  \\
NGC1751		        & LMC &	100	& $11.73\pm0.04$ & $0.79\pm0.04$ & $ 0.27\pm0.04$ &  4 & $0.65\pm0.06$ &  - & VI  &  \\
NGC1754		        & LMC &	100	& $11.57\pm0.04$ & $0.75\pm0.04$ & $ 0.15\pm0.04$ &  4 & $0.28\pm0.06$ & 12 & VII &  \\
NGC1777		        & LMC &	 38	& $12.80\pm0.02$ & $0.60\pm0.03$ & $ 0.17\pm0.03$ &  4 & $0.31\pm0.09$ &  8 & IVB &  \\
NGC1783		        & LMC &	 60	& $10.97\pm0.02$ & $0.63\pm0.02$ & $ 0.20\pm0.03$ & 15 & $0.30\pm0.03$ &  - & V   &  \\
NGC$1786_{\rm (u)}$	& LMC &	 60	& $10.16\pm0.04$ & $0.67\pm0.00$ & $ 0.06\pm0.02$ & 15 & $0.28\pm0.16$ &  5 & VII & 2\\
NGC$1786_{\rm (c)}$	& LMC &	 60	& $10.88\pm0.04$ & $0.74\pm0.00$ & $ 0.10\pm0.02$ & 15 &               &    & VII & 2\\
NGC1805				& LMC &	 60 & $10.63\pm0.03$ & $0.11\pm0.02$ & $-0.55\pm0.01$ & 15 & $0.32\pm0.02$ &  - & I   &  \\
NGC1806				& LMC &	 60	& $11.27\pm0.04$ & $0.26\pm0.02$ & $ 0.73\pm0.02$ & 15 & $0.25\pm0.04$ &  - & V   &  \\
NGC1818				& LMC &	 60	& $ 9.85\pm0.02$ & $0.18\pm0.02$ & $-0.46\pm0.00$ & 15 & $0.39\pm0.02$ &  - & I   &  \\
NGC1831				& LMC &	 60	& $11.18\pm0.02$ & $0.34\pm0.02$ & $ 0.13\pm0.02$ & 15 & $0.34\pm0.03$ &  8 & IVA &  \\
NGC1835				& LMC &	 60	& $10.16\pm0.05$ & $0.74\pm0.04$ & $ 0.11\pm0.03$ & 15 & $0.25\pm0.06$ & 12 & VII &  \\
NGC1841				& LMC &	 25	& \multicolumn{1}{l}{14.08} & \multicolumn{1}{l}{0.90}  & \multicolumn{1}{l}{\ \ 0.50}  & 14 & $0.62\pm0.09$ &  5 & VII &  \\
NGC1846				& LMC &	 60	& $11.40\pm0.02$ & $0.31\pm0.05$ & $ 0.74\pm0.03$ & 15 & $0.41\pm0.04$ &  - & VI  &  \\
NGC1847				& LMC &	 72	& $11.06\pm0.02$ & $0.20\pm0.02$ & $-0.33\pm0.03$ &  3 & $0.49\pm0.02$ &  - & I   &  \\
NGC1850				& LMC &	 60	& $ 9.36\pm0.06$ & $0.11\pm0.03$ & $-0.34\pm0.07$ & 15 & $0.33\pm0.01$ &  - & II  &  \\
NGC1856				& LMC &	 60	& $10.07\pm0.02$ & $0.34\pm0.01$ & $ 0.06\pm0.01$ & 15 & $0.65\pm0.06$ &  8 & IVA &  \\
NGC1860		   	   	& LMC &	 72	& $11.04\pm0.02$ & $0.14\pm0.02$ & $-0.39\pm0.03$ &  3 & $0.22\pm0.03$ &  - & I   &  \\
NGC1866		 	 	& LMC &	 60	& $ 9.89\pm0.01$ & $0.26\pm0.02$ & $-0.06\pm0.01$ & 15 & $0.28\pm0.06$ &  - & III &  \\
NGC1868				& LMC &	 62 & $11.57\pm0.02$ & $0.45\pm0.02$ & $ 0.15\pm0.03$ &  3 & $0.12\pm0.03$ &  8 & IVA &  \\
NGC1898				& LMC &	 60 & $11.52\pm0.06$ & $0.75\pm0.02$ & $-0.03\pm0.03$ & 15 & $0.22\pm0.06$ & 12 & VII &  \\
NGC1916				& LMC &	 44	& $10.38\pm0.02$ & $0.78\pm0.02$ & $ 0.18\pm0.03$ &  3 & $0.42\pm0.05$ &  - & VII &  \\
NGC1928				& LMC &	 62	& $11.88\pm0.02$ & $0.22\pm0.02$ & $-0.31\pm0.03$ &  3 & $0.20\pm0.05$ &  9 & VII &  \\
NGC1939				& LMC &	 38	& $11.78\pm0.05$ & $0.69\pm0.05$ & $ 0.09\pm0.05$ &  4 & $0.40\pm0.08$ &  9 & VII &  \\
NGC1978				& LMC &	 60	& $10.74\pm0.04$ & $0.78\pm0.04$ & $ 0.23\pm0.07$ & 15 & $0.76\pm0.05$ &  - & VI  &  \\
NGC1984				& LMC &	 50	& $ 9.99\pm0.04$ & $0.01\pm0.04$ & $-0.82\pm0.04$ &  4 & $0.36\pm0.02$ &  - & 0   &  \\
NGC1987				& LMC &	 60	& $12.18\pm0.01$ & $0.51\pm0.01$ & $ 0.18\pm0.02$ & 15 & $0.28\pm0.03$ &  - & IVB &  \\
NGC1994    			& LMC &	 50	& $ 9.78\pm0.04$ & $0.09\pm0.04$ & $-0.69\pm0.04$ &  4 & $0.41\pm0.02$ &  - & I   &  \\
NGC2004				& LMC &	 72	& $ 9.60\pm0.02$ & $0.13\pm0.02$ & $-0.71\pm0.03$ &  3 & $0.33\pm0.02$ &  - & I   &  \\
NGC2005				& LMC &	 25	& $11.57\pm0.00$ & $0.73\pm0.00$ & $ 0.20\pm0.00$ & 14 & $0.37\pm0.06$ & 12 & VII &  \\
NGC2011		   	   	& LMC &	 45	& $10.58\pm0.02$ & $0.00\pm0.02$ & $-0.72\pm0.03$ &  3 & $0.47\pm0.02$ &  - & I   &  \\
NGC2019		 	 	& LMC &	 60	& $10.95\pm0.01$ & $0.77\pm0.01$ & $ 0.16\pm0.02$ & 15 & $0.37\pm0.06$ & 12 & VII &  \\
NGC2031				& LMC &	 72	& $10.83\pm0.02$ & $0.26\pm0.02$ & $-0.07\pm0.03$ &  3 & $0.40\pm0.03$ &  - & III &  \\
NGC$2058_{\rm (v)}$	& LMC &	 60	& $11.85\pm0.04$ & $0.24\pm0.01$ & $-0.12\pm0.01$ & 15 & $0.39\pm0.02$ &  - & III & 3\\
NGC$2058_{\rm (G)}$	& LMC &	 60	& $10.73\pm0.03$ & $           $ & $            $ &    &               &    & III & 3\\
NGC2100				& LMC &	 60	& $ 9.60\pm0.04$ & $0.16\pm0.02$ & $-0.56\pm0.02$ & 15 & $0.80\pm0.02$ &  - & I   &  \\
NGC2107				& LMC &	 60	& $11.51\pm0.02$ & $0.38\pm0.03$ & $ 0.13\pm0.03$ & 15 & $0.36\pm0.04$ &  - & IVA &  \\
NGC2108				& LMC &	 62	& $12.32\pm0.01$ & $0.58\pm0.01$ & $ 0.22\pm0.02$ &  2 & $0.50\pm0.05$ &  - & IVB &  \\
NGC2121				& LMC &	 62	& $12.37\pm0.01$ & $0.84\pm0.01$ & $ 0.25\pm0.02$ &  2 & $0.22\pm0.06$ &  8 & VI  &  \\
NGC2134				& LMC &  60	& $11.05\pm0.04$ & $0.26\pm0.02$ & $-0.02\pm0.03$ & 15 & $0.62\pm0.03$ &  - & III &  \\
NGC2136				& LMC &	 60	& $10.54\pm0.02$ & $0.28\pm0.02$ & $-0.13\pm0.02$ & 15 & $0.58\pm0.02$ &  - & III &  \\
NGC2153				& LMC &	100	& $13.05\pm0.04$ & $0.69\pm0.04$ & $ 0.03\pm0.04$ &  4 & $0.27\pm0.05$ &  - & VII &  \\
NGC2154				& LMC &	 62	& $12.32\pm0.01$ & $0.68\pm0.01$ & $ 0.24\pm0.02$ &  2 & $0.39\pm0.03$ &  - & V   &  \\
NGC2155				& LMC &	 62	& $12.60\pm0.01$ & $0.81\pm0.01$ & $ 0.23\pm0.02$ &  2 & $0.06\pm0.03$ &  8 & VI  &  \\
NGC2156				& LMC &	 72	& $11.38\pm0.02$ & $0.12\pm0.02$ & $-0.07\pm0.03$ &  3 & $0.20\pm0.02$ &  - & III &  \\
NGC2157		   	   	& LMC &	 60	& $10.16\pm0.02$ & $0.19\pm0.02$ & $-0.16\pm0.01$ & 15 & $0.42\pm0.02$ &  - & III &  \\
\hline
\end{tabular}
{\it Notes to Table~\ref{tab:comp}}. See next page
\end{minipage}
\end{table*}

\setcounter{table}{3}
\begin{table*}
\centering
\begin{minipage}{160mm}
\caption{Continued}
\begin{tabular}{@{}lcrrrrrcrrr@{}}
\hline
ID & Galaxy & \multicolumn{1}{c}{D}& \multicolumn{1}{c}{$V$}  & \multicolumn{1}{c}{$(B-V)$} & \multicolumn{1}{c}{$(U-B)$}  & Ref. & $A_{V}$ & Ref. & SWB & Note\\
(1) & (2) & (3) & \multicolumn{1}{c}{(4)} & \multicolumn{1}{c}{(5)} & \multicolumn{1}{c}{(6)} & (7) & (8) & (9) &  \multicolumn{1}{c}{(10)} & (11) \\
\hline
NGC2159		& LMC &	 72	& $11.38\pm0.02$ & $0.28\pm0.02$ & $-0.14\pm0.03$ &  3 & $0.29\pm0.03$ &  - & III &  \\
NGC2162		& LMC &	 62	& $12.70\pm0.01$ & $0.68\pm0.01$ & $ 0.31\pm0.02$ &  2 & $0.09\pm0.06$ &  8 & V   &  \\
NGC2164		& LMC &	 60	& $10.34\pm0.01$ & $0.10\pm0.01$ & $-0.24\pm0.01$ & 6 & $0.33\pm0.02$ &  - & II  &  \\
NGC2172		& LMC &	 72	& $11.75\pm0.02$ & $0.18\pm0.02$ & $-0.16\pm0.03$ & 3 & $0.26\pm0.03$ &  - & III &  \\
NGC2173		& LMC &	150	& $11.88\pm0.05$ & $0.82\pm0.05$ & $ 0.28\pm0.05$ & 4 & $0.22\pm0.06$ &  8 & VI  &  \\
NGC2190		& LMC &	 61	& $12.94\pm0.02$ & $0.63\pm0.03$ & $ 0.29\pm0.03$ & 4 & $0.39\pm0.02$ &  - & V   &  \\
NGC2193		& LMC &	 38	& $13.42\pm0.02$ & $0.71\pm0.03$ & $ 0.20\pm0.03$ & 4 & $0.39\pm0.02$ &  - & V   &  \\
NGC2203		& LMC &	150	& $11.29\pm0.04$ & $0.77\pm0.04$ & $ 0.26\pm0.04$ & 4 & $0.39\pm0.02$ &  - & VI  &  \\
NGC2209		& LMC &	 62	& $13.15\pm0.01$ & $0.52\pm0.01$ & $ 0.36\pm0.02$ & 2 & $0.47\pm0.09$ &  8 & IVB &  \\
NGC2210		& LMC &	 60	& $10.94\pm0.03$ & $0.71\pm0.03$ & $ 0.11\pm0.01$ & 6 & $0.28\pm0.09$ &  5 & VII &  \\
NGC2213		& LMC &	 62	& $12.38\pm0.01$ & $0.71\pm0.01$ & $ 0.28\pm0.02$ & 2 & $0.19\pm0.09$ &  8 & V   &  \\
NGC2214		& LMC &	 60	& $10.93\pm0.01$ & $0.11\pm0.02$ & $-0.27\pm0.01$ & 6 & $0.39\pm0.02$ &  - & II  &  \\
NGC2231		& LMC &	 44	& $13.20\pm0.01$ & $0.67\pm0.01$ & $ 0.24\pm0.02$ & 2 & $0.39\pm0.02$ &  - & V   &  \\
NGC2249		& LMC &	 72	& $12.23\pm0.02$ & $0.43\pm0.02$ & $ 0.20\pm0.03$ & 3 & $0.03\pm0.06$ &  8 & IVB &  \\
NGC2257		& LMC &	 61 & $12.62\pm0.02$ & $0.62\pm0.03$ & $ 0.01\pm0.03$ & 4 & $0.00\pm0.00$ & 13 & VII &  \\
ESO121-003	& LMC &	 61	& $14.04\pm0.02$ & $0.87\pm0.03$ & $ 0.17\pm0.05$ & 4 & $0.10\pm0.05$ &  9 & VII &  \\
Hodge4		& LMC &	 38	& $13.33\pm0.02$ & $0.66\pm0.03$ & $ 0.16\pm0.05$ & 4 & $0.39\pm0.02$ &  - & V   &  \\
Hodge11		& LMC &	 62	& $11.98\pm0.01$ & $0.63\pm0.01$ & $ 0.00\pm0.02$ & 2 & $0.23\pm0.02$ & 10 & VII &  \\
Hodge14		& LMC &	 62	& $13.42\pm0.01$ & $0.72\pm0.01$ & $ 0.21\pm0.02$ & 2 & $0.25\pm0.06$ &  8 & V   &  \\
Kron3		& SMC &	 62	& $12.05\pm0.02$ & $0.69\pm0.01$ & $ 0.05\pm0.02$ & 1 & $0.00\pm0.06$ & 11 & VII &  \\
Lindsay1	& SMC &	 62	& $13.32\pm0.05$ & $0.75\pm0.00$ & $ 0.17\pm0.05$ & 1 & $0.19\pm0.06$ & 11 & VII &  \\
Lindsay113	& SMC &	 62	& $13.61\pm0.04$ & $0.73\pm0.02$ & $ 0.05\pm0.09$ & 1 & $0.00\pm0.06$ & 11 & VII &  \\
LW431		& LMC &	 38	& $13.67\pm0.02$ & $0.74\pm0.03$ & $ 0.13\pm0.05$ & 4 & $0.39\pm0.02$ &  - & VII &  \\
SL842		& LMC &	 38	& $14.15\pm0.02$ & $0.79\pm0.03$ & $ 0.11\pm0.05$ & 4 & $0.39\pm0.02$ &  - & VII &  \\\hline
\end{tabular}

{\it Notes on Table~\ref{tab:comp}:} Column~(1) is the cluster
designation, (2) gives the galaxy in which the object resides. The
diameter of the aperture used for the integral visual magnitude and
colour measurements is given in column~(3). Column~(4) gives
the $V$ magnitude and it's uncertainty, while (5) and (6) list
$(B-V)$ and $(U-B)$ colours and their corresponding uncertainties. These
values are not reddening corrected. References
to the sources of the integrated-light measurements are listed in
column~(7). Column~(8) presents $A_{V}$ information and corresponding uncertainties. References to the reddening information are given in column~(9). Preference is given to extinction estimates based on deep CMDs. In case these were not available, extinction values retrieved from the website of Magellanic Clouds Photometric Survey are provided (marked with dashes in column~(9)). Column~(10) shows the SWB type of
the object, for the LMC clusters the information comes from
\cite{bica_et_al_96}, SWB types of SMC clusters were recovered from
from the $s$-parameter calibration of \cite{elson_fall_85}. Finally some remarks to the objects or the photometry
are given in column~(11).

{\sc Notes on column~(11):} (1) The information about this object in
\cite{bica_et_al_96} and \cite{vdb_hagen_68} is completely
discrepant with each other. Both values are listed with lowercase "B" and "v" added in
parenthesis to the cluster designation.

(2) There is a foreground star superposed on the object. The flux from the star and
part of the cluster were measured in a 15$''$ diaphragm and were subtracted from the flux of the larger (D=60\arcsec) aperture encompassing the object to derive the final colours and magnitudes. 
The uncorrected and corrected values are listed in lowercase "u" and "c" in parenthesis
after the cluster designation, respectively. All the data is from
\cite{vdb_hagen_68}. The photometric uncertainties  for the uncorrected mesurements
are also adopted for the corrected ones. They should be considered a lower
limit.

(3) Measurements for NGC 2058 from \cite{goudfrooij06} and
\cite{vdb_hagen_68} are discrepant by more than one magnitude in
$V$. A possible explanation is misidentification of the object in the
earlier study. Inspection of a 14\arcmin~x~14\arcmin \ $V$ image
available in SIMBAD revealed several less luminous clusters in close proximity to the
object, which might have caused the overestimate of the $V$ magnitude
in \citeauthor{vdb_hagen_68}. The $V$ value from
\citeauthor{goudfrooij06} is also listed in the table. Lowercase "G"
and "v" are added in parenthesis to the cluster designation to
indicate the origin of the photometry (\cite{goudfrooij06} and
\cite{vdb_hagen_68} respectively).

{\it References:\/} (1)\cite{alcaino78} (2)\cite{bernard75} (3)\cite{bernard_bigay_74} (4)\cite{bica_et_al_96} (5)\cite{brocato96} (6)\cite{chiosi07} (7)\cite{crowl2001} (8)\cite{kerber07} (9)\cite{mg04} (10)\cite{mighell96}  (11)\cite{mighell98b} (12)\cite{olsen98} (13)\cite{testa95} (14)\cite{vdb81} (15)\cite{vdb_hagen_68} (16)\cite{walker92}

\end{minipage}
\end{table*}

%%%%%%%%%%%%%%%%%%%%%%%%%%%%%%%%%%%%%%%%%%%%%%%%%%%
{\bf Notes on individual objects} -- All the objects that required special
attention during the reduction process are commented on below.

{\bf NGC~265} -- This object is situated in a rich SMC star field and the
depth of the images decreases from $J$ to $K_s$. Analysis of the curves
of growth show that for aperture diameters larger than 140\arcsec\ this
trend causes some variation in the measured cluster colours. The photometry
is carried out with the full set of apertures, but results for sizes
exceeding 140\arcsec\ should be treated with caution.

{\bf NGC~1644} -- This faint compact cluster is situated close to the edge of
the atlas image.  The photometry is carried out up to 90\arcsec\ aperture
diameter. The curve of growth in $K_s$ shows evidence for
variations in the background level.  Results for
aperture sizes larger than 60\arcsec\ could be affected by these
variations. Due to the proximity of the image edge our automated procedure
for deriving the aperture centres does not provide reliable results.
Therefore the cluster centre was chosen ``by eye'', but we consider this
position to be reliable due to the compactness of the object and lack of
stellar contamination in the surrounding field.

{\bf NGC~1928} -- This is an old globular cluster
\citep[see][]{mg04}. Situated in the LMC bar region, this object
suffers from strong background/foreground contamination. Combined with
the limited depth of the cluster's image on the 2MASS atlas frames, this makes
the integrated photometry challenging. Due to the presence of
several relatively bright stars in the vicinity of the cluster, we
decided to use only the results based on the unresolved light from the
object to derive the aperture centres. Several bright stars in the
aperture set were subtracted after an analysis of their properties
based on the infrared colours explained in detail in Paper~I. The
resulting integrated-photometry curves of growth show residual effects
of the background removal and the NIR photometry should be treated
with caution in this case.

{\bf NGC~2058} -- This cluster is located close to the edge of the atlas
image. Photometry is carried out with a set of apertures up to D=80\arcsec.

{\bf NGC~2107} -- The cluster lies close to the edge of the atlas image.
In this case photometry was carried out with a set of aperture sizes up to
140\arcsec. The curves of growth indicate that we obtain good
sampling of the flux from the object.

{\bf NGC~2108} -- There are three luminous stars in the cluster. Their
colours are consistent with those expected for carbon stars. It is
noteworthy to mention that \cite{ferraro2004} found the same number of AGB
stars in their NIR photometric study of this object. The age estimates of
the cluster are also consistent with the presence of carbon stars, so they
are included in the final photometry.

{\bf NGC~2134} -- Several luminous stars are present within the cluster
area. They affect the centring, so we use the results from the unresolved
component. The colours of these stars are consistent with those expected
for carbon stars, and they are included in the final integrated photometry
measurements.

{\bf NGC~2154} -- Due to the proximity of the object to the edge of the
atlas image, the photometry is carried out with a set of apertures up to
110\arcsec\ in diameter.
\label{notes}

%%%%%%%%%%%%%%%%%%%%%%%%%%%%%%%%%%%%%%%%%%%%%%%%%%%
\subsection{Compiling a Catalog of Optical Cluster Photometry}

We conducted an extensive literature search for appropriate integrated
optical cluster colours to combine with our NIR measurements.  Our
compilation from the literature is presented in Table~\ref{tab:comp}. It
is based on integrated-light photoelectric observations and lists the $V$
magnitudes, $(B-V)$ and $(U-B)$ colours plus their uncertainties, reddening values towards the objects with their corresponding uncertainties are also included. The
photometry comes from different observers, so special care was taken to
ensure that the individual results are consistent. Recent independent CCD
datasets from \cite{goudfrooij06} and \cite{hunter03} were also used in
the consistency checks. The cases where discrepancies cannot be
readily explained
are listed as notes in the last column of Table~\ref{tab:comp}. As a rule,
we provide information about the largest aperture size available. This
reduces the effects of both the aperture centring (which could differ for
the optical and NIR data), the stochastic fluctuations of the background
stellar population and of the stars in the clusters themselves.  The
photometry of all SMC clusters is taken from the homogeneous dataset of
\cite{alcaino78}, and is also listed in Table~\ref{tab:comp}. Reddening information is compiled from a number of sources (indicated in the table) and preference is given to values based on deep CMDs.
\label{comp}

%%%%%%%%%%%%%%%%%%%%%%%%%%%%%%%%%%%%%%%%%%%%%%%%%%%
\section{Testing the Models}
In recent years significant improvements in modeling the properties of
Simple Stellar Populations have been achieved, and several independent
sets of SSP models have been published.  Here, we will focus our
attention on comparing the models by Vazdekis, Bruzual \& Charlot, 
Anders \& Fritze-v.\ Alvensleben, and Maraston with the integrated
photometry of Magellanic Cloud 
clusters presented in paper I and in this work.  Our main goals are
to: {\it (i)\/} determine which models best reproduce the
observed cluster colours; {\it (ii)\/} establish whether clusters of
different ages and metallicities can be accurately distinguished via a
combination of optical-NIR colours; and {\it (iii)\/} provide information that could help improve the model predictions.
Below, we first briefly discuss the colours we will use to compare observed
data with SSP model predictions as well as the NIR photometric system we
will adopt in that context. We then present the actual comparisons
between data and SSP model predictions for distinct age groups. The 
selection of clusters for each age group is described in Appendices~A
and B.  
\label{testmodels0}

\subsection{Choice of Optical-NIR colours}
To select a set of optical-NIR colours that are most applicable
for a proper and meaningful comparison between observed data and SSP model
predictions, we look for colours that do a good job at breaking the
well-known age-metallicity degeneracy while typically delivering
photometry with adequate signal-to-noise (S/N) ratios. For this purpose we
follow the work of \cite{puzia07} who showed that the colour
combination $B-J$ vs.\ $J-K$ provides very good age resolution
(through $B-J$) while $J-K$ is much more sensitive to metallicity than
to age \citep[except during a brief age interval after the AGB phase
transition where $J-K$ shows a modest age dependence; see
also][]{ferraro00}. 
In addition, we present $V-J$ vs. $J-K$
colour-colour diagrams since the $V$-band typically provides a higher
S/N in observations than the $B$ band.  

\subsection{Stochastic Effects in the Stellar Populations}
When a comparison between observations of star clusters and theoretical
predictions is performed, it is important to keep in mind that
models assume that clusters are sufficiently massive that 
all stages of stellar evolution are well sampled. Predictions of any model based on
these assumptions will match the observations only in the limit of a
sufficiently large number of observed stars.  The mass of real stellar systems thus drives the validity of 
comparison with model predictions.

The level of stochastic fluctuations which arise at different total
cluster masses has been addressed in a number of previous studies
\citep[e.g.][]{bruzual2002, lm2000, cl04}. Most notably,
\citeauthor{lm2000} calculate the minimum masses of a stellar
population with solar metallicity that ensure 
that the luminosity fluctuations are less than 10\% of the mean
luminosity 
(${\cal{M}}_{(10\%)}$),
$\sigma_{L}/L \le 10 \%$
(roughly corresponding to $\sigma = 0.1$ mag) for several photometric
passbands.  
\cite{cl04} define a ``Lowest Luminosity Limit'' (LLL) which requires
that the total luminosity of a modeled cluster be larger than the
contribution of the brightest star included in the isochrones, and
show that the highest LLL masses are derived for the $K$ band.
Any object which complies with these mass limits provides a
meaningful comparison for the entire range of optical-NIR colours in
our study. 
Below, we check whether our composite clusters are more massive than the
implied LLL masses by comparing the most luminous star in the isochrone
(at the mean age and metallicity of each composite cluster) with the
estimated cluster mass (described below). Using the on-line tool CMD
1.2l\footnote{available at:
http://stev.oapd.inaf.it/~lgirardi/cgi-bin/cmd} provided by L. Girardi, we
adopt a \cite{kroupa2001} IMF corrected for binaries (his Equation~6), and
find that all of our composite cluster easily satisfy the less stringent
LLL criterion. The LLL masses that we use are given in
Table~\ref{tab:mod}, and have been corrected for the difference in the
adopted lower mass limits ($0.01~M_{\odot}$ in \citeauthor{cl04} and
$0.1~M_{\odot}$ in this work). \cite{cl04} compare their values of LLL
with the ${\cal{M}}_{(10\%)}$ masses derived by \citeauthor{lm2000},
scaled for differences in the adopted IMF between the two works. These
scaled values of ${\cal{M}}_{(10\%)}$ are also presented in
Table~\ref{tab:mod}.~This ${\cal{M}}_{(10\%)}$ limit is closely matched by
composite cluster (age bin: $2-4.6$ Gyr) and surpassed by all others,
ensuring a robust comparison between our measurements and model
predictions.
\label{stoceffects}

%%%%%%%%%%%%%%%%%%%%%%%%%%%%%%%%%%%%%%%%%%%%%%%%%%%
\subsection{The Old Cluster Population}
Integrated $JHK_S$ magnitudes were measured within the apertures of the
optical photometry for all old clusters listed in Table~\ref{tab:all}. The
photometry is presented in Table~\ref{tab:ph}. We note that the data in the
table are {\it measured\/} magnitudes, not corrected for
reddening. A reddening correction {\it is\/} however applied when plotting
the objects on the model grids in
Figures~\ref{fig:oldvdat}\,--\,\ref{fig:agejk}. Reddening values
based on deep CMDs from the literature are used where available. For
the rest of the sample, reddening estimates based on photometric
information from the MCPS  are applied.

\setcounter{figure}{2}
\begin{figure} 
\centering
\includegraphics[bb=14 14 287 785,width=6.4cm]{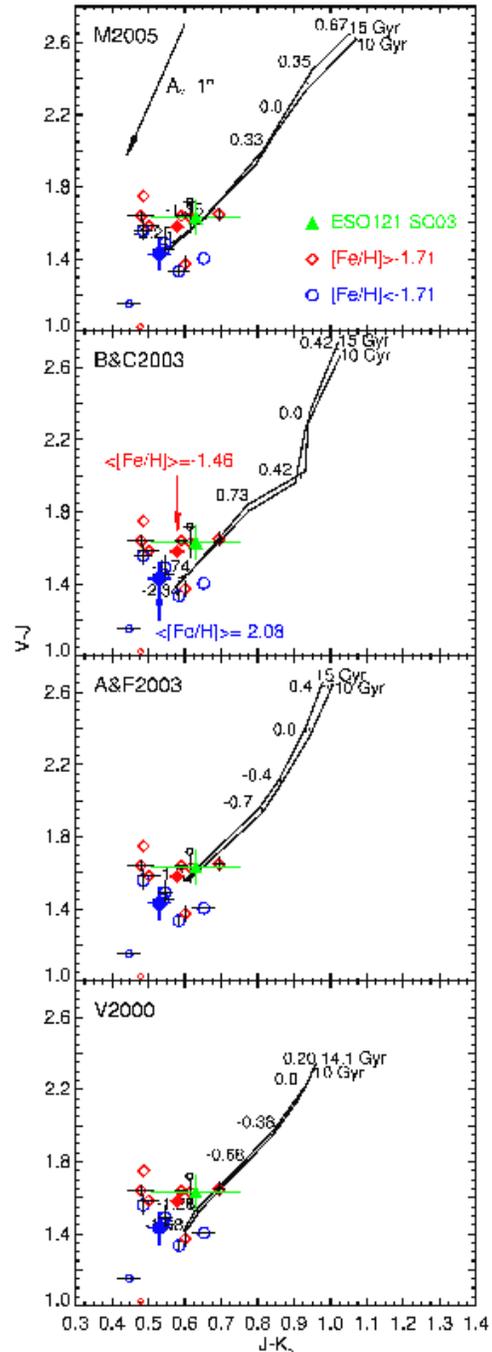}
\caption{$(V-J)$ vs. $(J-K_S)$ colour-colour diagrams for the old
clusters. Isochrones for 10 and 15 Gyr are plotted with solid
lines and metallicity values are marked along them. The panels present the
models of Maraston, Bruzual \& Charlot, Anders \& Fritze-v.\
Alvensleben and Vazdekis from top to bottom, respectively. The 
reddening vector 
is shown in the top panel. The two outliers are NGC~1928 (small red
diamond) and NGC~1939 (small blue circle). NGC~1916 (small black circle)
suffers from differential reddening (no age and metallicity
available).
The 9-Gyr-old cluster ESO121--SC03  is shown as a green triangle. Clusters with metallicities lower than the mean value for the sample are plotted as blue circles, the others are
denoted with red diamonds. The mean colours for each of
those sub-samples are shown as filled blue circle and red diamond
respectively. The corresponding mean [Fe/H] values are shown in the
second panel.}
\label{fig:oldvdat}
\end{figure}

\setcounter{figure}{3}
\begin{figure} 
\centering
\includegraphics[bb=14 14 287 785,width=6.4cm]{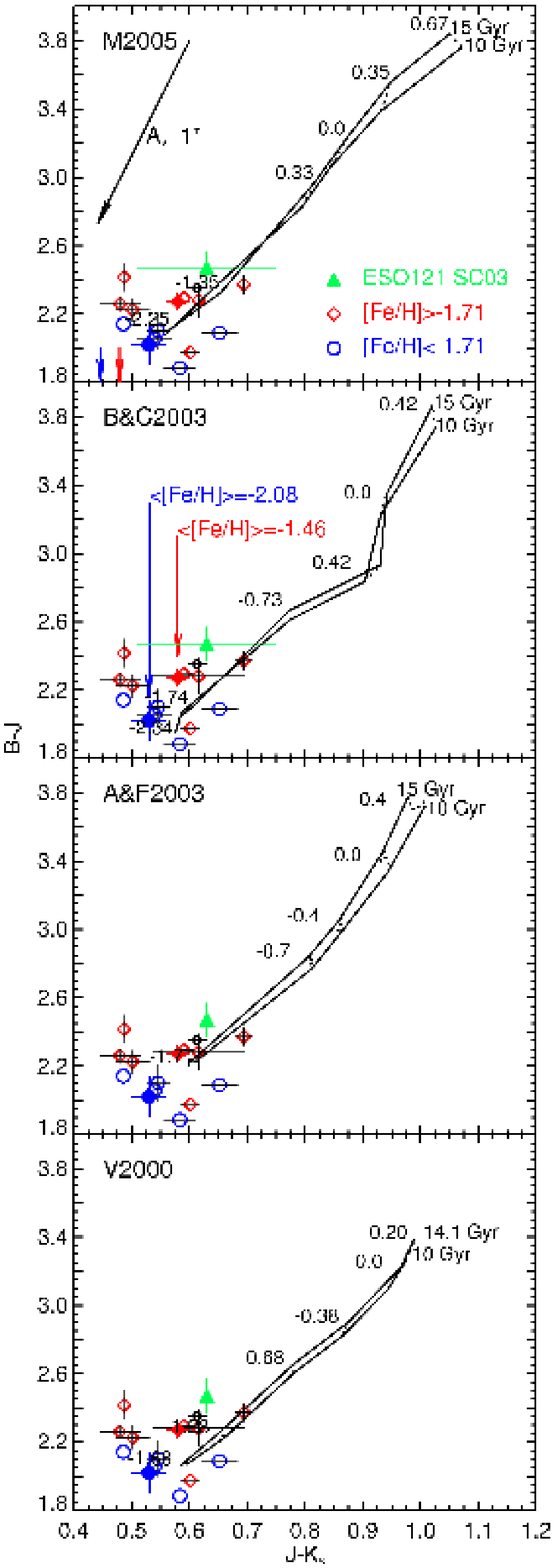}
\caption{$(B-J)$ vs. $(J-K_S)$ colour-colour diagrams for the old
clusters. As in Figure~\ref{fig:oldvdat} two different isochrones are
plotted and metallicity values are marked. The panels are organized in
the same order, with an extinction vector shown in the top panel. The two
clusters NGC~1928 and NGC~1939, fall outside of the diagrams, and
the directions towards them are marked with red and blue arrow
respectively. NGC~1916 is shown as a small black circle. }
\label{fig:oldbdat}
\end{figure}

In Figures \ref{fig:oldvdat} and \ref{fig:oldbdat} we show a comparison
between the model predictions and reddening corrected colours for the 15 old Magellanic
Cloud star clusters in our sample. The errorbars include uncertainties in the
photometry, reddening correction, and an estimate of the error due to the
statistical fluctuations within the field stellar population in the
vicinity of each cluster. 
The most significant outliers, NGC~1928 and NGC~1939
(the two points in the lower left part of Figure~\ref{fig:oldvdat}),
suffer from significant crowding due to their location in the LMC bar (see
the remarks about NGC~1928 in Section~\ref{notes}). This likely affected
their optical photometry measurements, which were retrieved
from the literature\footnote{E.g., \cite{elson_fall_85} derived S = 22 (Age
$\sim$ 50 Myr) for NGC~1928 based on the integrated-light colours from
\cite{vdb81}. This contradicts the results of \cite{mg04} based on high-resolution
{\it HST/ACS\/} imaging, demonstrating clearly that the field contamination
in the LMC bar region can significantly affect the integrated-light
measurements if not properly accounted for.}. However, the measured colours
for most (other) clusters are generally in good agreement with the model
predictions.

As mentioned in the previous section, it is possible that stochastic fluctuations in the number of
(e.g., giant) stars will cause a spread of cluster colours at any
given age and metallicity. In order to assess this effect, we estimate
the stellar mass which contributes to our integrated magnitudes for each  Magellanic Cloud cluster as follows. We use
our extinction-corrected $J$- band magnitude (the 2MASS 
and \citeauthor{bb88} photometric systems are closest in $J$ and the 2MASS
survey has its highest sensitivity in $J$), combined with the $J$ band $M/L$ ratios
predicted by the Maraston (2005) models 
which assume a Kroupa (2001) IMF.
These estimates of the stellar mass
contributing to each observed colour are presented in Table~\ref{tab:ph}.
For comparison, we include mass estimates from \cite{mclvdm05} which 
are determined from profile fitting (to determine the total cluster luminosity)
combined with $M/L_V$ ratios determined from SSP model fitting
of each cluster (assuming the BC03 models with a Chabrier-disk IMF).
While the two estimates are in 
reasonable agreement, our 
masses are systematically lower.  
We found that the main driver of this difference are the systematically lower 
total cluster luminosities compared to the values 
determined by \cite{mclvdm05}. This effect could be predicted 
since we do not extrapolate the cluster light outside of
the aperture size used for the optical photometry.
Secondary effects in the mass differences between the two works
come from differences in the adopted ages and metallicities, and hence
SSP $M/L$.

We divide the old cluster sample by metallicity into two subsamples,
accumulating enough mass in each bin to render the effects of stochastic fluctuations
negligible.  
Clusters with [Fe/H] $< -1.71$ are designated as ``metal-poor'' and
shown as open  
blue circles on Figures~\ref{fig:oldvdat} and \ref{fig:oldbdat}), and
those with [Fe/H]$>-1.71$ are referred to as ``metal-rich'' (depicted as
open red diamonds). The mean magnitudes and colours corresponding to the objects of the sub-sample are calculated from the sum of the fluxes of individual objects in each bin.

The metal-poor and metal-rich sub-samples
have mean [Fe/H] values of $-2.08$ and $-1.46$ dex, respectively,
and are presented along with the resulting metallicities (filled blue
circle for the metal-poor clusters and filled red diamond for the
metal-rich ones respectively). ESO121\,--\,SC03, the $\sim$\,9~Gyr cluster
in the LMC, is also shown on these figures as a green triangle. The metal-rich and metal-poor globular cluster colours clearly
separate in the optical-NIR colour-colour space, and generally follow the
model predictions. 

To provide a quantitative measure (and a summary) of the performance
of the different models in terms of fitting the average colours of the
various subsamples, we compiled the relevant information in
Table~\ref{tab:mod} (for {\it all\/} age groups).  The mean colours of
the composite clusters are listed there along with the interpolated
model predictions for the ages and metallicities corresponding to the
composite points (using linear interpolation between adjacent model
isochrones and iso-metallicity grid lines), as well as the measured
colour differences between the data and the model predictions (we will
hereafter refer to these colour differences as ``absolute colour
residuals'').  In Table~\ref{tab:mod} we also present the inferred
ages and metallicities of the composite clusters according to the
models. In some cases it is not possible to derive reliable stellar
population properties due to insufficient coverage of the model grids
(those are left blank in the table). For the old metal-rich cluster
sub-sample ($<$[Fe/H]$>=-1.46$) {\it all\/} models \citep{vazdekis99,
bc03, maraston05} infer ages exceeding the corresponding oldest model
isochrones. \citeauthor{maraston05} models with BHB are in a good
agreement with the mean composite cluster ages for the entire old
cluster sample and the metal-poor ($<$[Fe/H]$>=-2.08$) sub-sample,
although the inferred metallicities are a little lower than the mean
values for each composite point.

Inspection of Table~\ref{tab:mod} shows that the models of 
\cite{maraston05} (with BHB) provide the best overall match to the
observations of old clusters, 
especially in terms of estimating their ages,
while the \cite{bc03}
models formally do the best job of estimating SSP metallicities.

\label{oldclusters}

%%%%%%%%%%%%%%%%%%%%%%%%%%%%%%%%%%%%%%%%%%%%%%%%%%%%%%%%%%%%%%%%%%%%%%
\subsection{The 2 Gyr $\le$ Age $<$ 10 Gyr Cluster Population}

Nine LMC and six SMC clusters with adequate age and metallicity
measurements were selected in this age interval. Information on the
cluster ages, metallicities, and masses 
are compiled in Tables~B9 and B10. A comparison between our photometry
and the model tracks in $V-J$ vs.\ $J-K_S$ and $B-J$ vs.\ $J-K_S$
colour-colour diagrams are shown in Figures~\ref{fig:2_10vdat} and
\ref{fig:2_10bdat}, respectively. We further divide the $2-10$~Gyr
sample into two subsamples, with ages older than and younger than
4.6 Gyr. 
When compared with the \cite{maraston05} models,
these show that generally the younger subsample agrees with the
predicted location of the $2$~Gyr model, and that the older bin also
falls in the expected region of colour-colour space. 

Because this bin includes a large range of ages and some of the
individual clusters have stellar mass estimates lower than
$\cal{M}_{\it LLL}$ (see Section~\ref{stoceffects}), one can expect
some scatter among the individual measurements with respect to the
model predictions, which is indeed observed. Most of our mass
estimates in this age bin are systematically lower than those of
\cite{mclvdm05} (see \S~\ref{oldclusters} for possible
explanation). The composite clusters representing the entire sample in
this age range and the metal-poor older subpopulation (4.6 Gyr $\le$
Age $<$ 10 Gyr) both accumulate masses exceeding $\cal{M}_{\rm
(10\%)}$, so that the composite colours ought to yield meaningful
comparisons to the SSP model predictions. Unfortunately this is not
quite the case for the younger metal-rich subpopulation (2 Gyr $\le$
Age $<$ 4.6 Gyr).  
These clusters do however,
exceed $\cal{M}_{\it LLL}$ by a factor
of $\sim10$, so while we might expect some bias in the colors
it is certainly worth presenting information in this age
range in Table~\ref{tab:mod}. 
We point out that these results must
be treated with some caution (also noted in the table). Reliable ages
and metallicities are needed for more clusters in this particular age
interval in order to provide more accurate comparison between model
predictions and observed properties of the stellar populations.

All four sets of models reproduce the mass-weighted average colours
reasonably well in terms of quantitative colour differences (see
Table~\ref{tab:mod}).  Ages and metallicities inferred by {\it all\/}
SSP models for the three composite clusters in this age interval are
also listed in the table.  There is some degeneracy present in the
\cite{maraston05} models in this regime due to the intersection of the
10 Gyr and 4 Gyr isochrones in colour-colour space. Taking this into
account, it is not possible to infer a single value for the stellar
population's age or metallicity. Stellar population properties
measured with respect to different \citeauthor{maraston05} isochrones
are listed in Table~\ref{tab:mod}. There is good agreement between the
mean ages of the composite clusters and the ages infered by the model,
but the corresponding metallicities are typically lower. Generally all
models provide consistent stellar population properties between both
colour-colour diagrams typically showing somewhat lower ages and
higher metallicities. In some cases differences are observed in the
inferred properties of the stellar populations for some sub-samples in
this age interval (see Table~\ref{tab:mod}), when predictions are
based on $(V-J)$ vs.\ $(J-K_S)$ vs. the $(B-J)$ vs.\ $(J-K_S)$
colour-colour diagrams. 
Note that all models
(except \citeauthor{maraston05}) infer super-solar metallicities for
the (2 Gyr $\le$ Age $<$ 4.6 Gyr) subpopulation. Further investigation
is needed to determine whether 
it is a model ingredient or assumption that is responsible for
the mismatch, or whether this mismatch occurs
due to insufficient mass accumulated in the age bin.

Overall the \cite{maraston05} models provide the best inference about
the properties of the stellar populations in this age range.

\setcounter{figure}{4}
\begin{figure}
\centering
\includegraphics[bb=14 14 287 785,width=6.4cm]{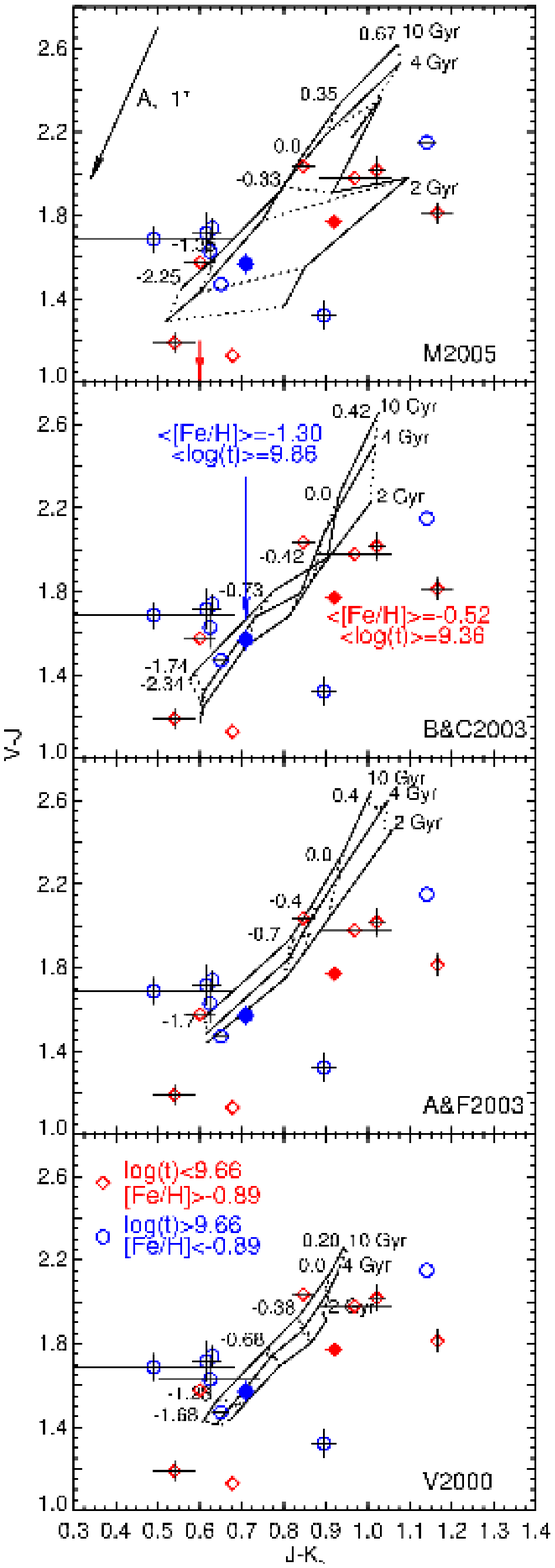}
\caption{$(V-J)$ vs. $(J-K_S)$ colour-colour diagrams for the clusters
  between 2 and 10 Gyr. The isochrones for three different ages are
  plotted with solid lines and metallicity values are marked along the
  model tracks for 10 Gyr. Black dotted lines stand for equal
  metallicity. The four panels show four SSP models as in
  Figure~\ref{fig:oldvdat}. A reddening vector for $A_{V}=1^{m}$ is
  shown in the 
  top panel. A legend to the individual symbols is provided in the
  bottom panel. The composite sub-samples are marked with filled
  symbols, along with their errors (colour-coded solid lines). The mean
  age and metallicity for each sub-sample are indicated in the second
  panel. The red arrow points to the position of {\bf Hodge 4},
  falling outside the boundaries of the plotted colour-colour space.} 
\label{fig:2_10vdat}
\end{figure}

\setcounter{figure}{5}
\begin{figure}
\centering
\includegraphics[bb=14 14 287 785,width=6.4cm]{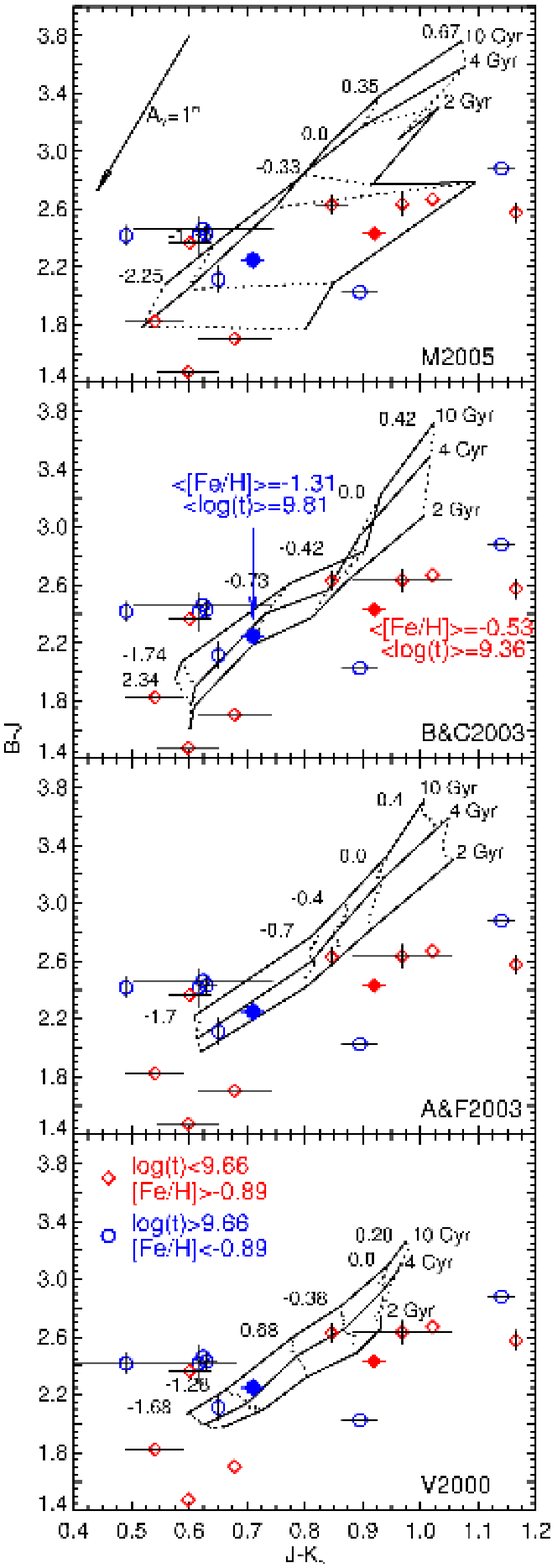}
\caption{As Figure\ref{fig:2_10vdat}, but for $(B-J)$ vs. $(J-K_S)$.}
\label{fig:2_10bdat}
\end{figure}

\label{2_10}

%%%%%%%%%%%%%%%%%%%%%%%%%%%%%%%%%%%%%%%%%%%%%%%%%%%%%%%%%%%%%%%%%%%%%%
\subsection{The 1 Gyr $\le$ Age $<$ 2 Gyr Cluster Population}

Eleven LMC and two SMC 
clusters with adequate age and metallicity
measurements from CMDs and/or spectroscopy of individual giant stars
were selected in this age interval (see Tables~B9 and B10).  Our
photometry and the model tracks in $V-J$ vs.\ $J-K_S$ and $B-J$ vs.\
$J-K_S$ colour-colour space are compared in Figures~\ref{fig:1_2vdat}
and \ref{fig:1_2bdat}.  This age interval covers a period after the
``AGB phase transition'', i.e., the onset of the AGB phase and
coincides with the development of the RGB. Therefore a significant
impact on the light output in the NIR (especially in the $K$-band) is
expected. Hence it is important to maximize the number of objects so
as to accumulate enough mass and decrease the effects of the
stochastic fluctuations in the stellar population. 
We added five clusters to our composite based 
on their recalibrated
`''S-parameter'' values (the S-parameter is an age indicator based
on integrated $U-B$ and $B-V$ colours, \citep[see][]{elson_fall_85,girardi95}).
Our recalibration of the S-parameter based on recent cluster age
determinations is described in detail in Appendix~A.
The total mass of the composite
cluster with ages between 1 and 2 Gyr is 
${\cal{M}}_{\rm tot} = 3.6 \times 10^5{\cal{M}}_{\odot}$, which exceeds ${\cal{M}}_{(10\%)}$.

Overall, the colours of clusters in this age range agree reasonably
well with the SSP model predictions.  Maraston models
most accurately reproduce
the age of the composite cluster, although it indicates
too low a metallicity. The latter is the reason why the Maraston model
has the largest colour residual for this age range in
Table~\ref{tab:mod}. The performance of the other three models in this
age range are similar to one another: They all indicate an age that is
somewhat too young and a metallicity that is somewhat too high, but
the absolute colour residuals are smaller than that of the Maraston
model. Quantitatively, the \citet{bc03} model comes out best in this
age range in terms of absolute colour residuals. 

\setcounter{figure}{6}
\begin{figure} 
\centering
\includegraphics[bb=14 14 287 785,width=6.4cm]{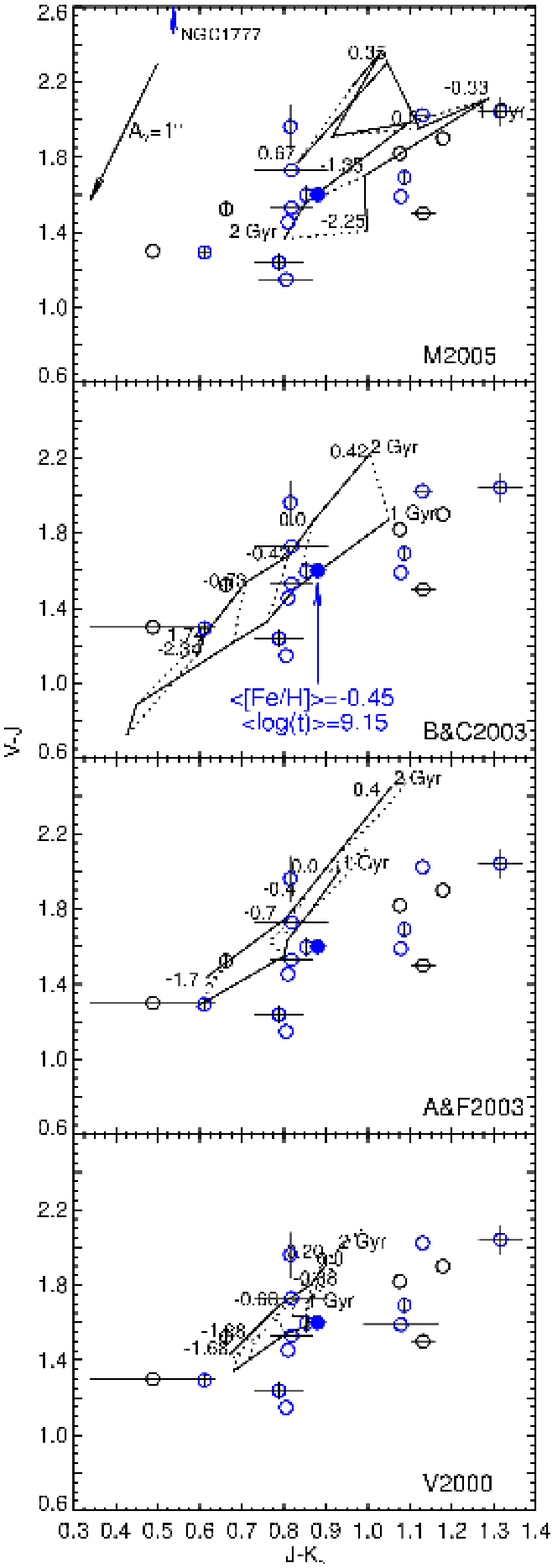}
\caption{$(V-J)$ vs. $(J-K_S)$ colour-colour diagrams for the clusters
between 1 and 2 Gyr. Isochrones for the different ages are plotted
with solid lines and metallicity values are marked along the 2 Gyr
isochrone. The four panels show four SSP models as in
  Figures~\ref{fig:oldvdat}\,--\,\ref{fig:2_10bdat}. The blue circles
  represent clusters with reliable age and metallicity estimates retrieved
  from the literature, while 
black circles denote the objects added to the sample on the basis of
our S-parameter recalibration (see Appendix A). The composite cluster is
marked by a filled circle, along with its error (solid blue lines). The mean age and
metallicity are indicated in the second panel.}
\label{fig:1_2vdat}
\end{figure}

\setcounter{figure}{7}
\begin{figure} 
\centering
\includegraphics[bb=14 14 287 785,width=6.4cm]{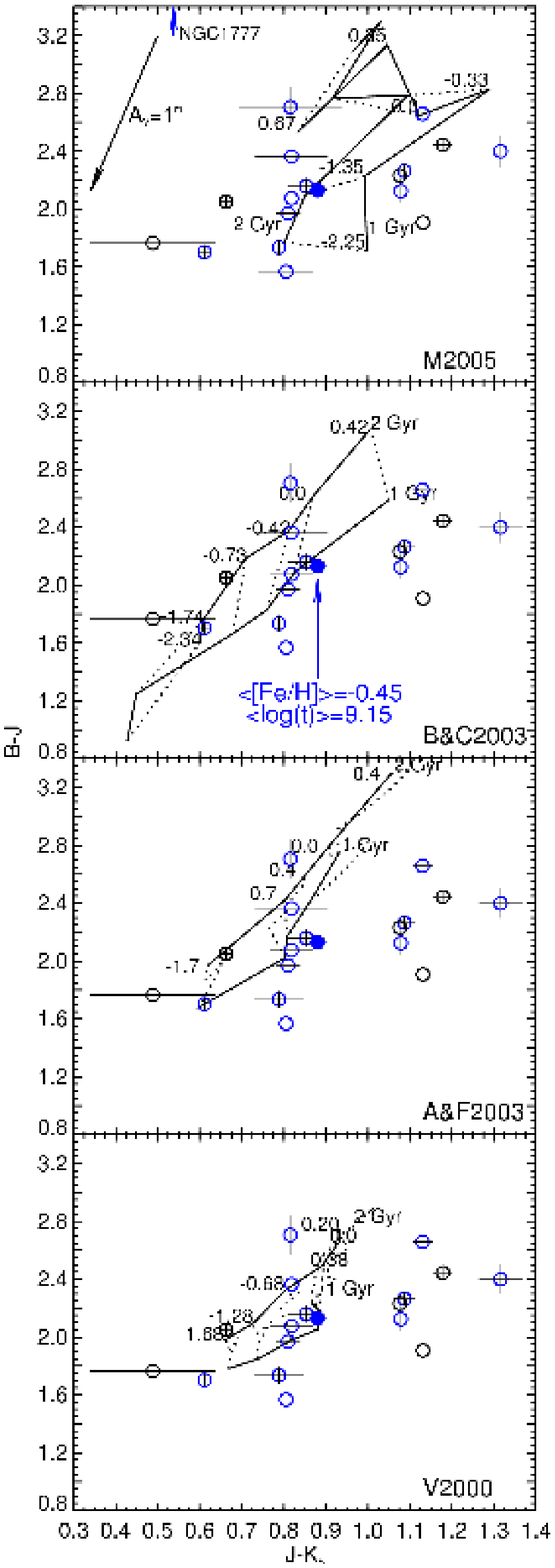}
\caption{Same as Figure~\ref{fig:1_2vdat}, but for $(B-J)$ vs. $(J-K_S)$} 
\label{fig:1_2bdat}
\end{figure}

\label{1_2}

%%%%%%%%%%%%%%%%%%%%%%%%%%%%%%%%%%%%%%%%%%%%%%%%%%%%%%%%%%%%%%%%%%%%%%

\subsection{The 200 Myr $\le$ Age $<$ 1 Gyr Cluster Population}

\setcounter{figure}{8}
\begin{figure} 
\centering
\includegraphics[bb=14 14 287 787,width=6.4cm]{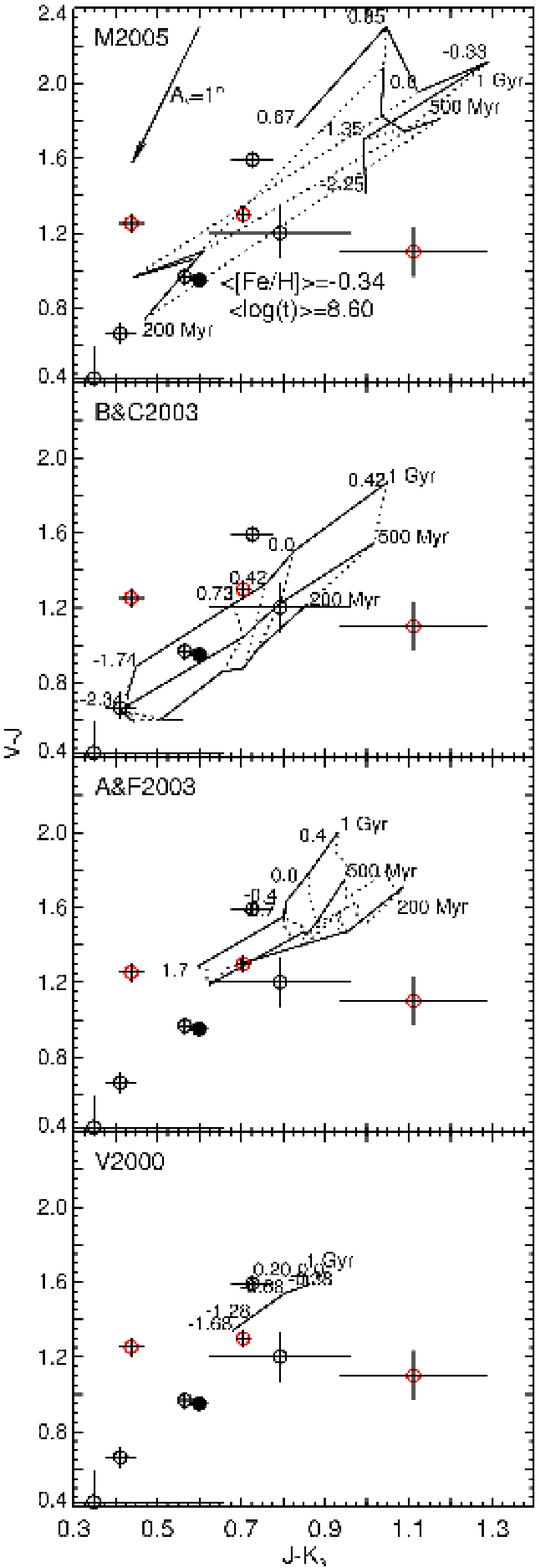}
\caption{$(V-J)$ vs. $(J-K_S)$ colour-colour diagrams for the clusters
between 200 Myr and 1 Gyr. The information (order of presentation of
the models, extinction) is shown as in the figures for the other age
bins. Three isochrones are plotted for each model (except for V2000
whose models do not include colour information for ages younger than 1
Gyr). The dotted lines show equal metallicities, and their values are marked 
along the 1 Gyr isochrone. The red circles represent clusters with
reliable age and metallicity estimates retrieved from the literature,
black circles denotes the objects added to the sample on the basis of
our S-parameter re-calibration. The composite cluster is marked with
filled black symbol, along with its error (solid lines). Its mean age and
metallicity are indicated at the top panel.}
\label{fig:02_1vdat}
\end{figure}

\setcounter{figure}{9}
\begin{figure} 
\centering
\includegraphics[bb=14 14 287 785,width=6.4cm]{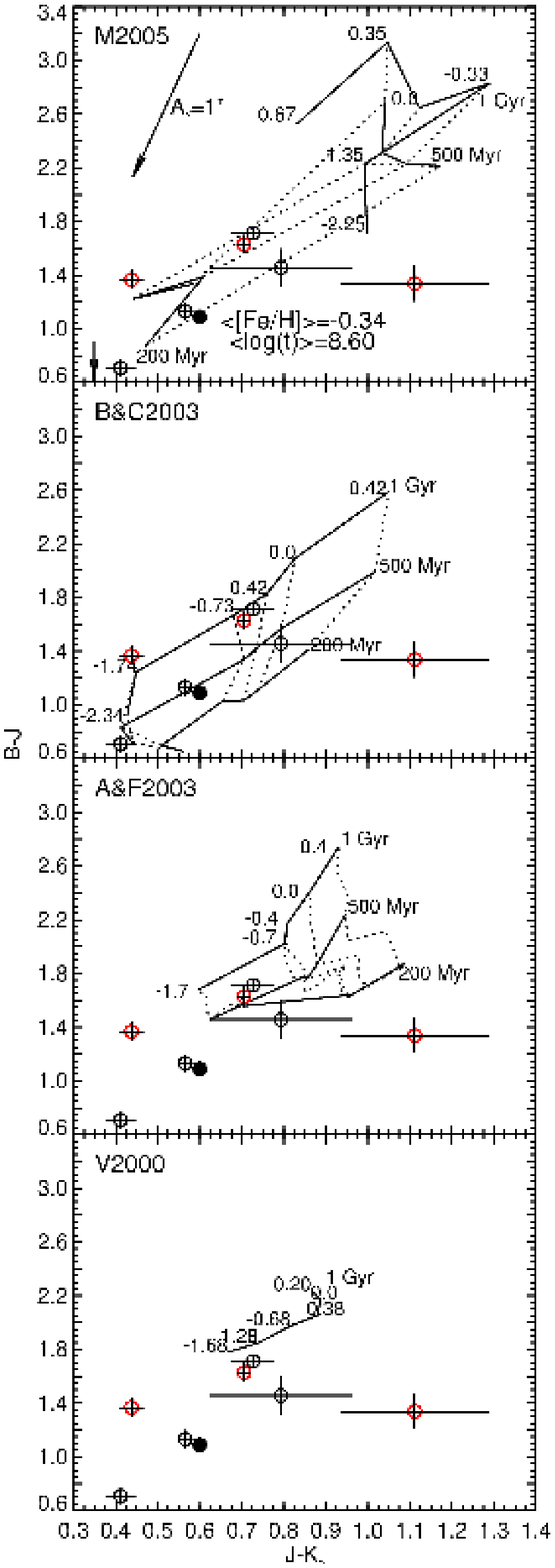}
\caption{$(B-J)$ vs. $(J-K_S)$ colour-colour diagrams for the clusters
  between 200 Myr and 1 Gyr. Generally the notes are the same as for the
  200 Myr - 1 Gyr $(V-J)$ vs. $(J-K_S)$ colour-colour diagram. The only
  difference is the black arrow in the top panel, pointing toward the
  position of NGC2156, which in this case is located out of the colour
  boundaries of the panels.} 
\label{fig:02_1bdat}
\end{figure}

In this age range, we identified only three clusters (2 in the LMC and
1 in the SMC) with reliable age and metallicity estimates based on
currently available deep CMDs
(See Table~\ref{tab:all}).  As in the previous
age bin, we added other clusters based on our recalibration of 
the S parameter
(see Appendix~A), resulting in the addition of 5 
objects to 
 this age bin. Model tracks in the
$V-J$ vs.\ $J-K_S$ and $B-J$ vs.\ $J-K_S$ colour-colour space are
presented along with our photometry results 
in
Figures~\ref{fig:02_1vdat} and \ref{fig:02_1bdat}.  The total mass of
the composite cluster of this age range is ${\cal{M}}_{\rm tot} = 1.1
\times 10^5 {\cal M}_{\odot}$, as listed in
Table~\ref{tab:mod}. Unfortunately this 
 barely 
exceeds
${\cal{M}_{\rm (10\%)}}$. More reliable age and metallicity estimates
are needed for Magellanic Cloud star clusters in this age interval to
be able to provide calibration data that could be crucial to improve
the treatment of the AGB phase and perform further tests of the
models.  
Overall, the \cite{bc03} 
and \cite{maraston05} models perform best in this age range.
The models of \cite{bc03} and \cite{maraston05} yield similar
metallicity estimates, which are {\em too low\/}. \cite{bc03}
overestimates slightly the mean age of the composite cluster,
\cite{maraston05} infer age that is younger. 
The \cite{bc03} models have the smallest colour residuals
in this age range.   

\label{02_1}

%%%%%%%%%%%%%%%%%%%%%%%%%%%%%%%%%%%%%%%%%%%%%%%%%%%%%%%%%%%%%%%%%%%%%%%%%%%%%%%%%%%%%%%%%%%%%
\subsection{The Complete Sample}

\setcounter{figure}{10}
\begin{figure} 
\centering
\includegraphics[bb=14 14 383 752,width=8.8cm]{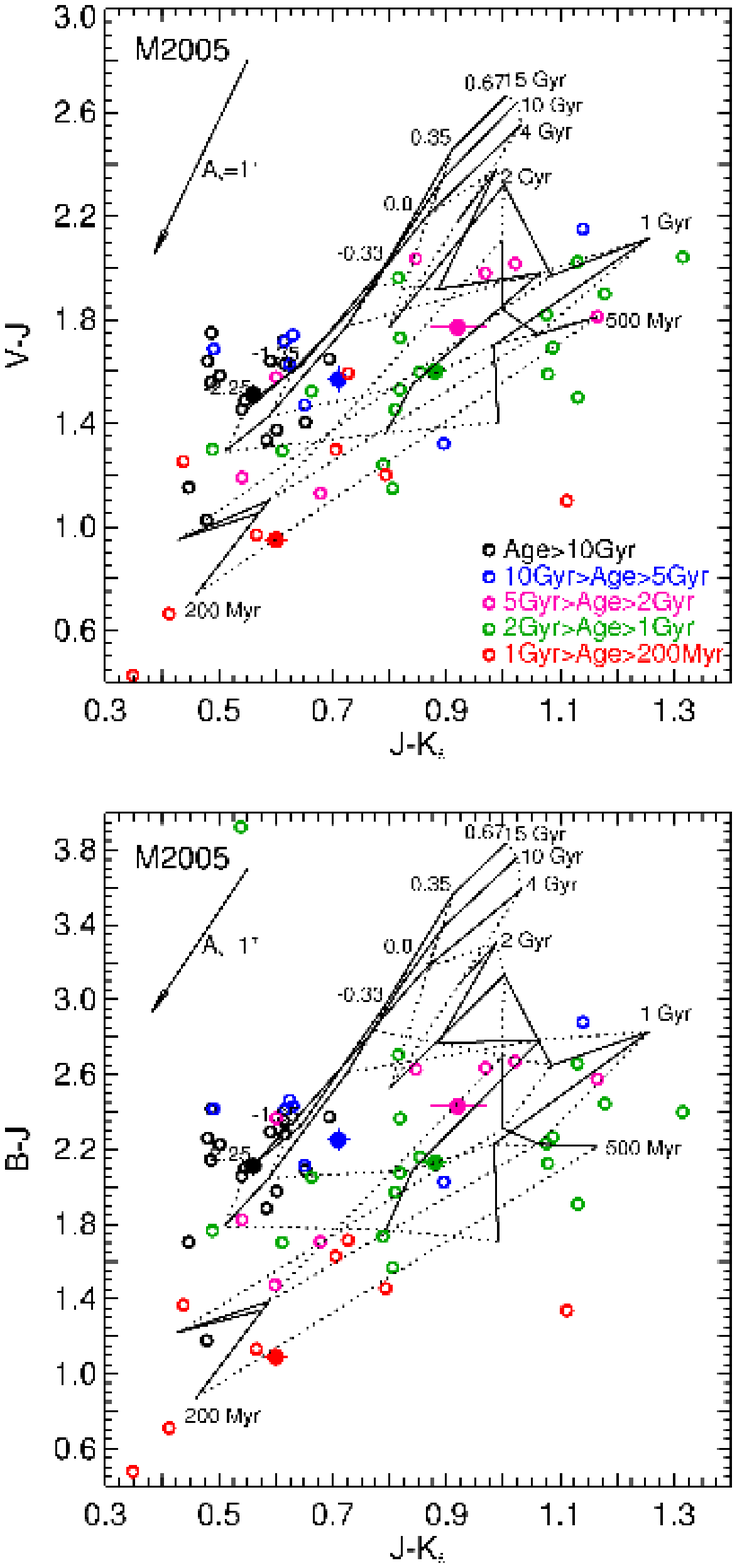}
\caption{Data of all 54 clusters from our test sample on top of the grid of Maraston 2005 models in the $(V-J)$ vs. $(J-K_{S})$ and $(B-J)$ vs. $(J-K_{S})$ colour-colour space. The individual points are colour-coded according to the age of the object (see the legend). The mean colours for each age bin are presented with filled circles. The error bars show the corresponding error of the mean colour.}
\label{fig:allobj}
\end{figure}

In Figure~\ref{fig:allobj} we present the entire cluster sample, from 200~Myr to
12~Gyr, studied in the previous sections.  One of the main attractions
for using a combination of reddening-corrected optical and NIR integrated colours is that
it is supposed to break the age-metallicity degeneracy.  Can it actually do
this?  First, we note the important caveat that in the Magellanic
Clouds, the typical cluster mass increases with age. This is 
partly
a statistical effect \citep[e.g.,][]{hunter03, whitmore07},
due to the fact that the younger age bins cover shorter times, and
fewer clusters formed originally in these shorter time intervals,
leading statistically to a 
somewhat
lower typical cluster mass at younger ages.
The large spread in intrinsic colours among young clusters is due in part to the low masses of
these clusters which naturally causes stochastic
fluctuations in the numbers of massive IR-luminous stars. Another contributor to the larger spread in colours is the fact that the IR-luminous stars in these objects (i.e., the AGB stars) have  shorter lifetimes and higher luminosities relative to their counterparts at older ages (i.e., RGB stars) .
Despite these effects, we find that
overall, clusters in different age bins do in fact, appear to occupy
different regions of colour-colour space, although with a relatively
large spread.  The solid points reflect the mean colour for
each age bin discussed in the previous sections, most of which 
have accumulated masses larger than 
$\cal{M}_{\rm(10\%)}$. 
The solid points in Figure~\ref{fig:allobj} suggest that the following
conclusions can be safely drawn for massive SSPs based solely on the
combination of $B-J$ and $J-K$ (or $V-J$ and $J-K$) colours: 
{\it (i)\/} ages older than roughly 9~Gyr can be
separated from those younger than $\approx 5$~Gyr; 
{\it (ii)\/}
ages of $\approx2$~Gyr can be adequately separated from those older than
$\approx5$~Gyr ; 
and {\it (iii)\/} finally,
ages younger than 1~Gyr separate nicely from those older than 1~Gyr.

\setcounter{table}{4}
\begin{landscape}
\begin{table}
\centering
\caption{Information about the composite clusters and model performance.}
\begin{minipage}{230mm}
\label{tab:mod}
\begin{tabular}{@{}cccrccccccrrrrrr@{}}
\hline
\multicolumn{3}{c}{COMPOSITE COLOURS (CC)} & \multicolumn{4}{c}{MODEL PREDICTIONS (MP)} & \multicolumn{3}{c}{CC -- MP}& \multicolumn{2}{c}{DISTANCE} & \multicolumn{4}{c}{POPULATION PROPERTIES}\\
$(B-J)_{0}$ & $(V-J)_{0}$ & $(J-K_{S})_{0}$ &  &  &  &  &  &  &  & & & \multicolumn{2}{c}{$(B-J)$ vs.} & \multicolumn{2}{c}{$(V-J)$ vs.}\\
$\sigma_{(B-J)_{0}}$  & $\sigma_{(V-J)_{0}}$  & $\sigma_{(J-K_{S})_{0}}$ & & & & & & & & & & \multicolumn{2}{c}{$(J-K_{s})$} & \multicolumn{2}{c}{$(J-K_{s})$} \\
 &  &  & {\bf Mod.}$^{\bf (a)}$ & $(B-J)$ & $(V-J)$ & $(J-K_{S})$ & $\Delta(B-J)$ & $\Delta(V-J)$ & $\Delta(J-K_{S})$ & $R_{B}$ & $R_{V}$ & $log(\tau)$ & [Fe/H] & $log(\tau)$ & [Fe/H]\\[3pt]
(1) & (2) & (3) & (4) & (5) & (6) & (7) & (8) & (9) & (10) & (11) & (12) & (13)\ \  & (14)\ \  & (15)\ \  & (16)\ \ \\
\hline
\multicolumn{16}{l}{{\bf Old globular clusters (Ages $\ge$ 10 Gyr):} 14 objects, $<[Fe/H]> = -1.71$, $log({\cal{M}}_{tot}) = 6.47^{+0.09}_{-0.11}$, $log({\cal{M}}_{LLL}) = 4.04$, $log({\cal{M}}_{(10\%)}) = 5.22$} \\[5pt]
 $2.11$ & $1.51$ & $0.56$ & {\bf AF03\ \ \ }     & $2.24$ & $1.57$ & $0.60$ & $-0.13$ & $-0.06$ & $-0.04$ & $4.77$ & $4.27$ & $ $ & $ $ & $ $ & $ $\\
 $0.05$ & $0.04$ & $0.01$ & {\bf BC03\ \ \ }     & $2.11$ & $1.44$ & $0.60$ & $\ \ 0.00$ & $+0.07$ & $-0.04$ & $4.00$ & $4.37$ & $>10.30$ & $-1.74$ & $>10.30$ & $-1.83$\\
               &               &               & {\bf M05\ \ \ }       & $2.20$ & $1.53$ & $0.60$ & $-0.09$ & $-0.02$ & $-0.04$ & $4.39$ & $4.03$ & $10.18$ & $-2.25$ & $10.18$ & $\sim -2.25$\\ [5pt]
\multicolumn{16}{l}{{\bf Old globular clusters (Ages $\ge$ 10 Gyr), [Fe/H] $<$ --1.71:} 5 objects$^{*}$, $<[Fe/H]> = -2.08$, $log({\cal{M}}_{tot}) = 6.01^{+0.11}_{-0.14}$, $log({\cal{M}}_{LLL}) = 4.00$, $log({\cal{M}}_{(10\%)}) = 5.18$} \\[5pt]
$2.02$ & $1.43$ & $0.53$ & {\bf BC03\ \ \ }  & $2.03$ & $1.42$ & $0.59$ & $-0.01$& $+0.01$ & $-0.06$ & $2.00$ & $2.00$ & $>10.30$ & $-2.10$ & $>10.30$ & $-2.34$\\ 
$0.12$ & $0.09$ & $0.03$ &  {\bf M05\ \ \ }    & $2.12$ & $1.49$ & $0.57$ & $-0.10$ & $-0.06$ & $-0.04$ & $1.57$ & $1.49$ & $\sim 10.10$ & $<-2.25$ & $10.10$ & $<-2.25$\\ [5pt]
\multicolumn{16}{l}{{\bf Old globular clusters (Ages $\ge$ 10 Gyr), [Fe/H] $\ge$ --1.71:} 7 objects$^{**}$, $<[Fe/H]> = -1.46$, $log({\cal{M}}_{tot}) = 6.24^{+0.08}_{-0.10}$, $log({\cal{M}}_{LLL}) = 4.07$, $log({\cal{M}}_{(10\%)}) = 5.26$} \\[5pt]
$2.27$ & $1.61$ & $0.58$ & {\bf V00\ \ \ }    & $2.16$ & $1.47$ & $0.61$ & $+0.11$ & $+0.14$ & $-0.03$ & $2.66$ & $3.81$ & $>10.25$ & $-1.50$ & $>10.25$ & $-1.28$\\
$0.05$ & $0.04$ & $0.02$ & {\bf AF03\ \ \ }  & $2.29$ & $1.60$ & $0.62$ & $-0.02$ & $+0.01$ & $-0.04$ & $2.04$ & $2.02$ & $>10.18$ & $<-1.70$ & $>10.18$ & $<-1.70$\\
              &               &              & {\bf BC03\ \ \ }  & $2.19$ & $1.53$ & $0.68$ & $+0.08$ & $+0.08$ & $-0.10$ & $5.25$ & $5.39$ & $>10.30$ & $-1.42$ & $>10.30$ & $-1.19$\\ 
              &               &               & {\bf M05\ \ \ }    & $2.30$ & $1.59$ & $0.63$ & $-0.03$ & $+0.02$ & $-0.05$ & $2.57$ & $2.55$ & $>10.18$ & $-1.66$ & $\sim 10.18$ & $-1.65$\\[5pt]
\multicolumn{16}{l}{{\bf 10 Gyr $>$ Age $\ge$ 2 Gyr sample:} 15 objects, $<[Fe/H]> = -0.89$,  $<log(t)>=9.66$, $log({\cal{M}}_{tot}) = 5.68^{+0.10}_{-0.13}$, $log({\cal{M}}_{LLL}) = 4.19$, $log({\cal{M}}_{(10\%)}) = 5.37$} \\[5pt]
$2.34$ & $1.67$ & $0.83$ & {\bf V00\ \ \ }    & $2.28$ & $1.61$ & $0.74$ & $+0.06$ & $+0.06$ & $+0.09$ & $9.12$ & $9.22$ & $9.30$ & $-0.54$ & $9.30$ & $-0.68$\\
$0.04$ & $0.03$ & $0.01$ & {\bf AF03\ \ \ }  & $2.36$ & $1.68$ & $0.73$ & $-0.02$ & $-0.01$ & $+0.10$ & $10.01$ & $10.01$ & $9.18$ & $-0.28$ & $9.18$ & $-0.70$\\
              &              &               & {\bf BC03\ \ \ } & $2.19$ & $1.53$ & $0.68$ & $+0.15$ & $+0.14$ & $+0.15$ & $15.46$ & $15.71$ & $9.22$ & $-0.12$ & $9.30$ & $-0.42$\\
              &              &               & {\bf M05}$^{\bf (b)}$ & $2.13$ & $1.48$ & $0.61$ & $+0.21$ & $+0.19$ & $+0.22$ & $22.62$ & $22.89$ & $9.71$ & $-0.65$ & $9.79$ & $-0.61$\\
              &              &               &                     &               &               &              &                 &                 &                 &               &              & $9.42$ & $-0.80$ & $9.54$ & $-0.78$\\[5pt]
\multicolumn{16}{l}{{\bf 10 Gyr $>$ Age $\ge$ 4.6 Gyr sample, [Fe/H] $<$ --0.88:} 7 objects, $<[Fe/H]> = -1.30$,  $<log(t)>=9.86$, $log({\cal{M}}_{tot}) = 5.55^{+0.11}_{-0.14}$, $log({\cal{M}}_{LLL}) = 4.11$, $log({\cal{M}}_{(10\%)}) = 5.22$} \\[5pt]
$2.25$ & $1.57$ & $0.71$ & {\bf V00\ \ \ }    & $2.18$ & $1.52$ & $0.66$ & $+0.07$ & $+0.05$ & $+0.05$ & $2.76$ & $2.69$ & $9.74$ & $-1.05$ & $9.48$ & $-1.08$\\
$0.06$ & $0.05$ & $0.02$ & {\bf AF03\ \ \ }  & $2.26$ & $1.59$ & $0.65$ & $-0.01$ & $-0.02$ & $+0.06$ & $3.01$ & $3.03$ & $9.48$ & $-0.96$ & $9.30$ & $-0.96$\\
              &              &               & {\bf BC03\ \ \ } & $2.08$ & $1.43$ & $0.62$ & $+0.17$ & $+0.14$ & $+0.09$ & $5.32$ & $5.30$ & $9.43$ & $-0.73$ & $9.38$ & $-0.73$\\ 
              &              &               & {\bf M05}$^{\bf (b)}$    & $2.23$ & $1.53$ & $0.62$ & $+0.02$ & $+0.04$ & $+0.09$ & $4.51$ & $4.57$ & $9.91$ & $-0.74$ & $9.90$ & $-0.75$\\ 
              &              &               &                          &                &              &              &                  &                 &            &                &               & $9.52$ & $-1.35$ & $9.51$ & $-1.35$ \\[5pt]
\multicolumn{16}{l}{{\bf 4.6 Gyr $>$ Age $\ge$ 2 Gyr sample, [Fe/H] $\ge$ --0.88:} 8 objects, $<[Fe/H]> = -0.52$,  $<log(t)>=9.36$, $log({\cal{M}}_{tot}) = 5.08^{+0.08}_{-0.09}$, $log({\cal{M}}_{LLL}) = 3.98$, $log({\cal{M}}_{(10\%)}) = 5.28$} \\[5pt]
$2.43$ & $1.77$ & $0.92$ & {\bf V00}$^{\bf (c)}$    & $2.31$ & $1.67$ & $0.76$ & $+0.12$ & $+0.10$ & $+0.16$ & $8.54$ & $8.67$ & $9.15$ & $0.20$ & $$ & $$ \\
$0.04$ & $0.03$ & $0.02$ & {\bf AF03\ \ \ }  & $2.56$ & $1.84$ & $0.83$ & $-0.13$ & $-0.07$ & $+0.09$ & $5.86$ & $5.07$ & $8.89$ & $0.40$ & $8.84$ & $0.00$ \\ 
             &               &               & {\bf BC03\ \ \ } & $2.40$ & $1.66$ & $0.77$ & $+0.03$ & $+0.11$ & $+0.15$ & $7.54$ & $8.35$ & $9.21$ & $0.21$ & $9.17$ & $0.20$ \\ 
             &               &               & {\bf M05}$^{\bf (d)}$    & $2.44$ & $1.74$ & $0.90$ & $-0.01$ & $+0.03$ & $+0.02$ & $1.03$ & $1.41$ & $9.50$ & $-0.59$ & $9.60$ & $-0.65$ \\
             &               &               &                          &               &               &               &                &               &                   &              &          & $9.37$ & $-0.61$ & $9.38$ & $-0.57$ \\
             &               &              &                          &               &               &               &                &               &                   &              &           &               &               & $8.62$ & $0.35$ \\[5pt]
\multicolumn{16}{l}{{\bf 2 Gyr $>$ Age $\ge$ 1 Gyr sample:} 17 objects$^{***}$, $<[Fe/H]> = -0.45$,  $<log(t)>=9.15$, $log({\cal{M}}_{tot}) = 5.55\pm0.06$ , $log({\cal{M}}_{LLL}) = 3.75$, $log({\cal{M}}_{(10\%)}) = 5.11$} \\[5pt]
$2.13$ & $1.60$ & $0.88$ & {\bf V00\ \ \ }    & $2.35$ & $1.74$ & $0.86$ & $-0.22$ & $-0.14$ & $+0.02$ & $7.40$ & $7.07$ & $9.00$ & $-0.21$ & $$ & $$\\
$0.03$ & $0.02$ & $0.02$ & {\bf AF03\ \ \ }  & $2.40$ & $1.75$ & $0.81$ & $-0.27$ & $-0.15$ & $+0.07$ & $9.06$ & $8.28$ & $8.88$ & $0.13$ & $8.80$ & $0.05$\\ 
              &              &               & {\bf BC03\ \ \ } & $2.24$ & $1.61$ & $0.83$ & $-0.11$ & $-0.01$ & $+0.05$ & $4.44$ & $2.55$ & $8.77$ & $0.15$ & $9.00$ & $0.15$\\
              &              &               & {\bf M05}$^{\bf (e)}$    & $2.64$ & $1.94$ & $1.14$ & $-0.51$ & $-0.34$ & $-0.26$ & $21.40$ & $21.40$ & $9.26$ & $-1.35$ & $9.26$ & $-1.35$\\ 
              &              &               &                          &               &              &               &                &                   &                  &               &             & $8.52$ & $0.24$ & $8.59$ & $0.00$\\
\hline
\end{tabular}
{\it Notes to Table~5:\/} See page 18.
\end{minipage}
\end{table}
\end{landscape}

\setcounter{table}{4}
\begin{landscape}
\begin{table}
\centering
\caption{Information about the composite clusters and model performance.}
\begin{minipage}{230mm}
\label{tab:mod}
\begin{tabular}{@{}cccrccccccrrrrrr@{}}
\hline
\multicolumn{3}{c}{COMPOSITE COLOURS (CC)} & \multicolumn{4}{c}{MODEL PREDICTIONS (MP)} & \multicolumn{3}{c}{CC -- MP}& \multicolumn{2}{c}{DISTANCE} & \multicolumn{4}{c}{POPULATION PROPERTIES}\\
$(B-J)_{0}$ & $(V-J)_{0}$ & $(J-K_{S})_{0}$ &  &  &  &  &  &  &  & & & \multicolumn{2}{c}{$(B-J)$ vs.} & \multicolumn{2}{c}{$(V-J)$ vs.}\\
$\sigma_{(B-J)_{0}}$  & $\sigma_{(V-J)_{0}}$  & $\sigma_{(J-K_{S})_{0}}$ & & & & & & & & & & \multicolumn{2}{c}{$(J-K_{s})$} & \multicolumn{2}{c}{$(J-K_{s})$} \\
 &  &  & {\bf Mod.}$^{\bf (a)}$ & $(B-J)$ & $(V-J)$ & $(J-K_{S})$ & $\Delta(B-J)$ & $\Delta(V-J)$ & $\Delta(J-K_{S})$ & $R_{B}$ & $R_{V}$ & $log(\tau)$ & [Fe/H] & $log(\tau)$ & [Fe/H]\\[3pt]
(1) & (2) & (3) & (4) & (5) & (6) & (7) & (8) & (9) & (10) & (11) & (12) & (13)\ \  & (14)\ \  & (15)\ \  & (16)\ \ \\
\hline
\multicolumn{16}{l}{{\bf 1 Gyr $>$ Age $\ge$ 200 Myr sample:} 8 objects, $<[Fe/H]> = -0.34$,  $<log(t)>=8.60$, $log({\cal{M}}_{tot}) = 5.04^{+0.05}_{-0.06}$, $log({\cal{M}}_{LLL}) = 3.60$, $log({\cal{M}}_{(10\%)}) = 5.05$} \\[5pt]
$1.09$ & $0.95$ & $0.60$ & {\bf AF03\ \ \ }  & $1.85$ & $1.59$ & $0.95$ & $-0.76$ & $-0.64$ & $-0.35$ & $25.83$ & $27.59$ & $ $ & $$ & $ $ & $$\\
$0.04$ & $0.03$ & $0.02$ & {\bf BC03\ \ \ } & $1.24$ & $1.01$ & $0.72$ & $-0.15$ & $-0.06$ & $-0.12$ & $7.08$ & $6.33$ & $8.64$ & $-0.89$ & $8.78$ & $-0.89$\\ 
              &              &               & {\bf M05\ \ \ }    & $2.10$ & $1.66$ & $1.01$ & $-1.01$ & $-0.71$ & $-0.41$ & $32.52$ & $31.31$ & $8.40$ & $-1.35$ & $8.41$ & $-1.35$\\
\hline
\end{tabular}

{\it Notes to Table~5:\/} Columns (1) -- (3) list the
weighted mean colours of the composite clusters, along with their corresponding errors. The number of clusters
combined in each composite, their mean ages and metallicities plus the
total accumulated mass is shown on the line above the colours.  Masses associated with the Lowest Luminosity Limit ${\cal{M}}_{\it LLL}$ and 10\% accuracy limit ${\cal{M}}_{\rm (10\%)}$ are presented for comparison. Columns
(4) -- (12) show the results of the model and data comparison. Column
(4) denotes the model: {\bf M05} stands for \cite{maraston05}, {\bf
  BC03} for \cite{bc03}, {\bf AF03} for \cite{af03}, and {\bf V00}
for \cite{vazdekis99}. Columns (5) through (7) give the
interpolated colours predicted by the models for the mean ages and
metallicities of the composite clusters. The corresponding differences
in colour space (data minus model predictions) are given in columns
(8) -- (10). Finally columns (11) and (12) give a distance between the
positions predicted by the models and the composite cluster locations
calculated as $R_{X} = \sqrt{(\Delta C1/\sigma_{C1})^{2}+(\Delta
  C2/\sigma_{C2})^{2}}$, where $\Delta Ci$ is the corresponding colour
difference and $\sigma_{Ci}$ is the uncertainty of the mean
colour of the composite cluster. $R_{V}$ stands for the distance in
the $(J-K_{S})$ vs. $(V-J)$ colour-colour space, and $R_{B}$ for the
$(J-K_{S})$ vs. $(B-J)$ distance. 
The composite colours and corresponding uncertainties are
calculated by summing the flux of the clusters in the corresponding subpopulation. The cumulative mass
in each composite cluster is calculated as the sum of the individual
cluster masses. 

The following objects were excluded from the corresponding composite clusters (marked with asterisks in the table) due to possible foreground/background contamination and/or small aperture diameters  of the optical photometry.

*\hspace{12pt}NGC1939 is excluded from the mean.

**\hspace{7pt}NGC1928 is excluded from the mean.

*** NGC1777 is excluded from the mean.

(a) We did not extrapolate in cases when the ages and metallicities of
the composite clusters were out of the parameter space covered by the
models. This is affecting the comparisons in the cases of "old" globular
clusters and for the youngest age bin. A rough estimate of the model
performance in these cases can be obtained from the figures, showing
the data for the corresponding age bins (see Fig.\ref{fig:oldvdat},  \ref{fig:oldbdat}, \ref{fig:1_2vdat} -- \ref{fig:02_1bdat}).

(b) In this case there is degeneracy in the \cite{maraston05} model predictions, in sense that 10 and 4 Gyr isochrones are overlapping in the colour-colour space. The composite-cluster ages and metallicities inferred by the models with respect to the 4 Gyr isochrone are listed in the table on the row below the model predictions for the 10 Gyr isochrone.

(c) Some degeneracy is present for the 1 Gyr isochrone of \cite{vazdekis99}, especially in the $(V-J)$ vs. $(J-K_{S})$ colour-colour space.

(d) The properties of the composite cluster for the younger sub-sample in the 10 Gyr $>$ Age $\ge$ 2 Gyr age bin are derived taking into account the degeneracy due to the intercept of the 10 and 4 Gyr isochrones. In the $(V-J)$ vs. $(J-K_{S})$ colour-colour diagram, the corresponding data also falls in the parameter space covered by the clusters with ages between 200 Myr and 500 Myr. Ages and metallicities inferred by the {\bf M05} model for each of these cases are listed in columns (13) -- (16) of the table: the first line (with respect to the 10 Gyr isochrone), second (4 Gyr isochrone) and third (200 -- 500 Myr case).

(e) Colour degeneracy between 1 -- 2 Gyr and 200 Myr -- 1 Gyr {\bf M05} models. The inferred stellar population properties are listed in the first (1 -- 2 Gyr case) and second (200 Myr -- 1 Gyr) line of columns (13) -- (16) in the table.

\end{minipage}
\end{table}
\end{landscape}

\label{sum}

\subsection{Age-Colour Comparison}

In this section we compare the observed cluster colours and the mean
sub-sample colours with model predictions as a function of age.
Integrated-light colour age evolution is a basic SSP model prediction and
Magellanic Cloud clusters are the only objects that allow these
predictions to be tested for intermediate ages. This information is shown
here to provide feedback to model builders by identifying ages at which
the discrepancies occur, hence pointing 
to the responsible stages of
stellar evolution and model ingredients that need further attention and
refinement. Results for several optical and NIR colours are presented in
Figures~\ref{fig:agebv} $-$ \ref{fig:agejk}.

The $(B\!-\!V)$ colour is presented in Figure~\ref{fig:agebv}. The
clusters in our test sample comply with the LLL criterion (see
Section~\ref{stoceffects}) in the optical and the distribution of the
individual points around the model tracks is reasonably tight. There is
good agreement between the properties of the composite clusters and the
model predictions. The correlation between age and optical colours is well
illustrated in this figure as well as the relatively poor metallicity
resolution.

Figures~\ref{fig:agebj}\,--\,\ref{fig:agejk} present combinations of
optical-NIR colours (except for $(J-K_{S})$ on
Figure~\ref{fig:agejk}). These illustrate that the majority of the
models tested in the present paper show a ``bump'' towards 
redder colors between
1 and 2 Gyr, which is likely due to the development of the RGB. The
exception appears to be the \cite{maraston05} model. However, the latter
shows a relatively pronounced effect of the AGB phase transition
(starting at a few hundred Myr). Note the steep increase of the
predicted colours, due to the increased flux in the NIR passbands.
This increased AGB contribution in the
\citeauthor{maraston05} models may be a reason why the 
development of the RGB is not as visually prominent as in the other models.

Also note that the colors of the composite and the individual clusters
in the youngest age bin are in much better agreement with the
predictions of the Maraston (2005) models if age estimates from
Girardi et al. are used (these are the ages which were available at
the time).  Our recalibration of the S-parameter described in Appendix
A shifts the ages in this interval to older values by ~0.2 dex,
introducing a slight discrepancy with the Maraston (2005) models,
but in good agreement with the Bruzual \& Charlot models. 
In
other words, the models of \citeauthor{maraston05} seem to reflect our
knowledge of the ages of Magellanic Cloud clusters 
prior to 2005. Our study may be used to make appropriate updates to the models.

\setcounter{figure}{11}
\begin{figure} 
\centering
\includegraphics[bb=14 14 348 745,width=8.2cm]{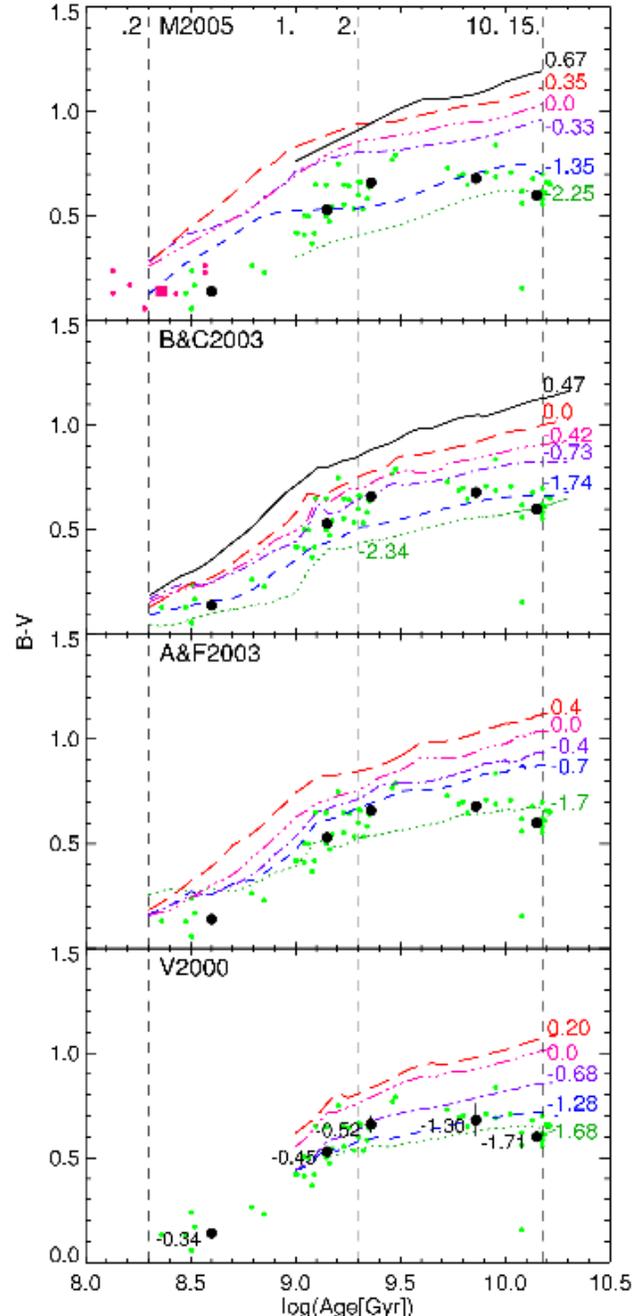}
\caption{$(B-V)$ colour predicted by the models of Maraston 2005, Bruzual \& Charlot 2003, Anders \& Fritze 2003 and Vazdekis 1999 (from top to bottom) as a function of age for different metallicities (labeled at the end of the colour-coded lines). Vertical dashed lines depict the boundaries of the age bins adopted in the present study. The reddening-corrected colours of the individual clusters from our sample are represented with small green dots. Solid black points show the mean colours of the sub samples, labeled by corresponding mean metallicities. The positions of the clusters in the youngest age bin when taking into account the ages derived by the S parameter calibration of Girardi et al. (1995) are shown as magenta dots on the top panel (the magenta square stands for the mean colour in this case). The prominent outlier in the oldest age bin is NGC 1928.}
\label{fig:agebv}
\end{figure}

\setcounter{figure}{12}
\begin{figure} 
\centering
\includegraphics[bb=14 14 348 745,width=8.2cm]{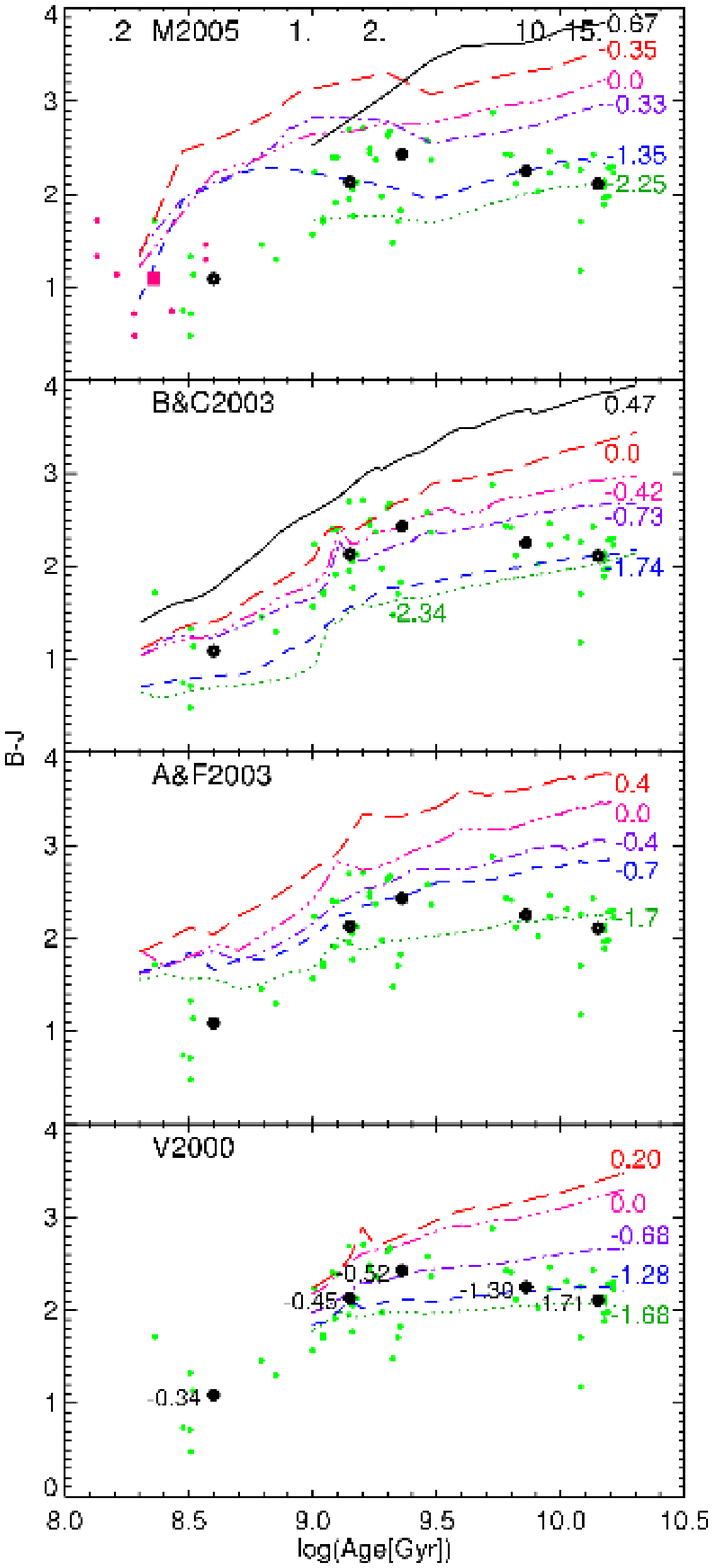}
\caption{The same as Figure~\ref{fig:agebv}, but for $(B-J)$ colour.}
\label{fig:agebj}
\end{figure}

\setcounter{figure}{13}
\begin{figure} 
\centering
\includegraphics[bb=14 14 348 745,width=8.2cm]{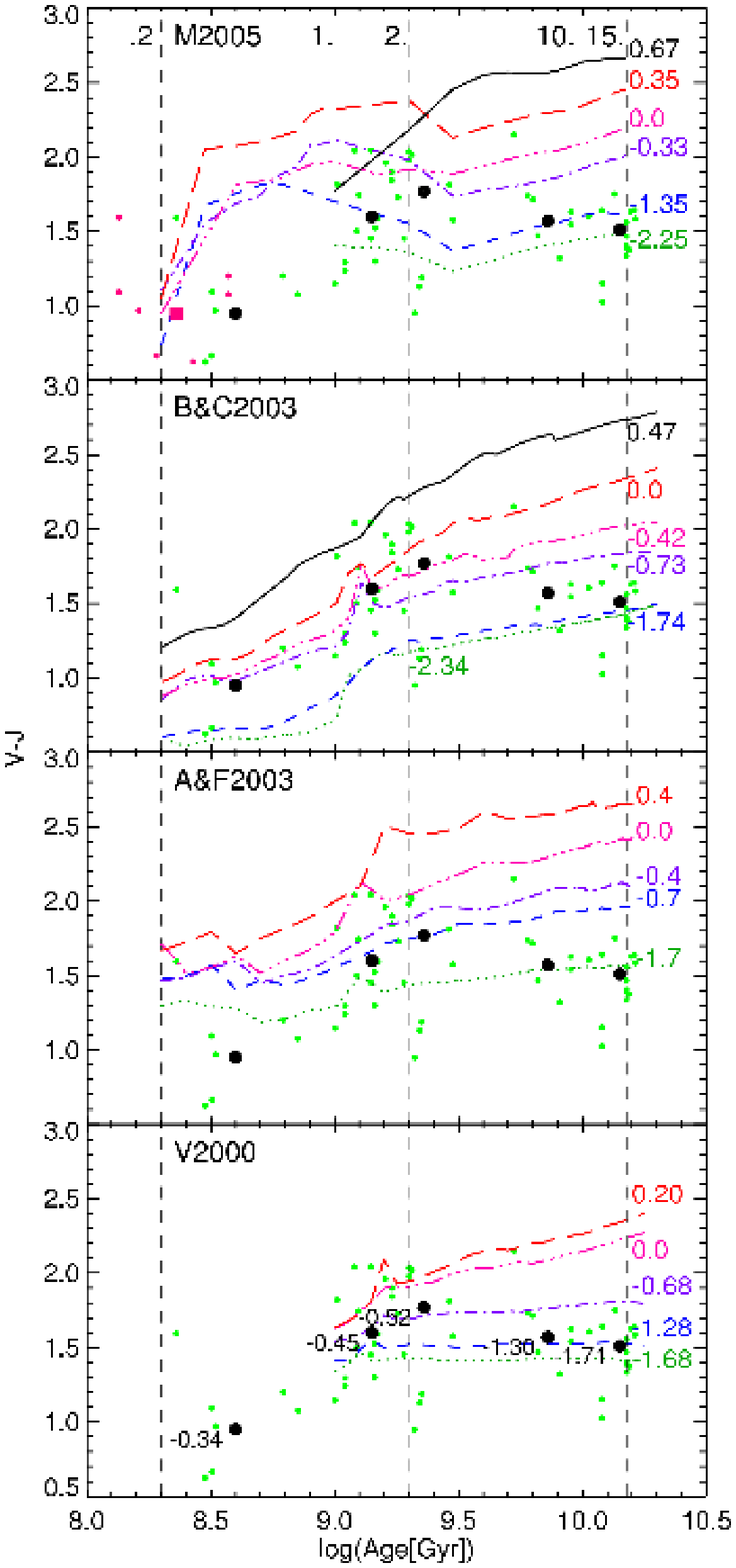}
\caption{The same as Figure~\ref{fig:agebv}, but for $(V-J)$ colour.}
\label{fig:agevj}
\end{figure}

\setcounter{figure}{14}
\begin{figure} 
\centering
\includegraphics[bb=14 14 348 745,width=8.2cm]{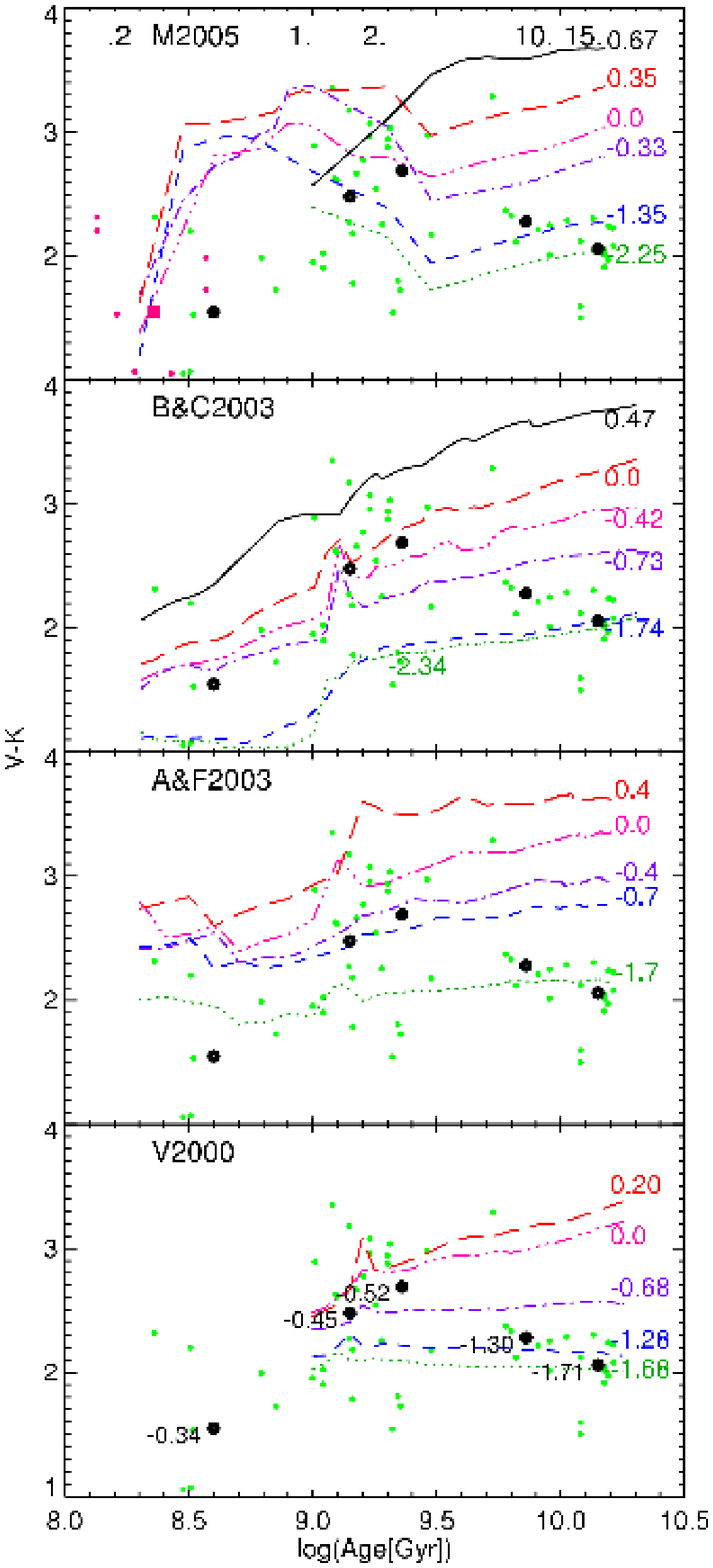}
\caption{The same as Figure~\ref{fig:agebv}, but for $(V-K_{s})$ colour.}
\label{fig:agevk}
\end{figure}

\setcounter{figure}{15}
\begin{figure} 
\centering
\includegraphics[bb=14 14 348 745,width=8.2cm]{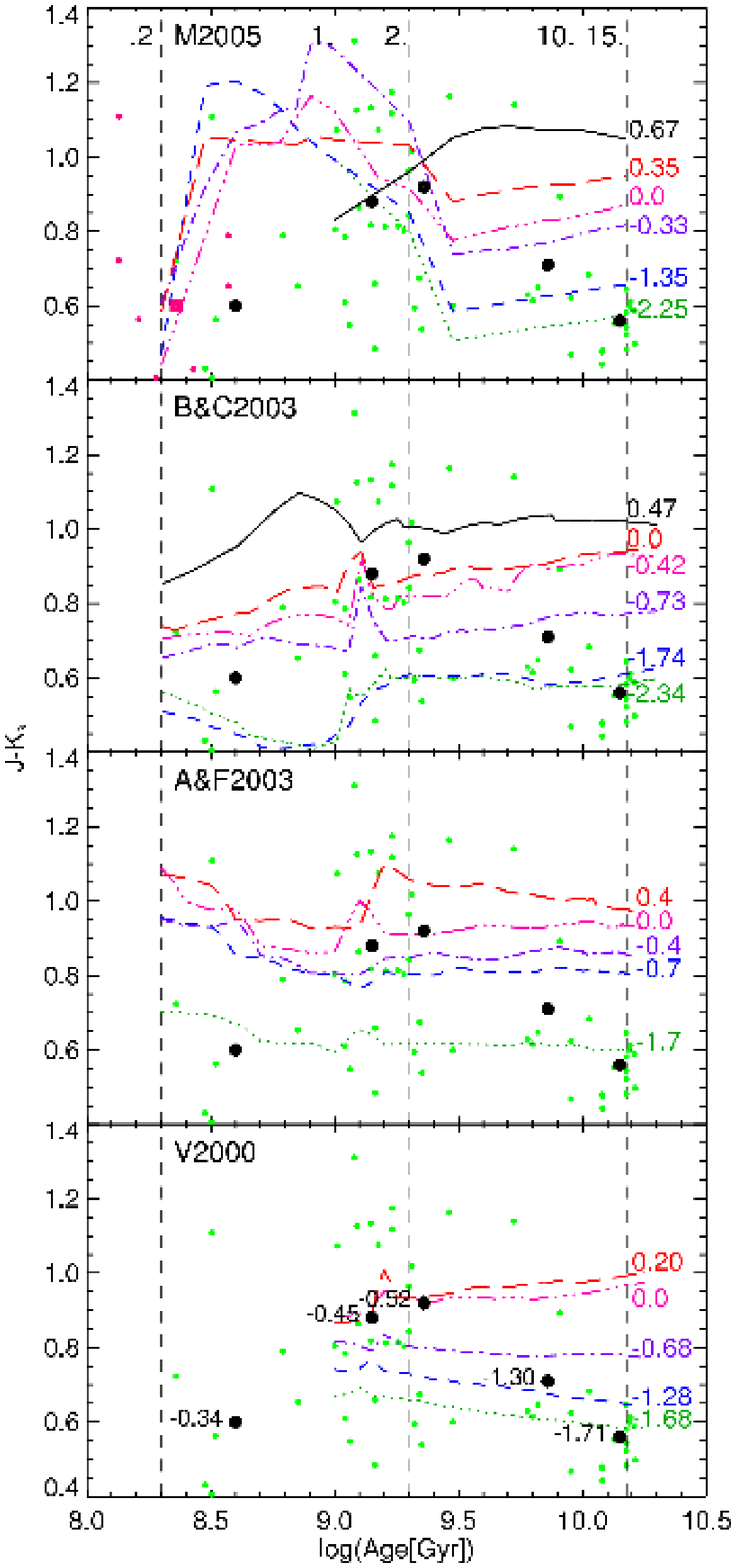}
\caption{The same as Figure~\ref{fig:agebv}, but for $(J-K_{s})$ colour.}
\label{fig:agejk}
\end{figure}

\section{Concluding Remarks}

We have presented new integrated $JHK_s$ 2MASS photometry for 9
Magellanic Cloud clusters, bringing our total sample (when combined
with the results of Paper I) to 54 clusters with reliable ages $\geq$200~Myr.
In addition, we compile integrated-light $B$
and $V$ photometric measurements, extinction estimates, and a
database of reliable age and metallicity determinations (mostly
recent results) from the literature for our sample clusters.  We divide the clusters into
different age (e.g., $\geq10$~Gyr, $3-9$~Gyr, $3-4$~Gyr, $1-2$~Gyr,
and 200~Myr$-$1~Gyr) and metallicity (when possible), and
quantify the observed spread in the intrinsic cluster colours in these
ranges. Care was taken to account for the spread of the
observational data around the model predictions due to the stochastic
fluctuations in the stellar populations of the clusters.  The smallest
spread in intrinsic colours is found for clusters with ages
$\ga 10$~Gyr, the colours of which are well-reproduced by all 
four sets of SSP model predictions.  The systematic shift between the model
predictions and the observed colours for a sample of old
Milky Way globular clusters reported by \cite{cohen07} is not
observed in our Magellanic Cloud cluster analysis.  The largest
spread in colour is found for clusters in the age range
$2-4$~Gyr.  We believe that much of the spread in the
colours for {\it individual\/} clusters younger than 10~Gyr results from stochastic
fluctuations in the numbers of infrared-luminous stars, since 
individual clusters tend to have less than ${\cal{M}}_{\rm (10\%)}$\footnote{stellar mass needed to decrease the the luminosity uncertainty due to stochastic effects in the stellar population to 10\%)}
contributing to the observed colours. 

Composite $(B-J)_0$, $(V-J)_0$, and $(J-K_s)_0$ cluster colours are
calculated for each age/metallicity interval, and compared with the
predictions of four widely used population synthesis models
\citep{maraston05, bc03, af03, vazdekis99}, in order to evaluate
their performance.  We interpolate the model grids to calculate the
offset or distance in colour-colour space between the model predictions
and the age and metallicity for our composite cluster colours.  All
four sets of models reproduce the colours of old ($\geq10$~Gyr)
Magellanic Cloud clusters quite well, with the \citet{maraston05} and 
\citet{bc03} models giving slightly better fits than the other two. 

In
the age range of 2\,--\,10 Gyr, the \citet{maraston05} models
have the largest separation in optical-NIR colour-colour space between
the 2~Gyr and 10~Gyr model tracks, which best reproduces our observed
composite colours in the $2-3$~Gyr and $3-9$~Gyr ranges.  While the
composite colour for $2-3$~Gyr-old clusters falls just off the grid 
for the other three models, actual quantitative distances between
the model predictions and composite cluster colours are not
significantly different among the four models. 
In the $1-2$~Gyr and $0.2-1$~Gyr age ranges, the \cite{bc03} models
generally give the best quantitative match to our composite Magellanic
Cloud cluster colours. 
Taking into account the inferred ages and metallicities, there is little difference between the \cite{bc03} and \cite{maraston05} model performance.  
The cluster colours fall off the \citet{af03} and \citet{vazdekis99} model predictions in the two youngest age ranges, largely due to their limited coverage at low metallicities.

Based on the comparisons presented in this work, it is found that each
model has strong and weak points when used to analyse the optical$+$NIR
colours of unresolved stellar populations. There is no model set that
clearly outperforms the others in all respects. Overall, the \cite{bc03}  and  typically yield the best quantitative match to our composite cluster colors. The \cite{maraston05} models are a close second. The same two models also yield the best match to the composite cluster ages and metallicities.

\section*{Acknowledgments}

The authors would like to thank the anonymous referee for useful comments
and suggestions that improved the paper. We thank D. Hunter for providing
access to her LMC and SMC optical photometry data that was instrumental
for the compilation of the optical photometry database used in the present
study. T.H.P. acknowledges support in form of a Plaskett Fellowship at the
Herzberg Institute of Astrophysics. This publication makes use of data
products from the Two Micron All-Sky Survey, which is a joint project of
the University of Massachusetts and the Infrared Processing and Analysis
Center/California Institute of Technology, funded by the National
Aeronautics and Space Administration and the National Science Foundation.
This publication makes use of SIMBAD astronomical database.

\appendix

\section[]{S parameter - age re-calibration}

There are currently very few CMD based ages for Magellanic Cloud
clusters in our youngest age bin ($200$~Myr-1~Gyr).  One possible
solution is to use the "S-parameter" calibration for LMC clusters, which would
allow us to select additional clusters for analysis in \S3.4.  The
S-parameter, introduced by \cite{elson_fall_85}, provides an empirical
relation between the age of a cluster and its integrated $(U-B)$ vs.
$(B-V)$ colours. \cite{girardi95} revised the S-parameter-age
calibration based on 24 clusters which had age estimates from high
quality {\it (at the time)\/} ground-based CCD observations.  They obtained the following
relation between S and cluster age (in log($\tau$)):

\begin{equation}
log(\tau / [years]) = (0.0733 \pm 0.0032)*S + 6.227 \pm 0.096
\label{eq:a1}
\end{equation}
With a rms dispersion in log($\tau$) equal to 0.137.

\setcounter{table}{5}
\begin{table}
\caption{Ages and S parameter values for the clusters from the Kerber et al. 2007 sample ( 25 $\le$ S $\le$ 40). }
\label{tab:s}
\begin{tabular}{@{}lcccc@{}}
\hline
& \multicolumn{2}{c}{Girardi et al.} & Kerber et al. & This paper \\
& \multicolumn{2}{c}{1995} & 2007 & \\
Cluster ID\ \ \ \ \  & S & log($\tau$) & log($\tau$) &  log($\tau$)\\
(1) & (2) & (3) & (4) & (5)\\
\hline
NGC1651   &  38  &  $9.00\pm0.14$  &  $9.30\pm0.03$  & $9.23\pm0.12$\\
NGC1718   &  40  &  $9.15\pm0.14$  &  $9.31\pm0.03$  & $9.38\pm0.12$\\
NGC1777   &  38  &  $9.00\pm0.14$  &  $9.06\pm0.04$  & $9.23\pm0.12$\\
NGC1831   &  32  &  $8.56\pm0.14$  &  $8.85\pm0.05$  & $8.79\pm0.12$\\
NGC1856   &  30  &  $8.42\pm0.14$  &  $8.47\pm0.04$  &  $8.65\pm0.12$\\
NGC1868   &  33  &  $8.64\pm0.14$  &  $9.05\pm0.03$  &  $8.87\pm0.12$\\
NGC2162   &  37  &  $8.93\pm0.14$  &  $9.10\pm0.03$  &  $9.16\pm0.12$\\
NGC2209   &  34  &  $8.71\pm0.14$  &  $9.08\pm0.03$  &  $8.94\pm0.12$\\
NGC2213   &  38  &  $9.00\pm0.14$  &  $9.23\pm0.04$  &  $9.23\pm0.12$\\
NGC2249   &  33  &  $8.64\pm0.14$  &  $9.00\pm0.03$  &  $8.87\pm0.12$\\
SL506         &  40  &  $9.15\pm0.14$  &  $9.35\pm0.03$  &  $9.38\pm0.12$\\[3pt]
NGC265    &  26   & $8.13\pm0.14$   &  $8.5\ \pm0.3\ $  &   $8.36\pm0.12$\\
\hline
\end{tabular}

{\it Notes to Table~\ref{tab:s}:} Column~(1) is the cluster ID. The S
parameter value and the corresponding age according to
\cite{girardi95} are listed in columns~(2) and (3). Column~(4) is the
log$(\tau)$ for the clusters from \cite{kerber07}. The age derived by
our calibration are given in column~(5).

\end{table}

\setcounter{figure}{16} 
\begin{figure} 
\includegraphics[bb=14 14 256 264,width=8.4cm]{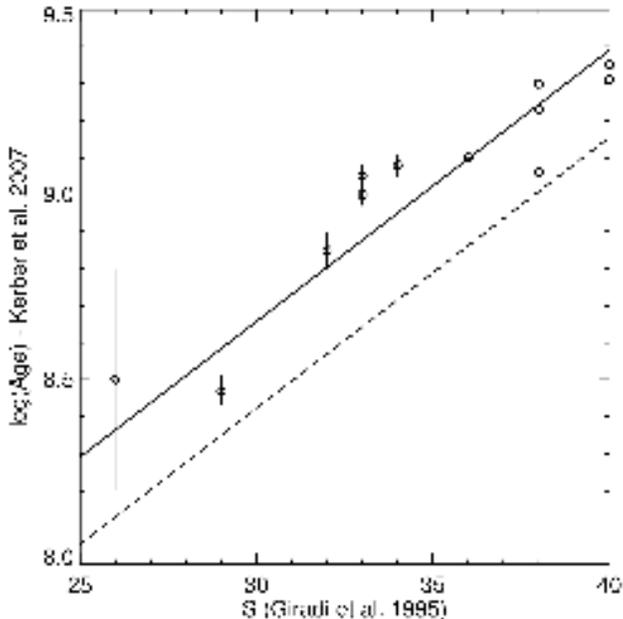} 
\caption{Ages from Kerber et al. 2007 and
Chiosi \& Vallenari 2007 as function of the S parameter. The dashed
line represent Girardi et al. 1995 calibration.  NGC265 (indicated
with a diamond) was added to improve the sampling in the 200 Myr - 1
Gyr age interval. The weighted mean offset of the data points with
respect of the original relation is shown with a solid line.}
\label{fig:s_logt} \end{figure}

We performed an independent analysis of the S-parameter using new age
estimates for Magellanic Cloud clusters derived from $HST$-based
CMDs by \cite{kerber07}.  These include the age range for which
we have few CMD-based ages ($\sim$ 0.3 Gyr to $\sim$ 1 Gyr).
The optical colours from Table~\ref{tab:comp} were used to derive S
parameter values according to the definition given by \cite{girardi95}. 
A comparison between the \citeauthor{kerber07} and \citeauthor{girardi95}
age estimates for 12 clusters shows that there is an offset, such
that the \citeauthor{kerber07} ages are systematically older. The uncertainty weighted mean offset of the \citeauthor{kerber07} ages
from the \citeauthor{girardi95} calibration is 0.235 in log($\tau$). 
We confirmed that the difference between the age estimates is {\it not\/}  caused by the adoption of different stellar isochrones \citep[][respectively for studies of \citeauthor{girardi95} and \citeauthor{kerber07}]{bertelli94,girardi02}, since they are indistinguishable from each other in the $B$ and $V$ bands which were used by \citet{kerber07}. \citeauthor{kerber07} have better data and apply a more robust technique for age determination, therefore we recalibrate the cluster ages with the appropriate offset:

\begin{equation}
log(\tau / [years]) = 0.0733*S + 6.458
\label{eq:a2}
\end{equation}

\noindent The standard deviation of \cite{kerber07} data around the
new relation is $\sigma$ = 0.123.

The objects selected to extend the sample based on the S parameter
values from equation A2 are listed in Table~\ref{tab:s_ext}. 

\setcounter{table}{6}
\begin{table}
\caption{Extention of the sample based on the S parameter.}
\label{tab:s_ext}
\begin{tabular}{@{}lcclcc@{}}
\hline
ID & S  & log($\tau$)& ID & S  & log($\tau$)\\
(1) & (2) & (3) & (1) & (2) & (3)\\
\hline
NGC1866	 	 & 27 & $8.43\pm0.12$ & NGC1644           & 37 & $9.16\pm0.12$\\
NGC2031	 	 & 26 & $8.36\pm0.12$ & NGC1783		 & 37 & $9.16\pm0.12$\\
NGC2107		 & 32 & $8.79\pm0.12$ & NGC1987		 & 35 & $9.01\pm0.12$\\
NGC2134		 & 28 & $8.50\pm0.12$ & NGC2108		 & 36 & $9.09\pm0.12$\\
NGC2156		 & 28 & $8.50\pm0.12$ & NGC2154 	 & 38 & $9.23\pm0.12$\\
\hline
\end{tabular}
{\it Notes to Table~\ref{tab:s_ext}:} Column~(1) is the cluster ID. Column~(2) lists the derived S parameter values, along with the calculated ages in column~(3).

\end{table}

%%%%%%%%%%%%%%%%%%%%%%%%%%%%%%%%%%%%%%%%%%%%%%%%%%%
\section[]{Defining the Test Sample} In order to test the performance
of SSP models we need to define a reliable test sample: Objects with
age and metallicity measurements based on high-quality data, covering
an appropriate parameter space. We prefer age and metallicity estimates
based on CMDs, particularly those taken with $HST$, but also include
metallicity estimates from individual stars, integrated spectroscopy 
and age estimates for the youngest two age bins from the S-parameter
(described in Appendix A).
In order to clarify possible age-related effects in the SSP models
performance, we decided to split the sample into several age bins:
{\it(i)} Globular clusters older than 10 Gyr; {\it(ii)} Clusters with
ages between 2 and 10 Gyr; {\it(iii)} Clusters older than 1 Gyr and
younger than 2 Gyr; {\it(iv)} Clusters with ages between 200 Myr and 1
Gyr.  The latter two age bins are where the effects of AGB and RGB stars on
the NIR integrated-light properties are most pronounced.

The intrinsic spread in integrated colours for clusters with a given
age and metallicity increases with decreasing mass
(\cite{bruzual2002}, due to stochastic fluctuations in the number of
massive stars.  Therefore, we consider clusters with masses exceeding $
{\cal{M}}_{\rm (10\%)}$ to be reliable test particles.  Since many clusters
(particularly in the youngest age bins) have mass estimates lower than
this value, we also present cumulative colours for each age bin, where the
measured colours correspond to masses exceeding this limit.

%%%%%%%%%%%%%%%%%%%%%%%%%%%%%%%%%%%%%%%%%%%%%%%%%%%
\subsection{The Old Cluster Population (Ages $\ge$ 10 Gyr)}

\setcounter{table}{7}
\begin{table*}
\centering
\begin{minipage}{170mm}
\caption{Information about the cluster test sample.}
\label{tab:all}
\begin{tabular}{@{}llclclcccccc@{}}
\hline
Cluster ID & \ \ \ \ [Fe/H] & & \ \ \ Age & & $E(B-V)$ & & $E(V-I)$ & & $A_{V\rmn{(CMD)}}$ & $A_{V\rmn{(MCPS)}}$ & Note\\
(1) & \ \ \ \ \ \ (2) & & \ \ \ \ (3) & & \ \ \ \ \ (4) & & (5) & & (6) & (7) & (8)\\
\hline
\multicolumn{5}{l}{\bf Old globular clusters (Ages $\ge$ 10 Gyr)}\\[3pt]
$LMC$\\[3pt]
NGC1466   & $-1.85\pm0.1\ $          & (22) &  $15\ \ \pm3$             & (11)  & $0.09\ \pm0.02$      & (22) &                        &      & $0.28\pm0.06$          & $0.39\pm0.02$ & O\\[2pt]
NGC1754   & $-1.42\pm0.15$          & (18) &  $15.6\pm2.2$             & (18)  & $0.09\ \pm0.02$      & (18) &                        &      & $0.28\pm0.06$          & $0.40\pm0.04$ & I\\[2pt]
NGC1786   & $-2.1\  \pm0.3\ $        & ( 2) &  $15\ \ \pm3$             & ( 2)  & $0.09\ \pm0.05$      & ( 2) & $0.12\pm0.05$          & ( 2) & $0.28\pm0.16$          & $0.62\pm0.04$ & I\\[2pt]
NGC1835   & $-1.62\pm0.15$          & (18) &  $16.2\pm2.8$             & (18)  & $0.08\ \pm0.02$      & (18) &                        &      & $0.25\pm0.06$          & $0.35\pm0.07$ & I\\[2pt]
NGC1841   & $-2.2\  \pm0.2$         & ( 2) &  $15\ \ \pm3$             & ( 2)  & $0.20\ \pm0.03$      & ( 2) & $0.28\pm0.03$          & ( 2) & $0.62\pm0.09$          & $0.39\pm0.02$ & O\\[2pt]
NGC1898   & $-1.18\pm0.16$          & (18) &  $13.5\pm2.2$             & (18)  & $0.07\ \pm0.02$      & (18) &                        &      & $0.22\pm0.06$          & $0.43\pm0.05$ & I\\[2pt]
NGC1916   &                         &      &                           &       &                      &      &                        &      &                        & $0.42\pm0.05$ & I\\[2pt]
NGC1928   & $-1.27\pm0.14$          & (13) &  $GC\pm2$                 & (13)  &                      &      & $0.08\pm0.02$          & (13) & $0.20\pm0.05$          & $0.43\pm0.03$ & I\\[2pt]
NGC1939   & $-2.10\pm0.19$          & (13) &  $GC\pm2$                 & (13)  &                      &      & $0.16\pm0.03$          & (13) & $0.40\pm0.08$          & $0.62\pm0.05$ & I\\[2pt]
NGC2005   & $-1.35\pm0.16$          & (18) &  $15.5\pm4.9$             & (18)  & $0.12\ \pm0.02$      & (18) &                        &      & $0.37\pm0.06$          & $0.47\pm0.04$ & I\\[2pt]
NGC2019   & $-1.23\pm0.15$          & (18) &  $16.3\pm3.1$             & (18)  & $0.12\ \pm0.02$      & (18) &                        &      & $0.37\pm0.06$          & $0.44\pm0.06$ & I\\[2pt]
NGC2210   & $-2.2\ \pm0.2$          & ( 2) &  $15\ \ \pm3$             & ( 2)  & $0.09\ \pm0.03$      & ( 2) & $0.12\pm0.03$          & ( 2) & $0.28\pm0.09$          & $0.39\pm0.02$ & O\\[2pt]
NGC2257   & $-1.7\ \pm0.2$          & (21) &  $10 - 14$                & (21)  & $\ \ \ \ \ \ 0.0 $   & (21) &                        &      & $0.00\pm0.00$          & $0.39\pm0.02$ & O\\[2pt]
Hodge11   & $-2.06 \pm0.2$          & (15) &  $15\ \ \pm3$             & (15)  & $0.075\pm0.005$      & (15) &                        &      & $0.23\pm0.02$          & $0.39\pm0.02$ & O\\[2pt]
Reticulum & $-1.66\pm0.12$          & (13) &  $GC\pm2$                 & (13)  &                      &      & $0.07\pm0.02$          & (13) & $0.18\pm0.05$          & $0.39\pm0.02$ & O\\[3pt]
$SMC$\\[3pt]
NGC121    & $-1.71\pm0.10$          & (16) &  $10.6\pm0.7$             & (16)  & $0.05\ \pm0.03$      & (16) &                        &      & $0.16\pm0.09$          & $0.18\pm0.01$ &  \\[5pt]
\multicolumn{5}{l}{\bf Cluster with 2 Gyr $\le$ Ages $<$ 10 Gyr}\\[3pt]
$LMC$\\[3pt]
NGC1651   & $-0.53\pm0.03$          & ( 9) &  $2.00\pm0.15$            & (12)  & \ \ $0.11\pm0.01$    & (12) &                        &      & $0.34\pm0.03$          & $0.35\pm0.05$ &  \\[2pt]
NGC1718   & $-0.80\pm0.03$          & ( 9) &  $2.05\pm0.15$            & (12)  & \ \ $0.10\pm0.03$    & (12) &                        &      & $0.31\pm0.09$          & $0.51\pm0.06$ &  \\[2pt]
NGC2121   & $-0.50\pm0.03$          & ( 9) &  $2.90\pm0.50$            & (12)  & \ \  $0.07\pm0.02$   & (12) &                        &      & $0.22\pm0.06$          & $0.53\pm0.04$ &  \\[2pt]
NGC2155   & $-0.50\pm0.05$          & ( 9) &  $3.00\pm0.25$            & (12)  & \ \  $0.02\pm0.01$   & (12) &                        &      & $0.06\pm0.03$          & $0.43\pm0.04$ &  \\[2pt]
NGC2193   & $-0.49\pm0.05$          & ( 9) &  $2.2\ \pm 0.5$           & (20)  &                      &      &                        &      &                        & $0.39\pm0.02$ &  \\[2pt]
SL663	  & $-0.54\pm0.05$          & ( 9) &  $3.15\pm0.40$            & (12)  & \ \  $0.07\pm0.02$   & (12) &                        &      & $0.22\pm0.06$          & $0.38\pm0.04$ &  \\[2pt]
SL842     & $-0.36\pm0.20$          & (19) &  $2.0        $            & ( 8)  &                      &      &                        &      &                        & $0.39\pm0.02$ &  \\[2pt]
Hodge4    & $-0.55\pm0.06$          & ( 9) &  $2.1\ \pm0.3$            & (23)  &                      &      &                        &      &                        & $0.39\pm0.02$ &  \\[2pt]
Hodge14   & $-0.45\pm0.10$          & (12) &  $2.25\pm0.15$            & (12)  & \ \  $0.08\pm0.02$   & (12) &                        &      & $0.25\pm0.06$          & $0.39\pm0.02$ &  \\[2pt]
ESO121-03 & $-0.97\pm0.10$          & (12) &  $8.3 - 9.8$              & (12)  &                      &      & $0.04\pm0.02$          & (13) & $0.10\pm0.05$          & $0.39\pm0.02$ &  \\[3pt]
$SMC$\\[3pt]
NGC339    & $-1.50\pm0.14$          & (16) &  $6.3\pm1.3$              & (16)  & \ \  $0.03\pm0.04$   & (16) &                        &      & $0.09\pm0.12$          & $0.18\pm0.01$ &  \\[2pt]
NGC361    & $-1.45\pm0.11$          & (16) &  $8.1\pm1.2$              & (16)  & \ \  $0.07\pm0.03$   & (16) &                        &      & $0.22\pm0.09$          & $0.17\pm0.02$ &  \\[2pt]
NGC416    & $-1.44\pm0.12$          & (15) &  $6.6\pm0.5$              & (15)  & \ \ $0.08\pm0.03 $   & (15) &                        &      & $0.25\pm0.09$          & $0.20\pm0.02$ &  \\[2pt]
Kron3     & $-1.12\pm0.12$          & ( 6) &  $6.0\pm1.3$              & (16)  & $-0.03\pm0.02^{*}$   & (16) &                        &      & $0.00\pm0.06$          & $0.18\pm0.01$ &  \\[2pt]
Lindsay1  & $-1.17\pm0.10$          & ( 6) &  $9.0\pm1.0$              & (16)  & \ \  $0.06\pm0.02$   & (16) &                        &      & $0.19\pm0.06$          & $0.18\pm0.01$ &  \\[2pt]
Lindsay113& $-1.44\pm0.16$          & ( 6) &  $5.3\pm1.3$              & (16)  &  \ \ $0.00\pm0.02$   & (16) &                        &      & $0.00\pm0.06$          & $0.18\pm0.01$ &  \\[5pt]
\multicolumn{5}{l}{\bf Cluster with 1 Gyr $\le$ Ages $<$ 2 Gyr}\\[3pt]
$LMC$\\[3pt]
NGC1644   & $-0.3        $          & ( *) & $1.45^{+0.46}_{-0.35} $   & ( *)  &                      &      &                        &      &                        & $0.39\pm0.02$ &  \\[2pt]
NGC1777   & $-0.60\pm0.10$          & (12) & $1.15\pm0.15$             & (12)  & $0.10\pm0.03$        & (12) &                        &      & $0.31\pm0.09$          & $0.39\pm0.02$ &  \\[2pt]
NGC1783   & $-0.3        $          & ( *) & $1.45^{+0.46}_{-0.35} $   & ( *)  &                      &      &                        &      &                        & $0.30\pm0.03$ &  \\[2pt]
NGC1868   & $-0.70\pm0.10$          & (12) & $1.10\pm0.10$             & (12)  & $0.04\pm0.01$        & (12) &                        &      & $0.12\pm0.03$          & $0.39\pm0.02$ &  \\[2pt]
NGC1978   & $-0.38\pm0.07$          & ( 7) & $1.9\ \pm0.1$             & (17)  &                      &      &                        &      &                        & $0.76\pm0.05$ &  \\[2pt]
NGC1987   & $-0.3        $          & ( *) & $1.02^{+0.33}_{-0.23} $   & ( *)  &                      &      &                        &      &                        & $0.28\pm0.03$ &  \\[2pt]

NGC2108   & $-0.3        $          & ( *) & $1.23^{+0.39}_{-0.30} $   & ( *)  &                      &      &                        &      &                        & $0.50\pm0.05$ &  \\[2pt]
NGC2154   & $-0.4        $          & ( *) & $1.70^{+0.54}_{-0.41} $   & ( *)  &                      &      &                        &      &                        & $0.39\pm0.03$ &  \\[2pt]
NGC2162   & $-0.46\pm0.07$          & ( 9) & $1.25\pm0.10$             & (12)  & $0.03\pm0.02$        & (12) &                        &      & $0.09\pm0.06$          & $0.39\pm0.02$ &  \\[2pt]
NGC2173   & $-0.42\pm0.03$          & ( 9) & $1.60\pm0.20$             & (12)  & $0.07\pm0.02$        & (12) &                        &      & $0.22\pm0.06$          & $0.39\pm0.02$ &  \\[2pt]

\hline
\end{tabular}

{\it Notes to Table~\ref{tab:all}:} See next page.

\end{minipage}
\end{table*}

\setcounter{table}{7}
\begin{table*}
\centering
\begin{minipage}{170mm}
\caption{Continued}
\label{tab:all}
\begin{tabular}{@{}llclclcccccc@{}}
\hline
Cluster ID & \ \ \ \ [Fe/H] & & \ \ \ Age & & $E(B-V)$ & & $E(V-I)$ & & $A_{V\rmn{(CMD)}}$ & $A_{V\rmn{(MCPS)}}$ & Note\\
(1) & \ \ \ \ \ \ (2) & & \ \ \ \ (3) & & \ \ \ \ \ (4) & & (5) & & (6) & (7) & (8)\\
\hline
\multicolumn{5}{l}{\bf Cluster with 1 Gyr $\le$ Ages $<$ 2 Gyr}\\[3pt]
$LMC$\\[3pt]
NGC2190   & $-0.12\pm0.20$          & (19) & $1.1          $           & ( 8)  & $           $        &      &                        &      & $           $          & $0.39\pm0.02$ &  \\[2pt]
NGC2203   & $-0.41\pm0.03$          & ( 9) & $1.8          $           & ( 8)  & $           $        &      &                        &      & $           $          & $0.39\pm0.02$ &  \\[3pt]
NGC2209   & $-0.50\pm0.10$          & (12) & $1.20\pm0.10$             & (12)  & $0.15\pm0.03$        & (12) &                        &      & $0.47\pm0.09$          & $0.39\pm0.02$ &  \\[2pt]
NGC2213   & $-0.52\pm0.04$          & ( 9) & $1.70\pm0.20$             & (12)  & $0.06\pm0.03$        & (12) &                        &      & $0.19\pm0.09$          & $0.40\pm0.04$ &  \\[2pt]
NGC2231   & $-0.52\pm0.04$          & ( 9) & $1.5          $           & ( 8)  & $           $        &      &                        &      & $           $          & $0.39\pm0.02$ &  \\[2pt]
NGC2249   & $-0.45\pm0.10$          & (12) & $1.00\pm0.10$             & (12)  & $0.01\pm0.02$        & (12) &                        &      & $0.03\pm0.06$          & $0.39\pm0.02$ &  \\[3pt]
$SMC$\\[3pt]
NGC152    & $-0.94\pm0.15$          & ( 4) & $1.4\pm0.2$               & ( 4)  & $0.05\pm0.01$        & ( 4) &                        &      & $0.16\pm0.03$          & $0.19\pm0.02$ &  \\[2pt]
NGC411    & $-0.68\pm0.07$          & ( 1) & $1.4\pm0.2$               & ( 1)  & $0.12\pm0.01$        & ( 4) &                        &      & $0.37\pm0.03$          & $0.17\pm0.02$ &  \\[5pt]
\multicolumn{5}{l}{\bf Cluster with 0.2 Gyr $\le$ Ages $<$ 1 Gyr}\\[3pt]
$LMC$\\[3pt]
NGC1831   & $-0.10\pm0.10$          & (12) & $0.71^{+0.09}_{-0.08} $   & (12)  & $0.11\pm0.01$        & (12) &                        &      & $0.34\pm0.03$          & $0.39\pm0.02$ &  \\[2pt]
NGC1856   & $-0.40\pm0.10$          & (12) & $0.30\pm0.03          $   & (12)  & $0.21\pm0.02$        & (12) &                        &      & $0.65\pm0.06$          & $0.22\pm0.03$ &  \\[2pt]
NGC1866   & $-0.50\pm0.10$          & (10) & $0.33^{+0.09}_{-0.07} $   & ( *)  &                      &      &                        &      &                        & $0.28\pm0.06$ &  \\[2pt]
NGC2031   & $-0.52\pm0.21$          & ( 5) & $0.23^{+0.07}_{-0.06} $   & ( *)  &                      &      &                        &      &                        & $0.40\pm0.03$ &  \\[2pt]
NGC2107   & $-0.2        $          & ( *) & $0.62^{+0.19}_{-0.15} $   & ( *)  &                      &      &                        &      &                        & $0.36\pm0.04$ &  \\[2pt]
NGC2134   & $-0.2        $          & ( *) & $0.32^{+0.10}_{-0.08} $   & ( *)  &                      &      &                        &      &                        & $0.62\pm0.03$ &  \\[2pt]
NGC2156   & $-0.2        $          & ( *) & $0.32^{+0.10}_{-0.08} $   & ( *)  &                      &      &                        &      &                        & $0.20\pm0.02$ &  \\[3pt]
SMC\\[3pt]
NGC265    & $-0.62^{+0.23}_{-0.61}$ & ( 3) & $0.32^{+0.32}_{-0.16} $   & ( 3)  &                      &      & $0.06^{+0.05}_{-0.04}$ & ( 3) & $0.19^{+0.15}_{-0.13}$ & $0.34\pm0.02$ &  \\[3pt]
\hline
\end{tabular}

{\it Notes to Table~\ref{tab:all}:}
Column~(1) - cluster designation. Columns (2) and (3) are the metallicity and the age of the cluster with corresponding references given in parenthesis. Reddening information based on CMDs plus corresponding references is listed in columns~(4), (5) and (6). $A_V$ retrieved from MCPS database is presented in column~(7). Finally, notes are given in column~(8)\\[5pt]
In column~(8) I and O stands for the positions of the objects in the LMC. I means  inner (bar) and O outer clusters. There are slight variations in this classification from author to author. Here we classify the objects as in \cite{mg04}.\\[5pt]
\cite{mg04} concluded that NGC1928, NGC1939 and Reticulum are coeval in age with the oldest Milky Way globular clusters within 2 Gyr. In the table this result is denoted as "$GC \pm 2$".\\[5pt]
ESO121-SC03 is the only object in the LMC age-metallicity gap. It is younger than the genuine old globular clusters, but significantly older than the intermediate-age massive clusters from this galaxy. It'll be compared with the model predictions alongside with the objects from the next age bin.\\[5pt]
{\it References:} (1)\cite{as99} (2)\cite{brocato96} (3)\cite{chiosi07} (4)\cite{crowl2001} (5)\cite{dirsch2000} (6)\cite{dh98} (7)\cite{ferraro06} (8)\cite{geisler97} (9)\cite{grocholski06} (10)\cite{hill00} (11)\cite{johnson99} (12)\cite{kerber07} (13)\cite{mg04} (14)\cite{mpg2006} (15)\cite{mighell96} (16)\cite{mighell98b} (17)\cite{mucciarelli07} (18)\cite{olsen98} (19)\cite{olszewski91} (20)\cite{rich2001} (21)\cite{testa95} (22)\cite{walker92} (23)\cite{woo03}

\end{minipage}
\end{table*}

\setcounter{table}{8}
\begin{table*}
\centering
\begin{minipage}{120mm}
\caption{2MASS integrated-light photometry and mass estimates.}
\label{tab:ph}
\begin{tabular}{@{}lrrrrll@{}}
\hline
Cluster ID & D\  \ & $J$\ \ \ \ \ \ \  & $H$ \ \ \ \ \ & $K_S$\ \ \ \ \  \ & $log(m)_{\rmn{NIR}}$ & $log(m)_{\rmn{Lit}}$ \\
(1) & (2) & (3)\ \ \ \ \ \ & (4)\ \ \ \ \ \  & (5)\ \ \ \ \ \  &\ \ \ \ (6) & \ \ \ \ (7)\\
\hline
\multicolumn{5}{l}{\bf Old globular clusters (Ages $\ge$ 10 Gyr)}\\[3pt]
$LMC$\\[3pt]
NGC1466      &  60 & $ 9.83\pm0.02$ & $ 9.39\pm0.01$ & $ 9.30\pm0.01$ & $5.18\pm0.07         $ & $5.24\pm0.04         $ \\[2pt]
NGC1754      & 100 & $ 9.74\pm0.06$ & $ 9.24\pm0.04$ & $ 9.08\pm0.05$ & $5.22^{+0.06}_{-0.08}$ & $5.07\pm0.05         $ \\[2pt]
NGC1786      &  60 & $ 8.57\pm0.01$ & $ 8.09\pm0.01$ & $ 8.00\pm0.01$ & $5.70^{+0.14}_{-0.22}$ & $5.57\pm0.05         $ \\[2pt]
NGC1835      &  60 & $ 8.34\pm0.01$ & $ 7.82\pm0.01$ & $ 7.71\pm0.01$ & $5.77\pm0.08         $ & $5.71\pm0.05         $ \\[2pt]
NGC1841      &  25 & $12.23\pm0.02$ & $11.69\pm0.02$ & $11.48\pm0.02$ & $4.28^{+0.05}_{-0.06}$ & $5.31^{+0.07}_{-0.06}$ \\[2pt]
NGC1898      &  62 & $ 9.52\pm0.02$ & $ 9.18\pm0.03$ & $ 9.00\pm0.03$ & $5.30^{+0.08}_{-0.09}$ & $5.35\pm0.06         $ \\[2pt]
NGC1916      &  44 & $ 8.36\pm0.01$ & $ 7.91\pm0.01$ & $ 7.68\pm0.01$ & $5.79^{+0.04}_{-0.05}$ & $5.77\pm0.05         $ \\[2pt]
NGC1928      &  62 & $10.71\pm0.09$ & $10.27\pm0.12$ & $10.20\pm0.15$ & $4.83^{+0.07}_{-0.09}$ &                        \\[2pt]
NGC1939      &  38 & $10.34\pm0.02$ & $ 9.89\pm0.02$ & $ 9.83\pm0.02$ & $5.01^{+0.06}_{-0.07}$ &                        \\[2pt]
NGC2005      &  25 & $ 9.93\pm0.01$ & $ 9.39\pm0.01$ & $ 9.27\pm0.01$ & $5.15\pm0.06         $ & $5.27\pm0.05         $ \\[2pt]
NGC2019      &  60 & $ 9.10\pm0.02$ & $ 8.65\pm0.02$ & $ 8.54\pm0.02$ & $5.49^{+0.05}_{-0.07}$ & $5.47\pm0.05         $ \\[2pt]
NGC2210      &  62 & $ 9.25\pm0.01$ & $ 8.72\pm0.01$ & $ 8.66\pm0.01$ & $5.43^{+0.09}_{-0.11}$ & $5.40\pm0.05         $ \\[2pt]
NGC2257      &  61 & $10.98\pm0.02$ & $10.60\pm0.03$ & $10.50\pm0.03$ & $4.71^{+0.03}_{-0.04}$ & $5.00^{+0.12}_{-0.07}$ \\[2pt]
Hodge11      &  62 & $10.48\pm0.02$ & $ 9.88\pm0.02$ & $ 9.86\pm0.02$ & $4.93\pm0.03         $ & $5.17^{+0.07}_{-0.06}$ \\[3pt]
$SMC$\\[3pt]
NGC121       &  62 & $ 9.52\pm0.01$ & $ 8.90\pm0.01$ & $ 8.81\pm0.01$ & $5.45^{+0.15}_{-0.21}$ & $5.57\pm0.04         $ \\[5pt]
\multicolumn{5}{l}{\bf Cluster with 2 Gyr $\le$ Ages $<$ 10 Gyr}\\[3pt]
$LMC$\\[3pt]
NGC1651      & 100 & $10.00\pm0.02$ & $ 9.23\pm0.02$ & $ 9.10\pm0.02$ & $4.43\pm0.04         $ & $4.53^{+0.11}_{-0.09}$ \\[2pt]
NGC1718      &  62 & $10.01\pm0.01$ & $ 9.13\pm0.01$ & $ 8.94\pm0.01$ & $4.44^{+0.08}_{-0.10}$ & $4.57\pm0.22         $ \\[2pt]
NGC2121      &  62 & $10.40\pm0.03$ & $ 9.73\pm0.03$ & $ 9.20\pm0.02$ & $4.50^{+0.08}_{-0.10}$ & $5.00^{+0.08}_{-0.07}$ \\[2pt]
NGC2155      &  62 & $10.98\pm0.02$ & $10.31\pm0.11$ & $10.37\pm0.03$ & $4.25^{+0.13}_{-0.18}$ & $4.56^{+0.09}_{-0.08}$ \\[2pt]
NGC2193      &  38 & $12.01\pm0.05$ & $11.36\pm0.04$ & $11.27\pm0.04$ & $3.64^{+0.03}_{-0.04}$ & $4.13\pm0.08         $ \\[2pt]
SL$663^{*}$  &  60 & $11.22\pm0.02$ & $11.08\pm0.04$ & $11.02\pm0.04$ & $4.18^{+0.08}_{-0.10}$ & $4.67^{+2.49}_{-0.45}$ \\[2pt]
SL842        &  38 & $11.89\pm0.07$ & $11.14\pm0.05$ & $10.86\pm0.05$ & $3.67^{+0.03}_{-0.04}$ & $3.91\pm0.10         $ \\[2pt]
Hodge4       &  38 & $12.10\pm0.02$ & $11.78\pm0.06$ & $11.44\pm0.05$ & $3.60^{+0.03}_{-0.04}$ & $5.31^{+1.91}_{-0.45}$ \\[2pt]
Hodge14      &  62 & $12.05\pm0.03$ & $11.37\pm0.03$ & $11.47\pm0.04$ & $3.61^{+0.07}_{-0.08}$ & $4.00\pm0.09         $ \\[2pt]
ESO121-03    &  61 & $12.34\pm0.09$ & $11.59\pm0.07$ & $11.70\pm0.08$ & $4.02^{+0.14}_{-0.18}$ &                        \\[3pt]
$SMC$\\[3pt]
NGC339       &  62 & $11.06\pm0.02$ & $10.70\pm0.02$ & $10.43\pm0.02$ & $4.57^{+0.28}_{-0.99}$ & $4.90\pm0.07         $ \\[2pt]
NGC361       &  62 & $10.76\pm0.02$ & $ 9.97\pm0.01$ & $ 9.83\pm0.02$ & $4.78^{+0.11}_{-0.15}$ & $5.30^{+0.10}_{-0.08}$ \\[2pt]
NGC416       &  62 & $ 9.77\pm0.01$ & $ 9.16\pm0.01$ & $ 9.08\pm0.01$ & $5.10^{+0.10}_{-0.13}$ & $5.21\pm0.05         $ \\[2pt]
Kron3        &  62 & $10.31\pm0.01$ & $ 9.69\pm0.01$ & $ 9.68\pm0.01$ & $4.85\pm0.03         $ & $5.15\pm0.06         $ \\[2pt]
Lindsay1     &  62 & $11.64\pm0.03$ & $11.26\pm0.03$ & $11.14\pm0.04$ & $4.45^{+0.09}_{-0.11}$ &                        \\[2pt]
Lindsay113   &  62 & $11.46\pm0.02$ & $10.49\pm0.01$ & $10.32\pm0.01$ & $4.35\pm0.03         $ &                        \\[5pt]
\multicolumn{5}{l}{\bf Cluster with 1 Gyr $\le$ Ages $<$ 2 Gyr}\\[3pt]
$LMC$\\[3pt]
NGC1644      &  60 & $11.31\pm0.11$ & $10.88\pm0.10$ & $10.76\pm0.10$ & $3.96^{+0.03}_{-0.04}$ &                        \\[2pt]
NGC1777      &  38 & $ 9.06\pm0.01$ & $ 8.59\pm0.01$ & $ 8.46\pm0.01$ & $4.52^{+0.08}_{-0.10}$ & $4.28\pm0.11         $ \\[2pt]
NGC1783      &  60 & $ 9.23\pm0.01$ & $ 8.62\pm0.01$ & $ 8.52\pm0.01$ & $4.77^{+0.04}_{-0.05}$ &                        \\[2pt]
NGC1868      &  62 & $10.19\pm0.01$ & $ 9.71\pm0.01$ & $ 9.56\pm0.01$ & $4.10^{+0.07}_{-0.09}$ & $4.33\pm0.18         $ \\[2pt]
NGC1978      &  60 & $ 8.74\pm0.02$ & $ 8.04\pm0.01$ & $ 7.81\pm0.02$ & $4.97\pm0.04$          &                        \\[2pt]
NGC1987      &  60 & $10.16\pm0.02$ & $ 9.48\pm0.01$ & $ 9.04\pm0.01$ & $4.21^{+0.04}_{-0.05}$ &                        \\[2pt]
NGC2108      &  62 & $10.46\pm0.02$ & $ 9.70\pm0.02$ & $ 9.25\pm0.02$ & $4.22^{+0.04}_{-0.05}$ &                        \\[2pt]
NGC2154      &  62 & $10.14\pm0.02$ & $ 9.36\pm0.01$ & $ 8.90\pm0.01$ & $4.30\pm0.04$          &                        \\[2pt]
NGC2162      &  62 & $10.89\pm0.04$ & $10.21\pm0.03$ & $10.01\pm0.03$ & $3.74^{+0.16}_{-0.26}$ & $4.02\pm0.15         $ \\[2pt]
NGC2173      & 150 & $ 9.76\pm0.10$ & $ 9.10\pm0.07$ & $ 8.91\pm0.07$ & $4.38^{+0.07}_{-0.09}$ & $4.70\pm0.07         $ \\[3pt]
\hline
\end{tabular}

{\it Notes to Table~\ref{tab:ph}:} See next page.
\end{minipage}
\end{table*}

\setcounter{table}{8}
\begin{table*}
\centering
\begin{minipage}{120mm}
\caption{2MASS integrated-light photometry and mass estimates.}
\label{tab:ph}
\begin{tabular}{@{}lrrrrll@{}}
\hline
Cluster ID & D\  \ & $J$\ \ \ \ \ \ \  & $H$ \ \ \ \ \ & $K_S$\ \ \ \ \  \ & $log(m)_{\rmn{NIR}}$ & $log(m)_{\rmn{Lit}}$ \\
(1) & (2) & (3)\ \ \ \ \ \ & (4)\ \ \ \ \ \  & (5)\ \ \ \ \ \  &\ \ \ \ (6) & \ \ \ \ (7)\\
\hline
\multicolumn{5}{l}{\bf Cluster with 1 Gyr $\le$ Ages $<$ 2 Gyr}\\[3pt]
$LMC$\\[3pt]
NGC2190      &  61 & $11.42\pm0.04$ & $10.79\pm0.04$ & $10.57\pm0.04$ & $3.64^{+0.03}_{-0.04}$ &                        \\[2pt]
NGC2203      & 150 & $ 9.28\pm0.07$ & $ 8.61\pm0.05$ & $ 8.40\pm0.05$ & $4.42^{+0.03}_{-0.04}$ &                        \\[2pt]
NGC2209      &  62 & $10.77\pm0.04$ & $ 9.90\pm0.04$ & $ 9.38\pm0.03$ & $3.84^{+0.06}_{-0.07}$ & $4.36^{+2.59}_{-0.31}$ \\[2pt]
NGC2213      &  62 & $10.40\pm0.02$ & $ 9.55\pm0.01$ & $ 9.25\pm0.01$ & $4.13^{+0.12}_{-0.17}$ & $4.30\pm0.10         $ \\[2pt]
NGC2231      &  44 & $11.33\pm0.08$ & $10.51\pm0.05$ & $10.19\pm0.04$ & $3.78^{+0.03}_{-0.04}$ & $4.36\pm0.12         $ \\[2pt]
NGC2249      &  36 & $11.06\pm0.05$ & $10.57\pm0.05$ & $10.25\pm0.04$ & $3.67^{+0.37}_{-....}$ & $4.03\pm0.20         $ \\[3pt]
$SMC$\\[3pt]
NGC152       &  62 & $10.78\pm0.02$ & $ 9.95\pm0.01$ & $ 9.62\pm0.01$ & $4.17^{+0.06}_{-0.07}$ & $4.56\pm0.09         $ \\[2pt]
NGC411       &  62 & $10.49\pm0.03$ & $ 9.84\pm0.03$ & $ 9.61\pm0.03$ & $4.28\pm0.04         $ & $4.47\pm0.10         $ \\[5pt]
\multicolumn{5}{l}{\bf Cluster with 0.2 Gyr $\le$ Ages $<$ 1 Gyr}\\[3pt]
$LMC$\\[3pt]
NGC1831      &  60 & $ 9.86\pm0.01$ & $ 9.34\pm0.01$ & $ 9.15\pm0.01$ & $4.16\pm0.04         $ & $4.59\pm0.18         $ \\[2pt]
NGC1856      &  60 & $ 8.98\pm0.02$ & $ 8.59\pm0.02$ & $ 8.44\pm0.02$ & $4.26\pm0.04         $ & $4.88\pm0.12         $ \\[2pt]
NGC1866      &  60 & $ 8.72\pm0.01$ & $ 8.28\pm0.01$ & $ 8.11\pm0.01$ & $4.34^{+0.07}_{-0.08}$ & $4.91\pm0.12         $ \\[2pt]
NGC2031      &  72 & $ 8.95\pm0.03$ & $ 8.28\pm0.03$ & $ 8.16\pm0.04$ & $4.45\pm0.04         $ & $4.48^{+0.06}_{-0.05}$ \\[2pt]
NGC2107      &  60 & $10.05\pm0.13$ & $ 9.43\pm0.08$ & $ 9.20\pm0.11$ & $4.01^{+0.04}_{-0.05}$ &                        \\[2pt]
NGC2134      &  60 & $ 9.94\pm0.03$ & $ 9.50\pm0.02$ & $ 9.43\pm0.02$ & $3.98^{+0.03}_{-0.04}$ &                        \\[2pt]
NGC2156      &  72 & $10.81\pm0.17$ & $10.43\pm0.22$ & $10.43\pm0.26$ & $3.59^{+0.04}_{-0.05}$ & $3.65\pm0.08         $ \\[2pt]
SMC\\
NGC265       &  62 & $10.90\pm0.13$ & $ 9.88\pm0.09$ & $ 9.76\pm0.12$ & $3.41^{+0.18}_{-0.30}$ &                        \\[3pt]

\hline
\end{tabular}

{\it Notes to Table~\ref{tab:ph}:} Column~(1) is the cluster
identification. Column~(2) gives the diameter of the used aperture (to
match the optical photometry). $J,H$ and $K_S$ integrated magnitudes
with corresponding errors are listed in columns~(3) through
(5). Column~(6) presents an estimate of the stellar mass which contributes
to the measured integrated colours.  These are typically lower than
the total cluster mass.  The mass estimates are based on the
cluster age, metallicity, observed J band magnitude, and the
model predictions of \cite{maraston05}. Column~(7) lists the total
 mass estimates of the objects in common between present study and
\cite{mclvdm05}.

* No integrated-light optical photometry was recovered from the literature for SL663. NIR measurements are presented for aperture diameter of 60\arcsec.

\end{minipage}
\end{table*}

\cite{olszewski96} listed 14 LMC clusters believed to be as old as the
oldest Galactic globulars.  More recent studies have established that
ESO 121-SC03 should be excluded from their original list, since it is
somewhat younger at 8.3-9.8 Gyr \citep{mpg2006}, (given that the typical age
of a Galactic globular cluster exceeds 10 Gyr).  On the other hand the
two "suspected" old globulars, NGC~1928 and NGC~1939, were confirmed
(see references below), giving a total of 15 ancient globular clusters
in the LMC.  NGC~121 is the only known old globular cluster in the SMC
\citep{mighell98b}.  The adopted age and metallicity for each cluster,
with associated references, are listed in Table~\ref{tab:all}.  The
table is supplemented with extinction information based on individual
cluster CMDs and values retrieved from MCPS \cite{zaritsky97}. The
estimated stellar mass which contributes to the integrated-light
measurements for each cluster (based on our 2MASS $J$ band
integrated-light photometry and mass-to-light ratios from
\cite{maraston05}) are listed in Column~7 of Table~\ref{tab:ph}, and
are lower than the total cluster masses.
Total cluster mass estimates from \cite{mclvdm05} are also given.

We believe that the following
ten old clusters are reliable test particles: 4 inner LMC objects
(NGC1786, NGC1835, NGC1898, NGC2019); one
outer LMC cluster (NGC2210), and NGC121 in the SMC. The
other three outer objects (NGC1841, NGC2257, Hodge11) have integrated
light measurements taken in apertures (set by the optical dataset)
which sample only a fraction of the total stellar population in each
cluster.   The stellar mass which contributes to the
integrated light measurements are lower than $
{\cal{M}}_{\rm (10\%)}$, and therefore these
clusters may have a larger spread in colour-colour space.

Here, we provide more details on the available cluster age and
metallicity information.  \cite{olsen98} studied NGC 1754, 1835, 1898,
1916, 2005 and 2019. Their F555W and F814W WFPC2 CMDs reach an
apparent magnitude $ V\approx25$, well below the MSTO point. The
metallicities were derived using the technique of \citep[][; hereafter
S94]{sarajedini94}. In most cases there is good agreement between the
photometric results of \cite{olsen98} and the spectroscopy of
individual cluster members of \cite{olszewski91}. For the clusters
showing large discrepancies (NGC 2005 and NGC 2019) the metallicities
in \citeauthor{olszewski91} are significantly lower, although
\citeauthor{olszewski91} do note that the measurements for these
clusters are uncertain.  \cite{grocholski06} recently estimated [Fe/H]
= $-1.31\pm0.05$ for NGC 2019 based on moderate resolution Ca II
Triplet spectroscopy of 5 cluster stars, in good agreement with the
result of \citeauthor{olsen98} ([Fe/H]=$-1.23\pm0.15$).  Age estimates
relative to those of MW GCs with similar metallicity were measured
according to the method of \cite{vdbd1990}.

Another three old LMC clusters (NGC1466, NGC2257, and Hodge11) were
observed with WFPC2 F555W and F814W filters by \cite{johnson99}. In
this case the authors did not attempt to derive their own metallicity
estimates, but adopted already published values instead.

\cite{mg04} published ACS Wide Field Channel (WFC) F555W and F814W
photometry for NGC 1928, 1939 (clusters located in the LMC bar
region), and for the remote outer cluster Reticulum. Employing the S94
method, \citeauthor{mg04} derived metallicities that are consistent
with earlier measurements. \citeauthor{grocholski06} also published
[Fe/H]=$-1.57\pm0.03$ for Reticulum, which is in very good agreement
with [Fe/H]=$-1.66\pm0.12$ given by \citeauthor{mg04}. The relative
ages of the three clusters with respect to MW clusters with a similar
metallicity were derived using the techniques of
\cite{vdbd1990}.

\label{test1}

%%%%%%%%%%%%%%%%%%%%%%%%%%%%%%%%%%%%%%%%%%%%%%%%%%%
\subsection{Cluster in the 2 Gyr $\le$ Age $<$ 10 Gyr interval} 

\setcounter{figure}{17}
\begin{figure*}
\centering
\begin{minipage}{180mm}
\includegraphics[bb=14 14 605 452,width=18.0cm]{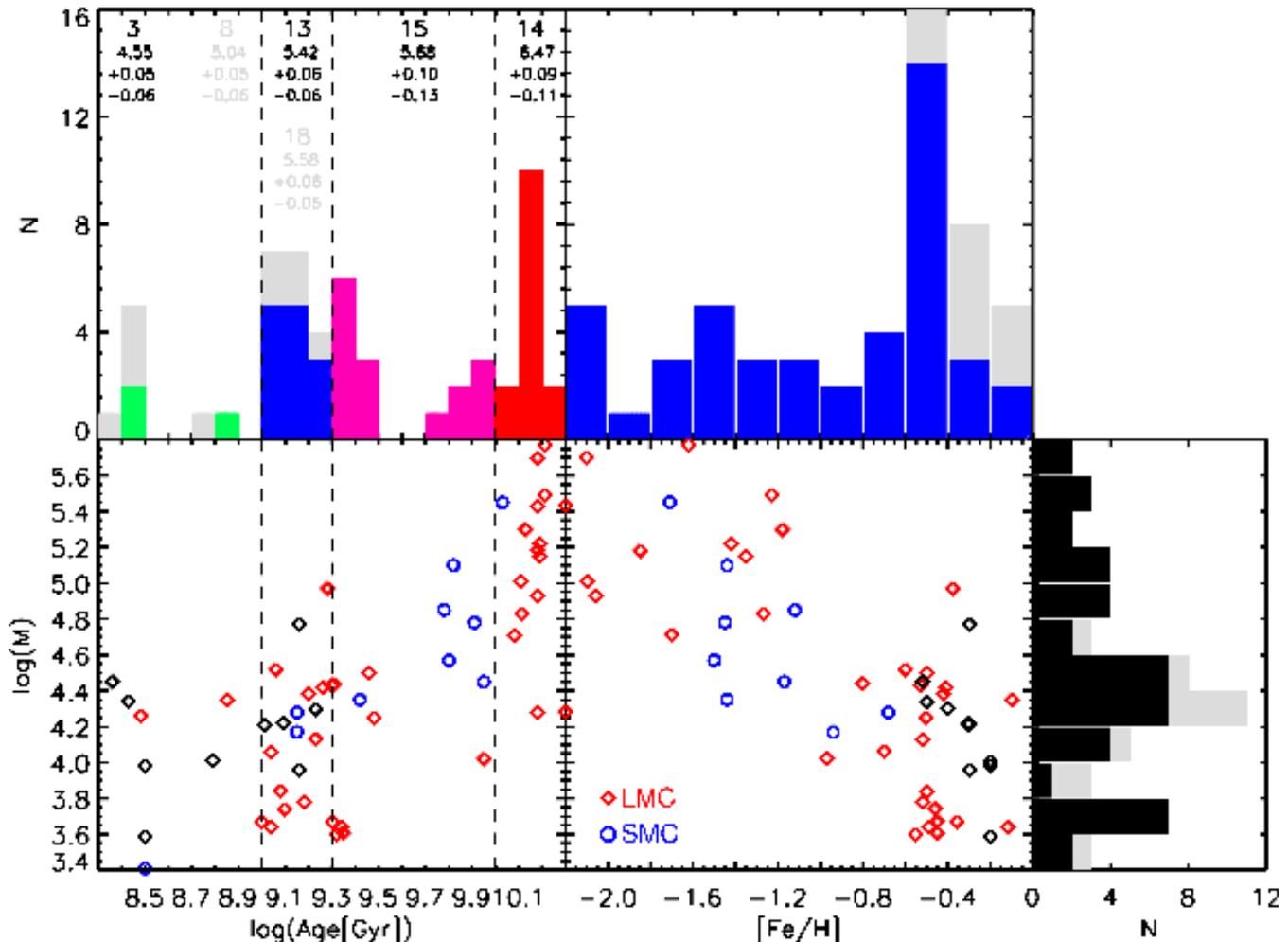}
\caption{Global properties of our test sample. The cumulative mass and number of objects in each age bin is shown on the age histogram. The objects from the S parameter extension are presented in black and their contribution to the histograms is shown in gray. Also in gray are given the number of objects and total mass for the age bins when the extended sample is taken into account.} 
\end{minipage}
\label{fig:sample}
\end{figure*}

A survey of the literature revealed 15 SMC/LMC clusters in this age
range. Their ages and metallicities alongside with the corresponding
references are listed in Table~\ref{tab:all}. Most of the LMC cluster
metallicity estimates comes from \cite{grocholski06}. Their results
are based on CaT spectroscopy of multiple stars in each cluster and
generally are in a good agreement with earlier studies
\citep[e.g.][]{olszewski91}. Cluster metallicities based on
spectroscopic measurements also agree with the CMD-based metallicity
estimates of \cite{kerber07}.  The latter work is the source of the
age and extinction estimates for the majority of the LMC objects.

All six of the SMC clusters included in the $2-10$~Gyr bin fall within
the age range between $5-8$~Gyr; no LMC clusters with these ages are known.
Most of the information about
these SMC clusters was retrieved from the papers of
\cite{mighell98a,mighell98b}, based on deep {\it HST} WFPC2
observations. Metallicities and reddening for the objects were derived
applying the S94 method. The age estimates listed in Table~\ref{tab:all}
are assuming an age of 9.0 Gyr for Lindsay1. Independent estimates of
\cite{alcaino03} confirm this value. Spectroscopic metallicities for
three SMC clusters were recovered from \cite{dh98}.

Integrated-light NIR photometry and mass estimates for the (2 Gyr
$\le$ Age $<$ 10 Gyr) sub-sample are presented in
Table~\ref{tab:ph}. We should note that the masses of the LMC objects
are in the order of $10^4{\cal{M}}_{\odot}$, therefore a relatively
large spread relative to the model predictions is not unexpected. Only
NGC416 has an estimated mass $>10^5{\cal{M}}_{\odot}$.

\label{test2}
%%%%%%%%%%%%%%%%%%%%%%%%%%%%%%%%%%%%%%%%%%%%%%%%%%%%%%%%%%%%%%%%%%%%%%
\subsection{Clusters in the 1 Gyr $\le$ Age $<$ 2 Gyr interval} As in
the previous section, the bulk of the LMC cluster properties in this
age range come from \cite{grocholski06} and \cite{kerber07}. The
adopted values are presented in Table~\ref{tab:all}. Occasionally we
supplement age information from \cite{geisler97} and
metallicity information from \cite{olszewski91}.  
The clusters in this age range have sampled stellar masses
around $10^4{\cal{M}}_{\odot}$, and again some spread is expected in
the individual data points. Only NGC~1978 is close to
$10^5{\cal{M}}_{\odot}$.

Some details on individual clusters in this age interval are provided
below.  \cite{ferraro06} showed that despite its large observed
ellipticity and suspected metallicity spread \citep{alcaino99,hill00},
NGC 1978 is not the product of merged clusters. They derived
metallicities for 11 gaint stars cluster in NGC~1978 from
 high-resolution UVES/FLAMES VLT spectroscopy. No
significant variations in the giant's metal abundance were found
(resulting mean [Fe/H] is listed in Table~\ref{tab:all}). We conclude
that NGC 1978 can be used as a test particle in our analysis. The
age of this object was derived by \cite{mucciarelli07} applying fit of
theoretical isochrones to {\it HST} ACS data.  Information for the two
SMC objects is retrieved from the works of \cite{as99} and
\cite{crowl2001}. Five additional objects were added using the calibration presented in Appendix~A.
\label{test3}

%%%%%%%%%%%%%%%%%%%%%%%%%%%%%%%%%%%%%%%%%%%%%%%%%%%%%%%%%%%%%%%%%%%%%%
\subsection{Clusters in the 200 Myr $\le$ Age $<$ 1 Gyr interval}
Based a search of the literature, we only identified three clusters in
the 200~Myr$-$1~Gyr age range, with age estimates based on deep CMDs.
In order to expand our test sample, we use the S-parameter
recalibration from Appendix A.  
information.  Based on equation A2, five more clusters were added to
this age bin. The properties of the sample are also summarized in
Table~\ref{tab:all}.

There are age and metallicity estimates for some of the objects from
this sub-sample available in the literature. \cite{dirsch2000}
presented data for six LMC clusters based on CCD Str{\" o}mgren
photometry. We adopt their metallicity estimate for NGC2031.

Information for several more clusters in the 200 Myr - 1 Gyr age
interval is available in the recent study of \cite{wolf07} The
authors use Bruzual-Charlot high-resolution stellar population
synthesis models to fit the SEDs and simultaneously estimate ages and metallicities
of globular clusters in the Magellanic Clouds and M31. Their age
results are in a good agreement with our S parameter ages.
\label{test4}

%%%%%%%%%%%%%%%%%%%%%%%%%%%%%%%%%%%%%%%%%%%%%%%%%%%%%%%%%%%
\section{Transformation of Model Colours to the 2MASS System}

\setcounter{figure}{18}
\begin{figure} 
\includegraphics[bb=14 14 256 328,width=8.4cm]{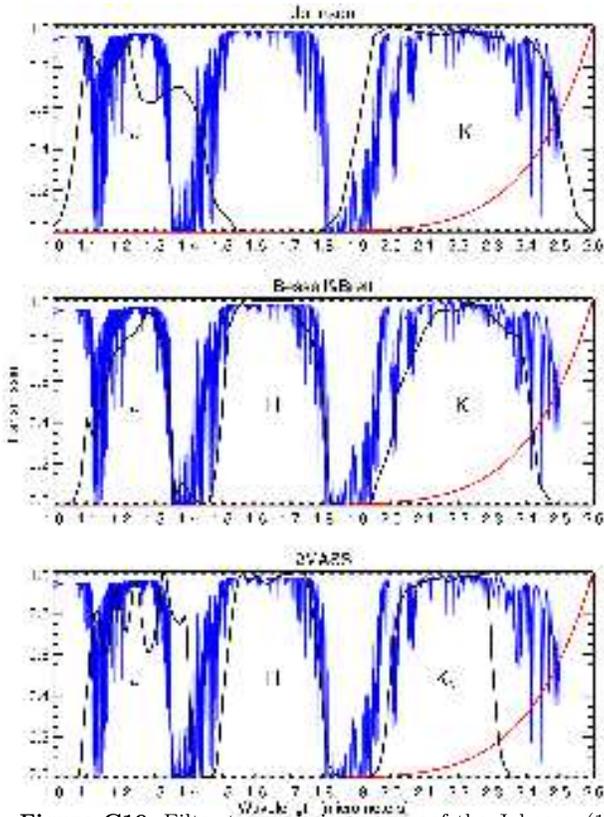}
\caption{Filter transmission curves of the Johnson (1965), Bessell \&
Brett (1988) and 2MASS photometric systems. The atmospheric
transmission for the south 2MASS facility (CTIO), is plotted in
blue. Red represents the thermal emission of the atmosphere at
$20\degr$C (scaled to reach unity at $2.6\mu$m for illustrational
purposes). It is obvious that the "$K$-short" $(K_{S})$ filter
significantly reduces the influence of the thermal background compared
to the Johnson and Bessel \& Brett systems. 
The 2MASS $J$ is broader than the atmospheric window, and the transmission variability was accounted for during the extensive calibration observations.
The $H$ filter was
introduced to the Johnson 1965 system somewhat later (Johnson 1968) and the
transmission curve was never published. The NIR colours for the
majority of the SSP models use the
filter transmissions of Bessell \& Brett (1988).}
\label{fig:nir_sys}
\end{figure}

\setcounter{figure}{19}
\begin{figure} 
\includegraphics[bb=14 14 256 336,width=8.4cm]{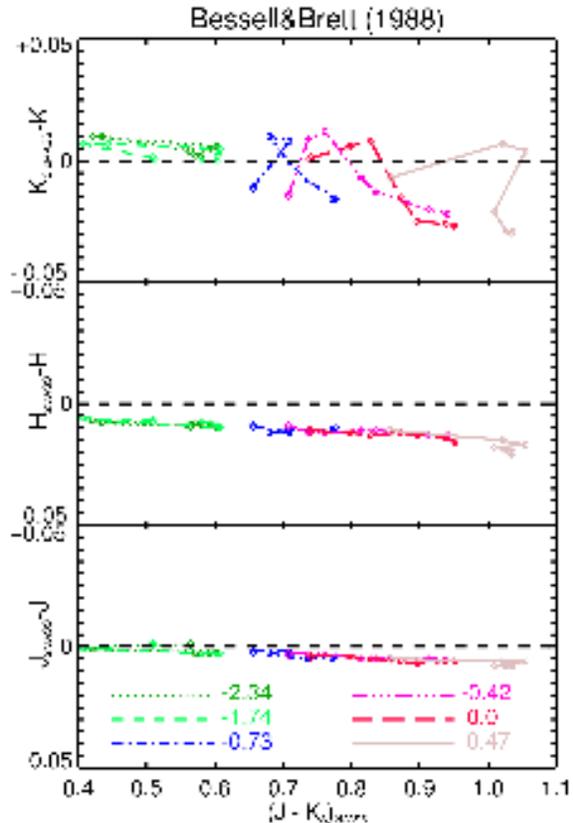}
\caption{The differences between 2MASS and Bessell \& Brett (1988) NIR
magnitudes as a function of ($J-K_{s})_{\rmn{2MASS}}$. The dependence of
the age and metallicity is also shown (diamonds represent age of 0.2
Gyr for each metallicity (colour coded) and the following circles on
the line correspond to 0.5, 1, 2, 5, 10, 15 Gyr respectively). Note
the scatter for $K_{s{\rmn{2MASS}}} - K$, illustrating the need of
independent fits for the different stellar populations.}
\label{fig:bb_trans}
\end{figure}

Today's SSP model predictions for NIR colours are provided in various
photometric systems. In order to avoid systematic offsets, observations
and model predictions should be compared in the same photometric system.

\cite{johnson65} defined a photometric system, in both the optical and
NIR, which is currently perhaps the most widely used. Unfortunately
however, the NIR passbands of \citeauthor{johnson65} are broader than
the atmospheric transmission windows. This can lead to substantial
variations in sky background levels (which can also vary on short
timescales). Hence there can be significant differences between
observations conducted in the original \cite{johnson65} filter set and
more recent NIR systems, which have been developed to fit within the
atmospheric windows and to decrease the thermal background at
longer wavelengths ($K$ band).  

\cite{bb88} examined the relations between the NIR photometric systems of
several different observatories and introduced a homogenized NIR system
based on the works of \cite{glass85} and \cite{johnson66}. The filter
transmission curves of the \cite{johnson65} and the \cite{bb88} systems
are shown in the top and middle panel of Figure~\ref{fig:nir_sys}. In the
bottom panel we plot the 2MASS filter system \citep{skrutskie2006}. The
estimated mean atmospheric transmission for CTIO taken from the online
2MASS All-sky Release Explanatory
Supplement\footnote{$\rmn{http://www.ipac.caltech.edu/2mass/releases/allsk
y/doc/}$ $\rmn{sec3\_1b1.tbl16.html}$} and the thermal emission of Earth's
atmosphere (blackbody with a temperature of $20\degr$C) are overplotted in
blue and red. Differences between the systems are clearly visible. The
2MASS photometric system by \citeauthor{skrutskie2006} appears to be least
affected by the Earth's atmosphere and by the thermal background. It is also
clear that observations in these different systems will result in
different NIR colours.

The 2MASS system seems to be a natural choice, given the extensive sky
coverage and the precise internal photometric calibration. 
Among the SSP models being considered here, only \cite{bc03}
provide 2MASS NIR colours. 
The other models use NIR passbands on the \citeauthor{bb88} photometric
system, although they are sometimes referred to as Johnson $JHK$
passbands. However it is clear that the system of \cite{bb88} has
different filter throughputs from \citeauthor{johnson65} (as illustrated
in Figure~\ref{fig:nir_sys}).  Given the advantages of the 2MASS
photometric system, we encourage the SSP model builders to provide
output in the 2MASS system. For the purposes of
the present work we transform the NIR colours predicted by the
models to the 2MASS system.  
To do so, we convolved spectral energy distributions (SEDs) provided by
\cite{bc03} with the filter transition curves from \cite{bb88} and
\citep{skrutskie2006} using the IRAF/STSDAS SYNPHOT package. Synthetic
NIR colours and magnitude differences were derived for a wide
age/metallicity parameter space. The 
differences between the synthetic 2MASS and \citeauthor{bb88} magnitudes
as a function of $(J-K_{S})_{\rmn{2MASS}}$ colour is presented in
Figure~\ref{fig:bb_trans}. It is obvious that different transformation
equations should be applied for the different populations as a function of
metallicity, especially for $K$ and $K_s$. We derived a set of
transformation equations adequate for our purposes:
\begin{equation}
J_{\rmn{2MASS}} = J_{\rmn{BB88}} + (a_{J}(J-K)_{\rmn{BB88}} + b_{J}) 
\end{equation}
\begin{equation}
H_{\rmn{2MASS}} = H_{\rmn{BB88}} + (a_{H}(J-K)_{\rmn{BB88}} + b_{H}) 
\end{equation}
\begin{equation}
K_{S \rmn{2MASS}} = K_{\rmn{BB88}} + (a_{K_{S}}(J-K)_{\rmn{BB88}} + b_{K_{S}}) 
\end{equation}

The transformation coefficients as function of metallicity are listed in
Table~\ref{tab:coef}. These transformations were used to recompute the
model colours. As an example the differences between the original and
the transformed model grids are shown in $(V-J)$ vs. $(J-K)$
colour-colour space in Figure~\ref{fig:modshift}. The grids for the 
\citet{bb88} {\it JHK\/} system used by the \citet{maraston05}, 
  \citet{af03}, and \citet{vazdekis99} models are shown in cyan and
  the corresponding grids using the 2MASS system are shown in
  black. We use the 2MASS system for the subsequent analysis.

\setcounter{figure}{20}
\begin{figure} 
\centering
\includegraphics[bb=14 14 256 718,width=6.4cm]{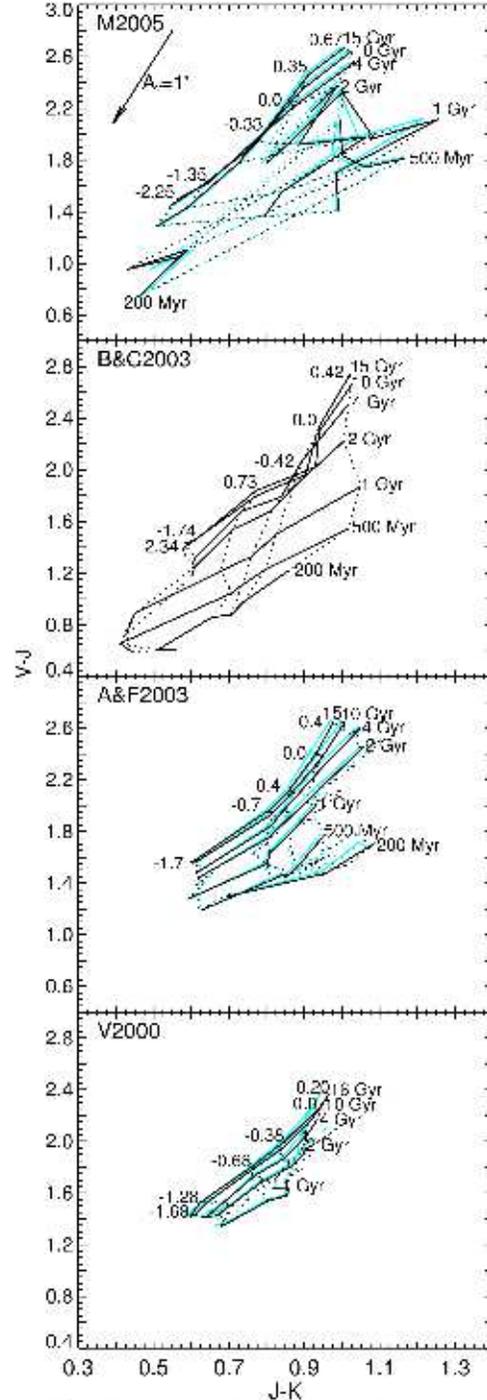}
\caption{Illustration of the variance of model grids computed in the
  photometric systems of Bessel \& Brett (1988; in cyan) and 2MASS (in
  black). $(V-J)$ vs.\ $(J-K)$ (resp. $(J-K_S)$ for 2MASS) for the
  models of Maraston (M2005), Anders \& Fritze (A\&F2003) and Vazdekis
  (V2000) are presented. The models of Bruzual \& Charlot (B\&C2003)
  were originally computed in the 2MASS photometric system. The ages
  are given on the right of the isochrones and the metallicities are
  labeled along the oldest isochrone for each model. A reddening
  vector corresponding to one magnitude of visual extinction is shown
  in the top panel.} 
\label{fig:modshift}
\end{figure}

\setcounter{table}{9}
\begin{table}
\centering
\caption{Transformation coefficients between 2MASS and Bessell \&
  Brett systems for Equations~(1), (2), and (3).} 
\label{tab:coef}
\begin{tabular}{@{}lllllll@{}}
\hline
[Fe/H] & $\ \ a_J$ & $\ \ b_J$ & $\ \ a_H$ & $\ \ b_H$ & $\ a_{K_S}$ & $\ b_{K_S}$ \\
\hline
-2.34 & -0.011& 0.005& -0.008& -0.004& -0.037&\ 0.026\\
-1.74 & -0.014& 0.006& -0.015&\ 0.000& -0.023&\ 0.016\\
-0.73 & -0.023& 0.012& -0.003& -0.009& -0.100&\ 0.067\\
-0.42 & -0.011& 0.005& -0.014& -0.001& -0.098&\ 0.072\\
\ 0.0   & -0.014& 0.007& -0.022&\ 0.006 & -0.198&\ 0.158\\
\ 0.47 & -0.008& 0.001& -0.033&\ 0.017 &\ 0.015 & -0.029\\
\hline
\end{tabular}
\end{table}
\label{transformation}

\bsp

\label{lastpage}

\end{document}